\documentclass[%
reprint,              
showpacs,
nofootinbib,
amsmath,amssymb,
aps,
pre,
]{revtex4-1}
\usepackage{graphicx,tikz,pgfplots,psfrag,subcaption}
\usepackage{dcolumn}
\usepackage{bm}
\usepackage{url,verbatim,filecontents}
\newcommand\localwidth{0.17\textwidth}
\newcommand\scalefactor{0.75}

\newcommand{\bC}{\mathbb{C}}
\newcommand{\spont}{\overline\varkappa}

\newcommand{\ds}{\;{\rm d}s} 
\newcommand{\dx}{\;{\rm d}x} 
\newcommand{\dd}[1]{\frac{\rm d}{{\rm d}#1}}
\newcommand{\ddt}{\dd{t}}

\newcommand{\nabs}{\nabla_{\!s}}
\newcommand{\matpartu}{\partial_t^\bullet}
\newcommand{\mat}[1]{\underline{\underline{#1}}\rule{0pt}{0pt}}
\newcommand{\id}{\rm id}
\newcommand{\Id}{\rm Id}
\renewcommand{\Re}{{\rm Re}}

\newcommand{\unitn}{\vec{\rm n}}
\newcommand{\TRc}{TR} 

\newcommand{\Ca}{{\rm Ca}}
\newcommand{\MoI}{{\rm M}}

\begin{document}

\title{Numerical computations of the dynamics of fluidic membranes and vesicles
}

\author{John W.\ Barrett}
\affiliation{Department of Mathematics, Imperial College London, 
London SW7 2AZ, UK}

\author{Harald Garcke}
\email{harald.garcke@ur.de}
\affiliation{Fakult\"at f\"ur Mathematik, Universit\"at Regensburg, 93040
 Regensburg, Germany}

\author{Robert N\"urnberg}
\affiliation{Department of Mathematics, Imperial College London, 
London SW7 2AZ, UK\rule{0pt}{0pt}}

\begin{abstract}
  Vesicles and many biological membranes are made of two monolayers of
  lipid molecules and form closed lipid bilayers. The dynamical
  behaviour of vesicles is very complex and a variety of forms and
  shapes appear. Lipid bilayers can be considered as a surface fluid
  and hence the governing equations for the evolution include the
  surface (Navier--)Stokes equations, which in particular take the membrane
  viscosity into account. The evolution is driven by forces stemming
  from the curvature elasticity of the membrane. In addition, the surface
  fluid equations are coupled to bulk \mbox{(Navier--)}Stokes equations.

  We introduce a parametric finite element method to solve this
  complex free boundary problem, and present the first three
  dimensional numerical computations based on the full (Navier--)Stokes
  system for several different scenarios. For example, the effects of the
  membrane viscosity, spontaneous curvature and area difference
  elasticity (ADE) are studied. In particular, it turns out, that even
  in the case of no viscosity contrast between the bulk fluids, the tank
  treading to tumbling transition can be obtained by increasing
  the membrane viscosity. Besides the classical tank treading and
  tumbling motions, another mode (called the transition mode in this paper,
  but   originally called the vacillating-breathing mode and subsequently 
  also called trembling, transition and swinging mode) 
  separating these classical modes appears and is studied by us numerically. 
  We also study 
  how features of equilibrium shapes in the ADE and spontaneous curvature 
  models, like budding behaviour or starfish forms, behave in a shear flow. 
\end{abstract}

 \pacs{87.16.dm, 87.16.ad, 87.16.dj}


\maketitle

\section{Introduction} 
Lipid membranes consist of a bilayer of molecules, which have
a hydrophilic head and two hydrophobic chains. These bilayers
typically spontaneously form closed bag-like structures, which
are called vesicles. It is observed that vesicles can attain
a huge variety of shapes and some of them are similar to the biconcave
shape of red blood cells. Since membranes play a fundamental role
in many living systems, the study of vesicles is a very active 
research field in different scientific disciplines, see
e.g.\ \cite{Seifert97,BaumgartHW03,NoguchiG05,McWhirterNG09}. It is
the goal of this paper to present a numerical approach to study the evolution of
lipid membranes. We present several computations showing quite
different shapes, and the influence of fluid flow on the membrane evolution.

Since the classical papers of \citet{Canham70} and
\citet{Helfrich73}, there has been a lot of work with the aim of describing
equilibrium membrane shapes with the help of elastic membrane
energies. \citet{Canham70} and \citet{Helfrich73}
introduced a bending energy for a non-flat membrane,
which is formulated with the help of the curvature
of the membrane. In the class of fixed topologies
the relevant energy density, in the simplest situation,
is proportional to the square of the
mean curvature $\varkappa$. The resulting energy functional is called the
Willmore energy. When computing equilibrium membrane shapes one has to take
constraints into account. Lipid membranes have a very small compressibility,
and hence can safely be modelled as locally incompressible. In addition,
the presence of certain molecules in the surrounding fluid, 
for which the membrane is
impermeable, leads to an osmotic pressure, which results in a constraint
for the volume enclosed by the membrane. The minimal
energetic model for lipid membranes consists of the Willmore
mean curvature functional together with enclosed volume and surface 
area constraints.
Already this simple model leads to quite different shapes including 
the biconcave red blood cell shapes, see \cite{Seifert97}. 

\citet{Helfrich73} introduced a variant of the Willmore energy,
with the aim of modelling a possible asymmetry of the bilayer membrane.
\citet{Helfrich73} studied the functional
$\int (\varkappa - \spont)^2$, where $\spont$ is a fixed constant,
the so-called spontaneous curvature. It is argued that
the origin of the spontaneous curvature is e.g.\ a different chemical
environment on both sides of the membrane. We refer
to \cite{MartensM08} and \cite{KamalMGH09} for a recent discussion,
and for experiments in situations which lead to spontaneous curvature effects
due to the chemical structure of the bilayer. 

Typically there is yet another asymmetry in the bilayer leading to
a signature in the membrane architecture. This results from the fact
that the two membrane layers have a different number of molecules.
Since the exchange of molecules between the layers is difficult,
an imbalance is conserved during a possible shape change. The
total area difference between the two layers is proportional to
$M=\int\varkappa$. Several models
have been proposed, which describe the difference in the total
number of molecules in the two layers with the help of the
integrated mean curvature. The bilayer coupling model, introduced by
Svetina and coworkers \cite{SvetinaOG82,SvetinaZ83,SvetinaZ89},
assumes that the area per lipid molecule is fixed and assumes that
there is no exchange of molecules between the two layers. Hence the
total areas of the two layers are fixed, and on assuming that the two layers are
separated by a fixed distance, one obtains, to the order of this distance,
that the area difference can be approximated by the integrated mean
curvature, see \cite{SvetinaOG82,SvetinaZ83,SvetinaZ89}. We note
that a spontaneous curvature contribution is irrelevant in the bilayer
coupling model as this would only add a constant to the energy as
the area and integrated mean curvature are fixed. 

\citet{MiaoSWD94} noted that in the bilayer coupling model
budding always occurs continuously which is inconsistent with
experiments. They hence studied a model in which the area of the two
layers are not fixed but can expand or compress under stress. Given a
relaxed initial area difference $\Delta A_0$, the total area
difference $\Delta A$, which is proportional to the integrated mean
curvature, can deviate from $\Delta A_0$. However, the total energy now
has a contribution that is proportional to $(\Delta A-\Delta A_0)^2$. 
This term describes the elastic area difference stretching
energy, see \cite{MiaoSWD94,Seifert97}, and hence one 
has to pay a price energetically to deviate from the relaxed area difference. 

It is also possible to combine the area difference elasticity model
(ADE-model) with a spontaneous curvature assumption, see 
\citet{MiaoSWD94} and \citet{Seifert97}. However, the
resulting energetical model is equivalent to an area difference elasticity 
model with a modified $\Delta A_0$, see \cite{Seifert97} for a more
detailed discussion.

It has been shown that the bilayer coupling model (BC-model) and the
area difference elasticity model (ADE-model) lead to a multitude of
shapes, which also have been observed in experiments with
vesicles. Beside others, the familiar discocyte shapes (including the ``shape''
of a red blood cell), stomatocyte shapes, prolate shapes and pear-like
shapes have been observed. In addition, the budding of membranes can
be described, as well as more exotic shapes, like starfish vesicles.
Moreover, higher genus shapes appear as global or local minima of the energies
discussed above. 
We refer to \cite{Seifert97,ZiherlS05,MiaoSWD94,%
SeifertBL91,WintzDS96} for more details on the possible shapes
appearing, when minimizing the energies in the ADE- and BC-models. 

Configurational changes of vesicles and membranes cannot be described
by energetical considerations alone, but have to be modelled with the
help of appropriate evolution laws. Several authors considered an
$L^2$--gradient flow dynamics of the curvature energies discussed
above. Pure Willmore flow has been studied in \cite{MayerS02,%
ClarenzDDRR04,Dziuk08,willmore,DuLRW05,FrankenRW13}, where the
last two papers use a phase field formulation of the Willmore
problem. Some authors also took other aspects, such
as constraints on volume and area \cite{willmore,BonitonP10,DuLW04}, 
as well as a constraint on the integrated mean
curvature \cite{willmore,DuLRW05}, into account. The effect of different
lipid components in an $L^2$--gradient flow approach of the curvature
energy has been studied in
\cite{ElliottS10,BartelsDNR12,MerckerM-CRH13,DuLRW09,BoedecLJ11,%
RahimiA12,RodriguesAMB15}.

The above mentioned works considered a global constraint on the
surface area. The membrane, however, is locally incompressible and hence
a local constraint on the evolution of the membrane molecules should
be taken into account. Several authors included the local
inextensibility constraint by introducing an inhomogeneous Lagrange
multiplier for this constraint on the membrane. This approach has been
used within the context of different modelling and computational
strategies such as the level set approach 
\cite{SalacM11,SalacM12,LaadhariSM14,DoyeuxGCPI13}, the phase field approach
\cite{BibenKM05,JametM07,AlandELV14}, the immersed boundary
method \cite{KimL10,KimL12,HuKL14}, the interfacial spectral boundary element 
method \cite{DodsonD11} and the boundary integral method \cite{FarutinBM14}. 

The physically most natural way to consider the local incompressibility
constraint makes use of the fact that the membrane itself can be
considered as an incompressible surface fluid. This implies that a
surface Navier--Stokes system has to be solved on the membrane. The
resulting set of equations has to take forces stemming from the surrounding
fluid and from the membrane elasticity into account. In total, bulk
Navier--Stokes equations coupled to surface Navier--Stokes equations 
have to be solved. As the involved Reynolds
numbers for vesicles are typically small one can often replace the
full Navier--Stokes equations by the Stokes systems on the surface and
in the bulk. The incompressibility condition in the bulk
(Navier--)Stokes equations naturally leads to conservation of the
volume enclosed by the membrane and the incompressibility condition on
the surface leads a conservation of the membrane's surface area. A
model involving coupled bulk-surface (Navier--)Stokes equations has
been proposed by \citet{ArroyoS09}, and it is this
model that we want to study numerically in this paper. 

Introducing forces resulting from membrane energies in fluid flow
models has been studied numerically before by different
authors, \cite{SalacM11,SalacM12,LaadhariSM14,JametM07,AlandELV14,HuKL14}. 
However, typically these authors studied simplified
models, and either volume or surface constraints were enforced by
Lagrange multipliers. In addition, either just the bulk or just the surface
(Navier--)Stokes equations have been solved. The only work considering
simultaneously bulk and surface Navier--Stokes equations are 
\citet{ArroyoSH10arxiv} and \citet{nsns,nsnsade}, where the
former work is restricted to axisymmetric situations. 
In the present paper we are going to make use of the numerical method
introduced in \cite{nsnsade}, see also \cite{nsns}. 

The paper is organized as follows. In the next section we precisely
state the mathematical model, consisting of the curvature elasticity
model together with a coupled bulk-surface (Navier--)Stokes system. In
Section~\ref{sec3} we introduce our numerical method which consists of
an unfitted parametric finite element method for the membrane
evolution. The curvature forcing is discretized and coupled to the
Navier--Stokes system in a stable way using the finite element method
for the fluid unknowns. Numerical computations in Section~\ref{sec4}
demonstrate that we can deal with a variety of different membrane
shapes and flow scenarios. In particular, we will study what influence the 
membrane viscosity, the area difference elasticity (ADE) and the spontaneous 
curvature have on the evolution of bilayer membranes in shear flow.
We finish with some conclusions.

\section{A continuum model for fluidic membranes} 

We consider a continuum model for the evolution of biomembranes and
vesicles, which consists of a curvature elasticity model for the membrane
and the Navier--Stokes equations in the bulk and on the surface. The model
is based on a paper by \citet{ArroyoS09}, where in addition we 
also allow the curvature energy model to be an
area difference elasticity model. We first introduce the curvature
elasticity model and then describe the coupling to the
surface and bulk Navier--Stokes equations. 

The thickness of the lipid bilayer in a vesicle is typically three to
four orders of magnitude smaller than the typical size of the
vesicle. Hence the membrane can be modelled as a two dimensional
surface $\Gamma$ in $\mathbb{R}^3$. Given the principal curvatures
$\varkappa_1$ and $\varkappa_2$ of $\Gamma$, one can define the mean curvature
\begin{equation*}
\varkappa = \varkappa_1+\varkappa_2
\end{equation*}
and the Gau{\ss} curvature 
\begin{equation*}
K = \varkappa_1\,\varkappa_2
\end{equation*}
(as often in differential geometry we choose to take
the sum of the principal curvatures as the mean curvature, instead of
its mean value). The classical works of \citet{Canham70} and
\citet{Helfrich73} derive a local bending energy, with the help
of an expansion in the curvature, and they obtain
\begin{equation}\label{elas1}
\int_\Gamma \left(\frac{\alpha}{2}\,\varkappa^2+\alpha_G\,K\right) \ds
\end{equation}
as the total energy of a symmetric membrane. The parameters 
$\alpha,\alpha_G$ have the dimension of energy and are called the
bending rigidity $\alpha$ and the Gaussian bending rigidity
$\alpha_G$. If we consider closed membranes with a fixed topology, the
term $\int_\Gamma K\ds$ is constant and hence we will neglect the
Gaussian curvature term in what follows. 

As discussed above, the total area difference $\Delta A$ of the two
lipid layers is, to first order, proportional to 
\begin{equation*} 
M(\Gamma) = \int_\Gamma \varkappa \ds\,.
\end{equation*}
Taking now into account that there is an optimal area difference
$\Delta A_0$, the authors in \cite{SeifertMDW91,WieseHH92,BozicSZW92}
added a term proportional to
\begin{equation*} 
(M(\Gamma)-M_0)^2
\end{equation*}
to the curvature energy, where $M_0$ is a fixed constant which is
proportional to the optimal area difference.

For non-symmetric membranes a certain mean curvature
$\spont$ can be energetically favourable. Then the
elasticity energy (\ref{elas1}) is modified to 
\begin{equation*} 
\int_\Gamma\left(\frac{\alpha}{2}\,(\varkappa-\spont)^2+\alpha_G\,K
\right)\ds\,.
\end{equation*}
The constant $\spont$ is called spontaneous
curvature. Taking into account that $\int_\Gamma\alpha_G\,K\ds$ does not
change for an evolution within a fixed topology class, the most general
bending energy that we use in this paper is given by
$\alpha \, E(\Gamma)$
with the dimensionless energy 
\begin{equation} \label{eq:EE}
E(\Gamma) = \frac{1}{2}\,\int_\Gamma
(\varkappa-\spont)^2\ds+\frac{\beta}{2}\,(M(\Gamma)-M_0)^2\,,
\end{equation}
where $\beta$ has the dimension $(\frac{1}{{\rm length}})^2$.

We now consider a continuum model for the fluid flow on the membrane
and in the bulk, inside and outside of the membrane. We assume that the
closed, time dependent membrane $(\Gamma(t))_{t\ge 0}$ lies inside a
spatial domain $\Omega\subset\mathbb{R}^3$. For all times the membrane
separates $\Omega$ into an inner domain $\Omega_-(t)$ and an outer
domain $\Omega_+(t)$. Denoting
by $\vec{u}$ the fluid velocity and by $p$ the pressure, 
the bulk stress tensor is given by
$\mat\sigma=2\,\mu\,\mat D(\vec{u})-p\,\mat\Id$, 
with $\mat D(\vec u) = \tfrac12\,(\nabla\,\vec u + (\nabla\,\vec u)^T)$ 
being the bulk rate-of-strain tensor. 
We assume that the Navier--Stokes system
\begin{equation*}
\rho\,(\vec{u}_t+(\vec{u}\,.\,\nabla)\,\vec{u}) - \nabla\,.\,\mat \sigma
=0\,,\quad
\nabla\,.\,\vec{u}=0
\end{equation*}
holds in $\Omega_-(t)$ and $\Omega_+(t)$. Here $\rho$ and $\mu$ are
the density and dynamic viscosity of the fluid, which can take different 
(constant) values $\rho_\pm$, $\mu_\pm$ in $\Omega_\pm(t)$. 
\citet{ArroyoS09} used the theory of interfacial fluid dynamics, which
goes back to \citet{Scriven60}, to introduce a relaxation
dynamics for fluidic membranes. In this model the fluid velocity is
assumed to be continuous across the membrane, the membrane is moved in 
the normal direction with the normal velocity of the bulk fluid and, 
in addition, the surface Navier--Stokes equations
\begin{equation*}
\rho_\Gamma\,\matpartu\,\vec{u}-\nabs\,.\,\mat\sigma_\Gamma =
[\mat\sigma]^+_-\,\vec\nu + \alpha\,\vec{f}_\Gamma\,,\quad
\nabs\,.\,\vec{u} =0
\end{equation*}
have to hold on $\Gamma(t)$. 
Here $\rho_\Gamma$ is the surface material density,
$\matpartu$ is the material derivative and $\nabs
,\nabs\,.$ are the gradient and divergence operators on the
surface. The surface stress tensor is given by 
\begin{equation*}
\mat\sigma_\Gamma=2\,\mu_\Gamma\,\mat D_s(\vec{u})-p_\Gamma\,
\mat{\mathcal{P}}_\Gamma\,,
\end{equation*}
where $p_\Gamma$ is the surface pressure, $\mu_\Gamma$ is the surface shear
viscosity, $\mat{\mathcal{P}}_\Gamma$ is the projection onto the tangent
space and 
\begin{equation*}
\mat{\mathcal{D}}_s(\vec{u}) = \tfrac{1}{2}\,\mat{\mathcal{P}}_\Gamma\, 
(\nabs\,\vec{u}+(\nabs\,\vec{u})^T) \,\mat{\mathcal{P}}_\Gamma
\end{equation*}
is the surface rate-of-strain tensor. Furthermore, the term
$[\mat\sigma]^+_-\,\vec\nu=\mat\sigma_+\,\vec{\nu}-\mat\sigma_-\,\vec{\nu}$
is the force exerted by the bulk on the membrane, where $\vec{\nu}$ 
denotes the exterior unit normal to $\Omega_-(t)$.
The remaining term $\alpha\,\vec{f}_\Gamma$ denotes the
forces stemming from the elastic bending energy. These forces are given
by the first variation of the bending energy $\alpha\,E(\Gamma(t))$, 
see \cite{Seifert97,ArroyoS09}.
It turns out that
$\vec{f}_\Gamma$ points in the normal direction, i.e.\ $\vec{f}_\Gamma =
f_\Gamma\,\vec{\nu}$, and we obtain, see \cite{Willmore65,Seifert97}, 
\begin{align*} 
f_\Gamma &=
-\Delta_s\,\varkappa-(\varkappa-\spont)\,|\nabs\,\vec{\nu}|^2+
\tfrac{1}{2}\,(\varkappa-\spont)^2\,\varkappa\nonumber\\
&\quad +\beta\,(M(\Gamma)-M_0)\,(|\nabs\,\vec{\nu}|^2-\varkappa^2)
\quad\text{on }\Gamma(t)
\,.
\end{align*}
Here $\Delta_s$ is the surface Laplace operator, $\nabs\,\vec{\nu}$ is
the Weingarten map and $|\nabs\,\vec{\nu}|^2=\varkappa^2_1+\varkappa^2_2$. 
Assuming e.g.\ no-slip boundary conditions on $\partial\Omega$, 
the boundary of $\Omega$, we obtain that 
the total energy can only decrease, i.e.\
\begin{align}
& \ddt \left(\int_\Omega\frac{\rho}{2}\,|\vec{u}|^2 \dx + 
\frac{\rho_\Gamma}{2}\,\int_\Gamma|\vec{u}|^2 \ds+\alpha\,E(\Gamma)\right) 
\nonumber \\
&\, =-2\left(\int_\Omega
  \mu\,|\mat D(\vec{u})|^2\dx + \mu_\Gamma
\int_\Gamma |\mat D_s(\vec{u})|^2\ds\right)\leq 0\,.\label{EnI}
\end{align}

We now non-dimensionalize the problem. We choose a time scale $\tilde t$, 
a length scale $\tilde x$ and the resulting velocity scale 
$\tilde u = \tilde x/\tilde t$. Then we define the bulk and surface Reynolds 
numbers 
\begin{equation*}
\Re = {\tilde x\,\rho_+\,\tilde u}/{\mu_+}
\quad\text{and}\quad
\Re_{\Gamma} = {\tilde x\,\rho_\Gamma \,\tilde u}/{\mu_\Gamma}\,,
\end{equation*}
the bulk and surface pressure scales
\begin{equation*}
\tilde p = {\mu_+}/{\tilde t}
\quad\text{and}\quad
\tilde p_\Gamma = {\mu_+\,\tilde x}/{\tilde t} = \mu_+\,\tilde u\,,
\end{equation*}
and 
\begin{align*}
\rho^* & = \rho / \rho_+ = 
\begin{cases} 
1 & \text{in $\Omega_+$} \\
\rho_- / \rho_+ & \text{in $\Omega_-$}
\end{cases}
\,,\\ 
\mu^* & = \mu/\mu_+= 
\begin{cases} 
1 & \text{in $\Omega_+$} \\
\Lambda 
& \text{in $\Omega_-$}
\end{cases}
\,,\quad \Lambda=\mu_- / \mu_+\,, \\ 
\mu_\Gamma^* & = {\mu_\Gamma} / ({\mu_+\,\tilde x})\,,
\end{align*}
as well as the new independent variables $\widehat x = x/\tilde x$, 
$\widehat t=t/\tilde t$. 
For the unknowns 
\begin{equation*}
\vec{\widehat u} = {\vec u}/{\tilde u}\,\,,\,\, \widehat
p = {p}/{\tilde p}\,\,,\,\, \widehat p_\Gamma = {p_\Gamma}/{\tilde p_\Gamma}
\end{equation*}
we now obtain the following set of equations (on dropping the
$\widehat{\phantom{u}}$-notation for the new variables for ease of exposition)
\begin{align}\label{surfnd}
&\Re\,\rho^*\,
(\vec u_t+(\vec u\,.\,\nabla)\,\vec u)-\mu^*\,\Delta\, \vec u +\nabla\, p
= 0\quad \text{in } \Omega_\pm(t)\,,
\nonumber\\
&\Re_\Gamma\,\mu_\Gamma^*\,
\matpartu\,\vec u-\nabs\,.\left( 2\,\mu_\Gamma^*\,
\mat D_s(\vec u) - p_\Gamma\, \mat {\mathcal{P}}_\Gamma \right) \nonumber\\
&\hspace{6mm}
= \left[2\,\mu^*\,\mat D(\vec u)-p\,\mat\Id \right]^+_- \vec\nu 
+ \alpha^*\, {\vec f^*_\Gamma}
\quad \text{on } \Gamma(t)\,,
\end{align}
with $\vec f^*_\Gamma = f^*_\Gamma\, \vec\nu$, 
\begin{align} \label{eq:fGamma}
f^*_\Gamma &=-\Delta_s\,\varkappa
-(\varkappa-\spont^*)\,|\nabs\,\vec\nu|^2
+\tfrac12\,(\varkappa-\spont^*)^2\,\varkappa \nonumber \\
&\qquad
+\beta^*\,(M(\Gamma)-M^*_0)\,(|\nabs\,\vec\nu|^2-\varkappa^2)
\quad\text{on }\Gamma(t)\,,
\end{align}
$\alpha^* = {\alpha}/{(\mu_+\,\tilde u\, \tilde x^2)}$ and
$\spont^* = \tilde x\,\spont$,
$M^*_0=M_0/\tilde x$, $\beta^* = \tilde x^2\,\beta$.
We remark that the Reynolds numbers for the two regions in the bulk are given
by $\Re$ and $\Re\,\rho^* / \mu^*$, respectively, and that they will in
general differ in the case of a viscosity contrast between the inner and outer
fluid. 
In addition to the above equations, we of course also
require that $\vec u$ has zero divergence in the bulk and that the surface
divergence of $\vec u$ vanishes on $\Gamma$. 

Typical values for the bulk dynamic viscosity $\mu$ are around
$10^{-3}-10^{-2} \frac{\rm kg}{\rm s\,m}$, 
see \cite{ArroyoS09,McWhirterNG09,Faivre06}, 
whereas the surface
shear viscosity typically is about $10^{-9}-10^{-8} \frac{\rm kg}{\rm s}$, see
\cite{AbreuLSS14,McWhirterNG09,RodriguesAMB15}. 
The bending modulus $\alpha$ is typically $10^{-20}-10^{-19}
\frac{\rm kg\,m^2}{\rm s^2}$, see
\cite{Faivre06,AbreuLSS14,RodriguesAMB15}.

The term $\mu_\Gamma^* = \mu_\Gamma / ({\mu_+\,\tilde x})$ in (\ref{surfnd}) 
suggests to choose the length scale
\begin{equation*}
\tilde x = {\mu_\Gamma}/{\mu_+}
\quad \iff \quad \mu_\Gamma^* = 1\,.
\end{equation*}
As $\alpha^* = {\alpha}/{(\mu_+\,\tilde u\,\tilde x^2)} =
{\alpha\,\tilde t}/{(\mu_+\,\tilde x^3)}$ appears in (\ref{surfnd}),
we choose the time scale
\begin{equation*}
\tilde t = {\mu_+\,\tilde x^3}/{\alpha}\,.
\end{equation*}
Choosing 
\begin{equation*}
\mu_\Gamma = 5\cdot 10^{-9}\frac{\rm kg}{\rm s}\,\,,\,\, \mu_+ = 10^{-3}
  \frac{\rm kg}{\rm s\,m}\,\,,\,\, \alpha = 10^{-19} \frac{\rm
    kg\,m^2}{\rm s^2}\,,
\end{equation*}
see e.g.\ \cite{ArroyoS09}, we obtain the length scale $5\cdot 10^{-6}\rm
m$ and the time scale $1.25 \rm s$, which are typical scales in
experiments. 
With these scales for length and time together with values of $\sim
10^3 \rm kg/m^3$ for the bulk density and $\sim 10^{-6} \rm kg/m^2$ 
for the
surface densities, we obtain for the bulk and surface Reynolds numbers
\begin{equation*}
\Re \approx 10^{-5}\quad \text{and}\quad 
\Re_\Gamma \approx 10^{-8}\,,
\end{equation*}
and hence we will set the Reynolds numbers to zero in this paper. We note
that it is straightforward to also consider positive Reynolds numbers in 
our numerical algorithm, see \cite{nsns,nsnsade} for details.
Together with the other
observations above, we then obtain the following reduced set of equations.
\begin{align}\label{eq:scaling1}
&-\mu^*\,\Delta\, \vec u +\nabla\, p = 0
\quad\text{in } \Omega_\pm(t)\,,
\nonumber\\
&-2\,\nabs\,.\,\mat D_s(\vec u) +\nabs\,.\,(p_\Gamma\,\mat {\mathcal{P}}_\Gamma)
\nonumber \\ & \hspace{1.3cm}
= \left[ 2\,\mu^*\,\mat D(\vec u) -p\,\mat\Id \right]^+_- \vec\nu 
+ \alpha^*\, \vec f_\Gamma^* \quad\text{on } \Gamma(t)\,.
\end{align}
A downside of the scaling used to obtain (\ref{eq:scaling1}) is that the 
surface viscosity no longer appears as an independent parameter. However,
studying the effect of the surface viscosity, e.g.\ on the tank treading 
to tumbling  
transition in shearing experiments, is one of the main focuses of this paper. 
It is for this reason that we also consider the following alternative scaling,
when suitable length and velocity scales are at hand.
For example, we may choose the length scale $\tilde x$ based on the (fixed)
size of the membrane and a velocity scale $\tilde u$ based on appropriate 
boundary velocity values. In this case we obtain from (\ref{surfnd}), 
for small Reynolds numbers, the following set of equations
\begin{align}\label{eq:scaling2}
&-\mu^*\,\Delta\, \vec u +\nabla\, p
= 0\quad \text{in } \Omega_\pm(t)\,,
\nonumber\\
&-\nabs\,.\left( 2\,\mu_\Gamma^*\,
\mat D_s(\vec u) - p_\Gamma\, \mat {\mathcal{P}}_\Gamma \right) \nonumber\\
&\hspace{1.3cm}
= \left[2\,\mu^*\,\mat D(\vec u)-p\,\mat\Id \right]^+_- \vec\nu 
+ \alpha^*\, \vec f_\Gamma^*
\quad \text{on } \Gamma(t)\,.
\end{align}
Note that here three non-dimensional parameters remain: $\mu_\Gamma^*$, 
$\Lambda$ and $\alpha^*$.
Here $\mu_\Gamma^*$ compares the surface shear viscosity to the bulk shear
viscosity, $\Lambda$ is the bulk viscosity ratio and $\alpha^*$ is an 
inverse capillary number, which describes the ratio
of characteristic membrane stresses to viscous stresses. Clearly, the system
(\ref{eq:scaling1}) corresponds to (\ref{eq:scaling2}) with
$\mu_\Gamma^*=1$. Hence from now on, we will only consider the scaling
(\ref{eq:scaling2}) in detail.

Of course, the system (\ref{eq:scaling2}) needs to be
supplemented with a boundary condition for $\vec u$ or $\mat\sigma$, and with
an initial condition for $\Gamma(0)$. 
For the former we partition the boundary $\partial\Omega$ of
$\Omega$ into $\partial_1\Omega$, where we prescribe a fixed velocity 
$\vec u = \vec g$,
and $\partial_2\Omega$, where we prescribe the stress-free condition
$\mat\sigma\,\unitn = \vec 0$, with $\unitn$ denoting the outer normal
to $\Omega$.

We note that our non-dimensionalization may be different to others presented 
in the literature. 
Often the length scale $a = ( \frac{\mathcal{A}(0)}{4\,\pi})^\frac12$ is
chosen, where $\mathcal{A}(0)$ denotes the surface area of the vesicle at time 
zero, see e.g.\ \cite{SpannZS14}. Our length scale $\tilde x$ may lead to
simulations with $\mathcal{A}(0) = 4\,\pi\,S^2$, with $S>0$,
so that our non-dimensional parameters 
in (\ref{eq:scaling2}) correspond to the non-dimensional values for a fixed 
length scale $\tilde x = a$ as follows:
\begin{equation} \label{eq:Spann}
\mathcal{M} = \frac{\mu_\Gamma}{\mu_+\,a} = \frac{\mu_\Gamma^*}S\,,\
a\,\spont = S\,\spont^*\,,\
\Ca = 
\frac{\mu_+\,\tilde u\,a^3}{\tilde x\,\alpha} = \frac{S^3}{\alpha^*}\,.
\end{equation}

\section{Numerical approximation}\label{sec3}
The numerical computations in this paper have been performed with a
finite element approximation introduced by the authors in
\cite{nsnsade}. The approach discretizes the bulk and surface
degrees of freedom independently. In particular, the surface
mesh is not a restriction of the bulk mesh. The bulk degrees of
freedoms $\vec u$ and $p$ are discretized with the lowest order 
Taylor--Hood element, P2--P1,
in our numerical computations. The evolution of the membrane is
tracked with the help of parametric meshes $\Gamma^h$, which are
updated by the fluid velocity. Since the membrane surface is locally
incompressible, it turns out that the surface mesh has good mesh
properties during the evolution. This is in contrast to other fluid
problems with interfaces in which the mesh often deteriorates during the
evolution when updated with the fluid velocity, see
e.g.\ \cite{Bansch01}. 

The non-dimensionalized elastic forcing by the membrane curvature energy, 
$\vec f_\Gamma^*$ in (\ref{eq:scaling2}), is
discretized with the help of a weak formulation by
\citet{Dziuk08}, which is generalized by \citet{nsnsade} to
take spontaneous curvature and area difference elasticity effects into
account. A main ingredient of the numerical approach is the fact that
one can use
a weak formulation of (\ref{eq:fGamma}) that can be discretized in a
stable way. In fact, 
defining $A^* = \beta^*\,(M(\Gamma)-M_0^*)$ and
$\vec y = \vec\varkappa + (A^*-\spont^*)\,\vec\nu$ the following
identity, which has to hold for all $\vec\chi$ on $\Gamma$,
characterizes $\vec f_\Gamma^*$:
\begin{align*}
& \left\langle\vec f_\Gamma^*, \vec\chi\right\rangle = 
\left\langle \nabs\,\vec y, \nabs\,\vec\chi\right\rangle
+ \left\langle \nabs\,.\,\vec y,\nabs\,.\,\vec\chi\right\rangle\\&\
-2\left\langle(\nabs\,\vec y)^T, 
\mat D_s(\vec\chi)\,(\nabs\, \vec\id)^T\right\rangle
+(A^*-\spont^*)\left\langle\vec\varkappa,[\nabs\,\vec\chi]^T\,\vec\nu
\right\rangle\\ & \
-\tfrac12\left\langle[|\vec\varkappa-\spont^*\,\vec\nu|^2-2\,(\vec y
\,.\, \vec\varkappa)]\,\nabs\,\vec\id,\nabs\,\vec\chi\right\rangle\\&\ 
-A^*\left\langle(\vec\varkappa \,.\, \vec\nu)\,\nabs\,\vec\id,
\nabs\,\vec\chi\right\rangle .
\end{align*}
Here $\langle \cdot,\cdot\rangle$ is the $L^2$--inner product on
$\Gamma$, and
$\nabs\,\vec y = \left( \partial_{s_j}\, y_i \right)_{i,j=1}^3$
with $(\partial_{s_1}, \partial_{s_2}, \partial_{s_3})^T = \nabs$.
Roughly speaking the above identity shows that $\vec f_\Gamma^*$ 
has a divergence structure. We remark here that similar
divergence structures have been derived with the help of Noether's
theorem, see \cite{CapovillaG04,Lengeler15arxiv}. 

The numerical method of \citet{nsnsade} has the
feature that a semi-discrete, i.e.\ continuous in time and discrete in
space, version of the method obeys a discrete analog of the energy
inequality (\ref{EnI}). In addition, this semi-discrete version has
the property that the volume enclosed by the
vesicle and the membrane's surface area are conserved exactly. After
discretization in time these properties are approximately fulfilled to
a high accuracy, see Section~\ref{sec4}. The fully discrete system is
linear and fully coupled in the unknowns. The overall system is
reduced by a Schur complement approach to obtain a reduced system in
just velocity and pressure unknowns. For this resulting linear system
well-known solution techniques for finite element discretizations for
the standard Navier--Stokes equations can be used, see 
\citet{fluidfbp}.

\section{Numerical computations}\label{sec4}

In shearing experiments the inclination angle of the vesicle in the shear
flow direction is often of interest. Here we will always consider shear flow 
in the $x_1$ direction with $x_3$ being the flow gradient direction.
Precisely, if $\overline\Omega = [-L,L]^2 \times [-W,W]$, then we
prescribe the inhomogeneous Dirichlet boundary condition
$\vec g(\vec x) = (x_3, 0, 0)^T$ on the top and bottom boundaries
$\partial_1\Omega = [-L,L]^2 \times \{\pm W\}$.
Assuming the vesicle's centre of mass is at the origin, 
then $\mat\MoI = \int_{\Omega_-(t)} |\vec x|^2\,\mat\Id - \vec x \otimes \vec x
\dx$ denote the vesicle's moment of inertia tensor. Let $\vec p$, with
$|\vec p|=1$ and $p_1 \geq 0$, be the
eigenvector corresponding to the smallest eigenvalue of $\mat\MoI$. Then
the vesicle's inclination angle is defined by
$\theta = \arg (p_1 + {\rm i}\,p_3) \in (-\pi/2,\pi/2]$, 
where $\arg : \bC \to (-\pi,\pi]$.
For later use we also note that the deformation parameter $D$ is defined by
$(b - c) / (b + c)$, where $b,c$ are the major and minor semiaxes of an 
ellipsoid with the same moment of inertia tensor,
see e.g. \cite{RamanujanP98}. Hence, in 2d,
$D = (\lambda_{\max}^\frac12 - \lambda_{\min}^\frac12) /
(\lambda_{\max}^\frac12 + \lambda_{\min}^\frac12)$, where $\lambda_{\max}$
and $\lambda_{\min}$ are the two eigenvalues of $\mat\MoI$.

The inclination angle $\theta$ is important for the classification of 
different types of dynamics in the shear flow experiments that we will
present. The classical deformation dynamics for vesicles
are the tank treading (TT) and the tumbling (TU) motions.
In the tank treading motion the vesicle adopts a constant inclination angle in the 
flow, while the surface fluid rotates on the membrane surface. 
This motion is observed for
small viscosity contrasts between the inner and the outer fluid and,
as we will see later, at low surface membrane viscosity. At large viscosity
contrasts or large membrane viscosity the tumbling motion occurs.
In the tumbling regime the membrane rotates as a whole, and the 
inclination angle oscillates in the whole interval $(-\pi/2,\pi/2]$. In
the last ten years a new dynamic regime for vesicles in shear flow has
been identified. In this regime the inclination angle is neither constant
nor does it oscillate in the whole interval $(-\pi/2,\pi/2]$. The
dynamics are characterized by periodic oscillations of the inclination
angle $\theta$ such that $\theta\in[-\theta_0,\theta_0]$ for a
$\theta_0$ in the open interval $(0,\pi/2)$. This regime was
first predicted theoretically by \cite{Misbah06} and subsequently 
observed experimentally in \cite{KantslerS06}.
Later this regime has been studied by different groups, see
e.g.\ 
\cite{ZabuskySDKS11,NoguchiG07,YazdaniB12,FarutinM12,AbreuLSS14,BibenFM11,%
SpannZS14} 
for more details. In \cite{Misbah06} this motion was called
vacillating-breathing and later the same motion was also called
trembling, transition mode or swinging. 
Following \cite{YazdaniB12} we will refer to this new regime
as the transition (TR) mode.

In our numerical simulations we will only consider the scaling
(\ref{eq:scaling2}). 
For all the presented simulations we will state the reduced volume as a
characteristic invariant. 
It is defined as
$\mathcal{V}_r = 6\,\pi^\frac12\,\mathcal{V}(0) / \mathcal{A}^\frac32(0)$,
see e.g.\ \cite{WintzDS96}. Here $\mathcal{V}(t)$ and $\mathcal{A}(t)$ denote the volume 
of the discrete inner phase and the discrete surface area, respectively, 
at time $t$.
Moreover, if nothing else is specified, then our numerical simulations are 
for no-slip boundary conditions, i.e.\ $\partial_1\Omega=\partial\Omega$ and 
$\vec g = \vec 0$. 
In all our experiments it holds that 
$\spont^*\,\beta^*=0$,
and we will only report the values of $\spont^*$ and $\beta^*$ for simulations
where they are nonzero.
Here we recall, as stated in the introduction, that the energy 
\begin{equation} \label{eq:Estar}
E^*(\Gamma) = \frac{1}{2}\,\int_\Gamma
(\varkappa-\spont^*)^2\ds+\frac{\beta^*}{2}\,(M(\Gamma)-M^*_0)^2
\end{equation}
for $\spont^*\,\beta^* \not=0$ is equivalent to (\ref{eq:Estar}) 
with $\spont^*=0$, the same value of $\beta^*>0$, and a modified value of 
$M_0^*$.
Finally, we stress that our sign convention for curvature is such that spheres
have negative mean curvature.

\subsection{2d validation}
\begin{figure}
\center
\begin{tikzpicture}[scale = \scalefactor]
\begin{axis}[legend pos=north west,
xlabel=$\mathcal{A}_r$,ylabel=$\theta/\pi$,%
xmin=0.5,xmax=1,ymin=0,ymax=0.25,%
ytick={0.05,0.1,...,0.25},%
yticklabel style={/pgf/number format/fixed},%
]
\addplot table[y expr=\thisrow{theta}/3.14159] {SM12a.dat};
\legend{$\theta$}
\end{axis}
\end{tikzpicture}
\caption{(Color online) 
A plot of $\theta / \pi$ against $\mathcal{A}_r$ for
$L=20$, $W = 5$, $\Lambda = 1$, $\alpha^* = 0.01$, $\Re = 10^{-3}$;
compare with \cite[Fig.~1]{SalacM12}.}
\label{fig:SM12}
\end{figure}
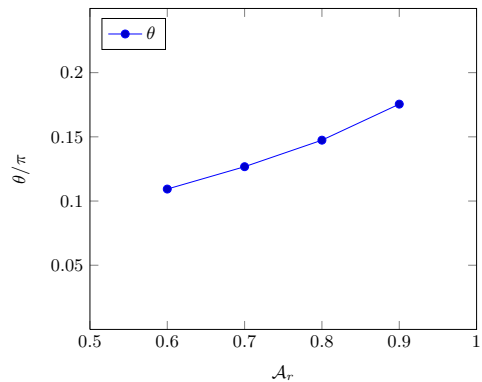

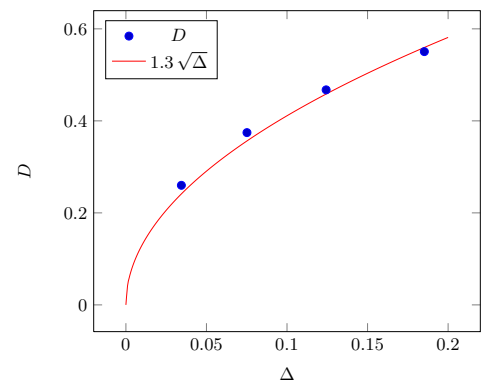
\begin{figure}
\center
\begin{tikzpicture}[scale = \scalefactor]
\begin{axis}[legend pos=north west,
xlabel=$\Delta$,ylabel=$D$,%
xticklabel style={/pgf/number format/fixed},%
only marks
]
\addplot table[x expr=2*(1-sqrt(\thisrow{nu})) / (3.14159 *sqrt(\thisrow{nu})),%
y expr=\thisrow{D}] {SM12a.dat};
\addplot[domain=0:0.2,smooth,samples=200,color=red] {1.3 * sqrt(x)};
\legend{$D$, $1.3\,\sqrt{\Delta}$}
\end{axis}
\end{tikzpicture}
\caption{(Color online)
A plot of $D$ against $\Delta$ for
$L=20$, $W = 5$, $\Lambda = 1$, $\alpha^* = 0.01$, $\Re = 10^{-3}$;
compare with \cite[Fig.~2]{SalacM12}.}
\label{fig:SM12a}
\end{figure}

\begin{figure}
\center
\begin{tikzpicture}[scale = \scalefactor]
\begin{axis}[legend pos=north west,
xlabel=$\mathcal{A}_r$,ylabel=$\Lambda_C$,%
xmin=0.5,xmax=1,ymin=3,ymax=9,%
xtick={0.6,0.7,...,0.9},%
ytick={3,4,...,9},%
]
\addplot table {SM12b.dat};
\legend{$\Lambda_C$}
\end{axis}
\end{tikzpicture}
\caption{(Color online)
A plot of $\Lambda_C$ against $\mathcal{A}_r$ for
$L=20$, $W = 5$, $\alpha^* = 0.01$, $\Re = 10^{-3}$;
compare with \cite[Fig.~3]{SalacM12}.}
\label{fig:SM12b}
\end{figure}
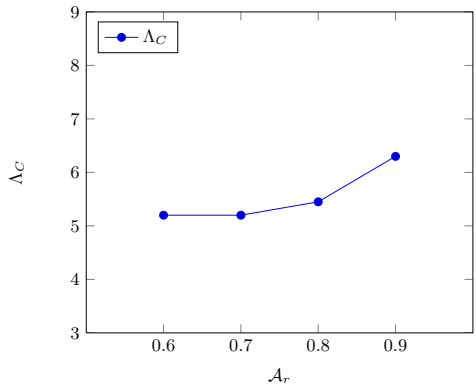

In order to validate our numerical method, we reproduce some numerical results
from \cite{SalacM12,KaouiKH12}, where we always consider a domain 
$\overline\Omega = [-L,L]\times[-W,W]$.
As these works consider Navier--Stokes flow in
the bulk, we consider (\ref{surfnd}) with $\Re = 10^{-3}$, $\Re_\Gamma = 0$,
$\mu_\Gamma^* = \spont^* = \beta^* = 0$
and vary $\Lambda$. For the comparison with Figures~1--3 in 
\cite{SalacM12} we also set $\alpha^* = 10^{-2}$.
Moreover, we consider vesicles with reduced
areas $\mathcal{A}_r = \frac{4\,\pi\,\mathcal{A}(0)}{P^2(0)} \in 
\{0.6,0.7,0.8,0.9\}$, and with $a = \frac{P(0)}{2\,\pi} = 1$, so that 
perimeter and area are given by $P(0) = 2\,\pi$ and 
$\mathcal{A}(0) = \mathcal{A}_r\,\pi$. 
At first, for $\Lambda=1$, we try to recreate \cite[Fig.~1]{SalacM12}.
To this end, we set $L = 20$ and $W=5$, and use
stress-free boundary conditions left and right, rather than 
periodic boundary conditions in the $x_1$-direction on the square domain
$[-5,5]^2$ as used in \cite{SalacM12}.
We obtain the results in Figure~\ref{fig:SM12}, where we plot $\theta / \pi$
against $\mathcal{A}_r$, 
which show a good agreement with \cite[Fig.~1]{SalacM12}.
Similarly, in trying to recreate \cite[Fig.~2]{SalacM12} we also compute
the deformation parameter $D$, and plot $D$ against the excess length parameter
$\Delta = 2\,(1 - \mathcal{A}_r^\frac12) / (\pi\,\mathcal{A}_r^\frac12)$.
We obtain the results in Figure~\ref{fig:SM12a}, which show good
agreement with \cite[Fig.~2]{SalacM12}.
In Figure~\ref{fig:SM12b} we plot the critical
viscosity ratio $\Lambda_C$ for the TT to TU transition against the reduced
area $\mathcal{A}_r$.
It should be noted that our numerical method produces larger values of
$\Lambda_C$ than reported in \cite[Fig.~3]{SalacM12}.

\begin{figure}
\center
\begin{tikzpicture}[scale = \scalefactor]
\begin{axis}[legend pos=south west,xlabel=$\Lambda$,
minor tick num=4,%
]
\addplot table[y expr=\thisrow{theta}/0.523598776] {KKH12a.dat};
\addplot table[y expr=\thisrow{V}/0.5] {KKH12a.dat};
\addplot table[y expr=2*3.14159/\thisrow{period}] {KKH12b.dat};
\legend{$\theta$, $V$, $\omega$}
\end{axis}
\end{tikzpicture}
\caption{(Color online)
A plot of $\theta / \frac\pi6$, $V / \frac12$ 
and $\omega / \frac1{2\,\pi}$ against $\Lambda$ for 
$\mathcal{A}_r = 0.8$, $L = 11.55$, $W=3.85$, $\alpha^*=2$;
compare with \cite[Fig.~1]{KaouiKH12}.}
\label{fig:KKH12}
\end{figure}
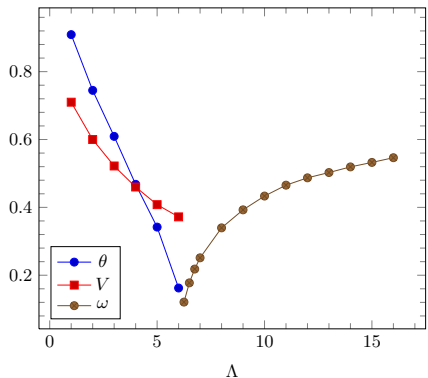

Moreover, in trying to recreate \cite[Fig.~1]{KaouiKH12}, we also ran with 
$\Re = 0.05$, $L = 11.55$ and $W=3.85$, so that the restriction parameter
$\chi$ as defined in \cite{KaouiKH12} is $\chi = 0.26$.
However, we note that periodic boundary conditions in the $x_1$-direction are
used in \cite{KaouiKH12}, with the length $L$ of the domain not clearly stated.
We obtain the results in Figure~\ref{fig:KKH12}, where apart from $\theta$,
normalized by $\frac\pi6$, we also show the membrane tank treading velocity 
$V = \frac1{P(0)}\,\int_{\Gamma} |\vec u| \ds$, normalized by $\frac12$,
and the tumbling frequency $\omega$, normalized by $\frac1{2\,\pi}$ (note that
the frequency in \cite[Fig.~1]{KaouiKH12} is said to be normalized by 
$\frac1{4\,\pi}$). 
It should be noted that qualitatively our results agree well with 
\cite[Fig.~1]{KaouiKH12}, but our numerical method produces a smaller value of
$\Lambda_C$ than reported in \cite[Fig.~1]{KaouiKH12}.

Overall we are satisfied that our numerical method performs well. The observed
differences with existing results in the literature can be explained by 
differences in the length of the domain, different boundary conditions and
different numerical methods used.

\subsection{3d validation}
\begin{figure}
\begin{tikzpicture}[scale = \scalefactor]
\begin{axis}[%
legend pos=north west,%
xlabel=$1/\alpha^*$ (Ca),%
xtick={0,1,...,10},%
ytick={1,1.5,...,7},%
minor tick num=4,%
ylabel=$\Lambda$,%
scatter/classes={%
TU={mark=square*,green},%
TRc={mark=diamond*,red},%
TT={mark=*,blue}}]
\legend{TU, \TRc, TT}
\addplot[scatter,only marks,%
scatter src=explicit symbolic]%
table[meta=motion] {
 Ca    lambda  motion
  1       1      TT
  1       1.5    TRc
  1       2      TRc
  1       2.5    TRc
  1       3      TU

  2       1      TT
  2       1.5    TRc
  2       2      TRc
  2       3      TRc
  2       3.5    TRc
  2       4      TU

  3       1.5    TT
  3       2      TRc
  3       2.5    TRc
  3       3.5    TRc
  3       4      TRc
  3       4.5    TU

  5       1.5    TT
  5       2      TRc
  5       2.5    TRc
  5       4      TRc
  5       4.5    TRc
  5       5      TU

  7       1.5    TT
  7       2      TRc
  7       2.5    TRc
  7       4      TRc
  7       4.5    TRc
  7       5      TU

 10       1.5    TT
 10       2      TRc
 10       2.5    TRc
 10       4.5    TRc
 10       5      TRc
 10       5.5    TU
}; 
\end{axis}
\end{tikzpicture}
\caption{(Color online)
Analogue of the phase diagram from \cite[Fig.\ 8]{SpannZS14}
for the domain $\overline\Omega = [-3,3]^3$, starting with a prolate shape
with $\mathcal{V}_r = 0.8$ and the longest axis in the $x_3$ direction.}
\label{fig:phase8b_prolate}
\end{figure}
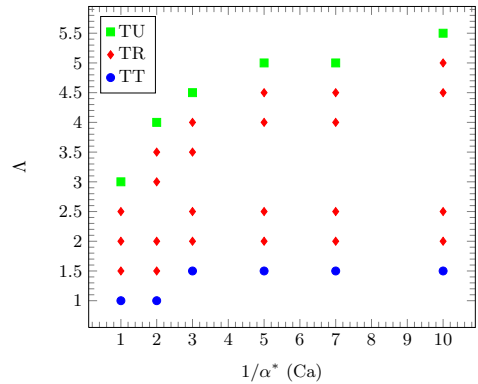

For a similar validation in 3d we compare our method to some numerical results
from \cite{SpannZS14}, where Stokes flow in an infinite domain is considered. 
In order to reproduce the phase diagram in
\cite[Fig.\ 8]{SpannZS14}, which also contains numerical results from
\cite{BibenFM11}, we let $\overline\Omega = [-3,3]^3$ and choose
the initial shape of the interface to be a prolate vesicle
with a reduced volume of $\mathcal{V}_r = 0.8$ and a surface area of 
$\mathcal{A}(0) = 4\,\pi$, so that $S=1$.
The results from our algorithm are shown in 
Figure~\ref{fig:phase8b_prolate}.
Due to finite size effects, and the different
boundary conditions, we observe different critical
values for the phase transitions compared to
\cite[Fig.\ 8]{SpannZS14}. However, qualitatively our numerical method
produces similar results.

\subsection{Effect of surface viscosity} 
\begin{figure*}
\begin{subfigure}[t]{0.35\textwidth}
\vskip 0pt
\begin{tikzpicture}[scale = \scalefactor]
\begin{axis}[%
legend pos=north west,%
xlabel=$1/\alpha^*$ (Ca),%
ymax=3.25,%
xtick={0,1,...,10},%
ytick={0,0.5,...,3},%
minor tick num=4,%
ylabel=$\mu_\Gamma^*~~(\mathcal{M})$,%
scatter/classes={%
TU={mark=square*,green},%
TRc={mark=diamond*,red},%
TT={mark=*,blue},%
PL={mark=o,black,mark size={5pt}}}]
\legend{TU, \TRc, TT}
\addplot[scatter,only marks,%
scatter src=explicit symbolic]%
table[meta=motion] {phasediagram6.dat};
\end{axis}
\end{tikzpicture}
\caption{Phase diagram.}
\label{fig:sub6a}
\end{subfigure}
\begin{subfigure}[t]{0.6\textwidth}
\vskip 0pt
\renewcommand\localwidth{0.25\textwidth}
\begin{tabular}{cccc}
\includegraphics[angle=-0,width=\localwidth]{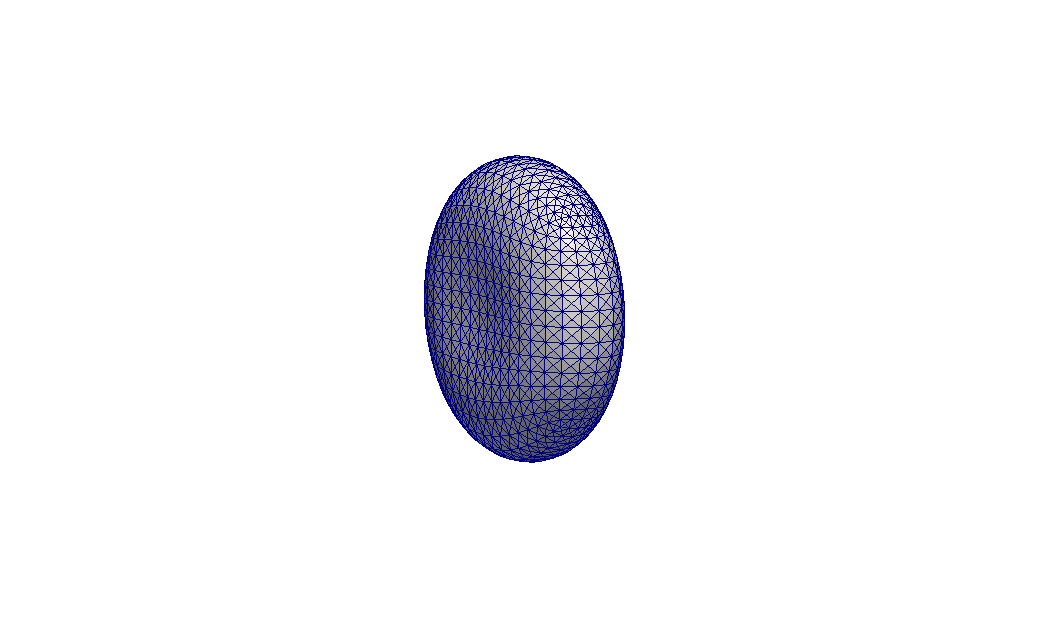} 
&
\includegraphics[angle=-0,width=\localwidth]{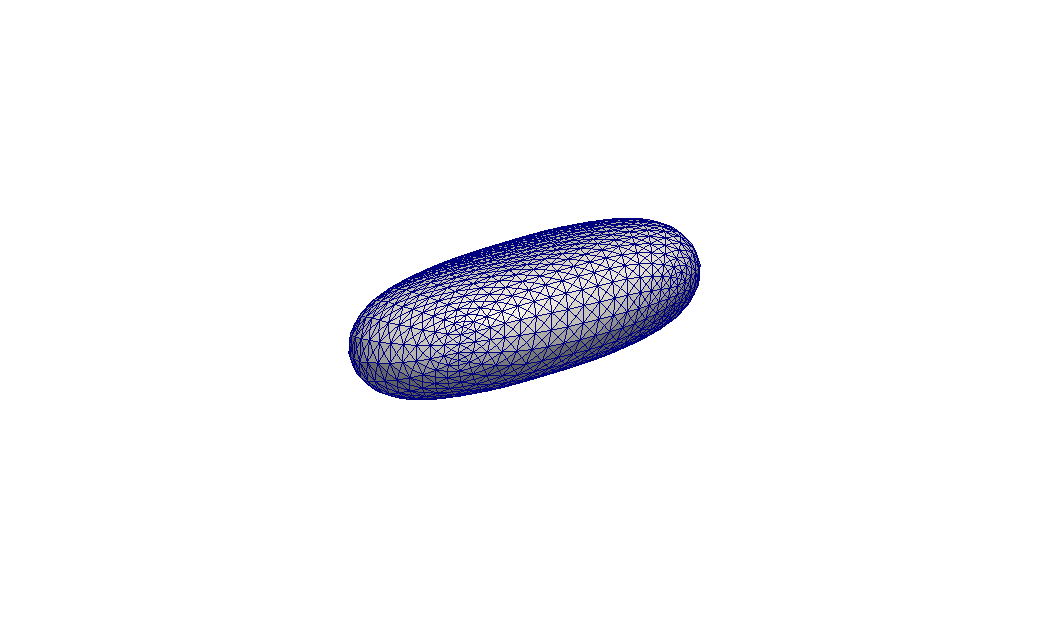} 
&
\includegraphics[angle=-0,width=\localwidth]{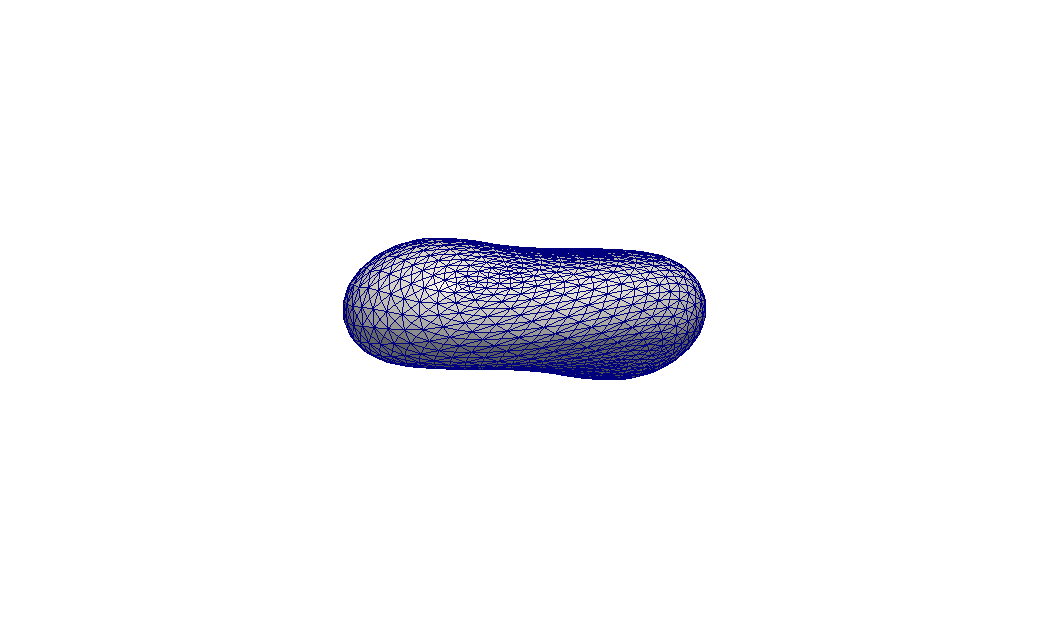} 
&
\includegraphics[angle=-0,width=\localwidth]{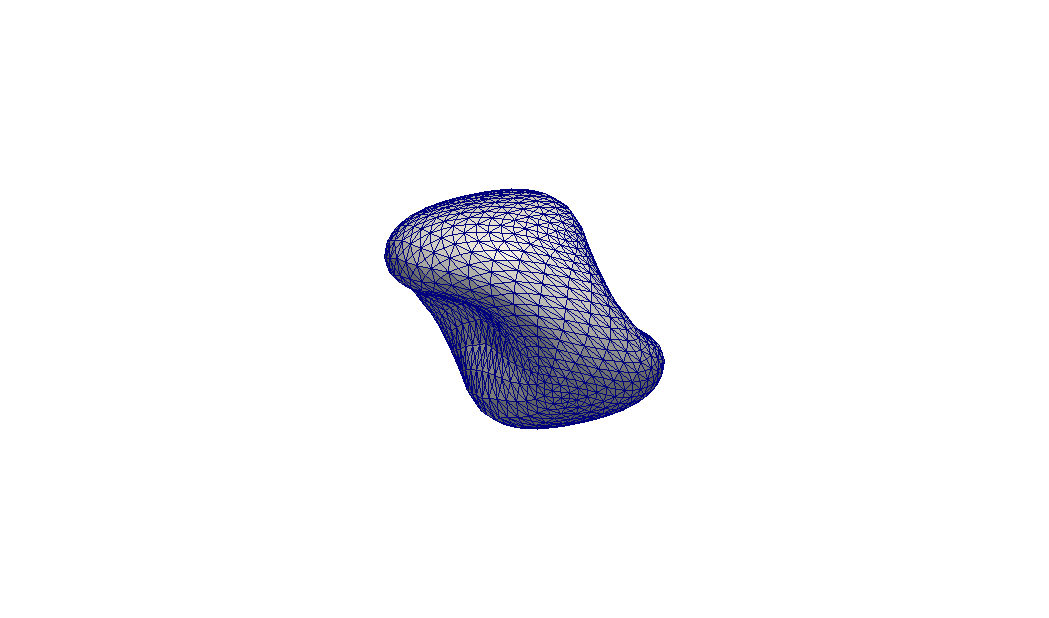} 
\\
\includegraphics[angle=-0,width=\localwidth]{figures/3dshear8_r0_m1_m3_t0} 
&
\includegraphics[angle=-0,width=\localwidth]{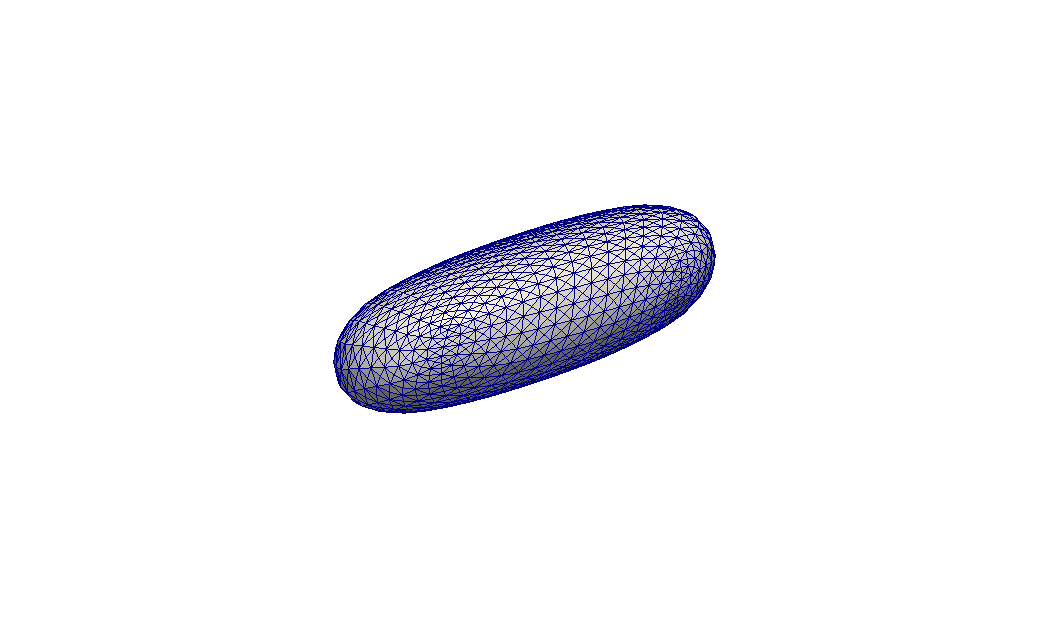} 
&
\includegraphics[angle=-0,width=\localwidth]{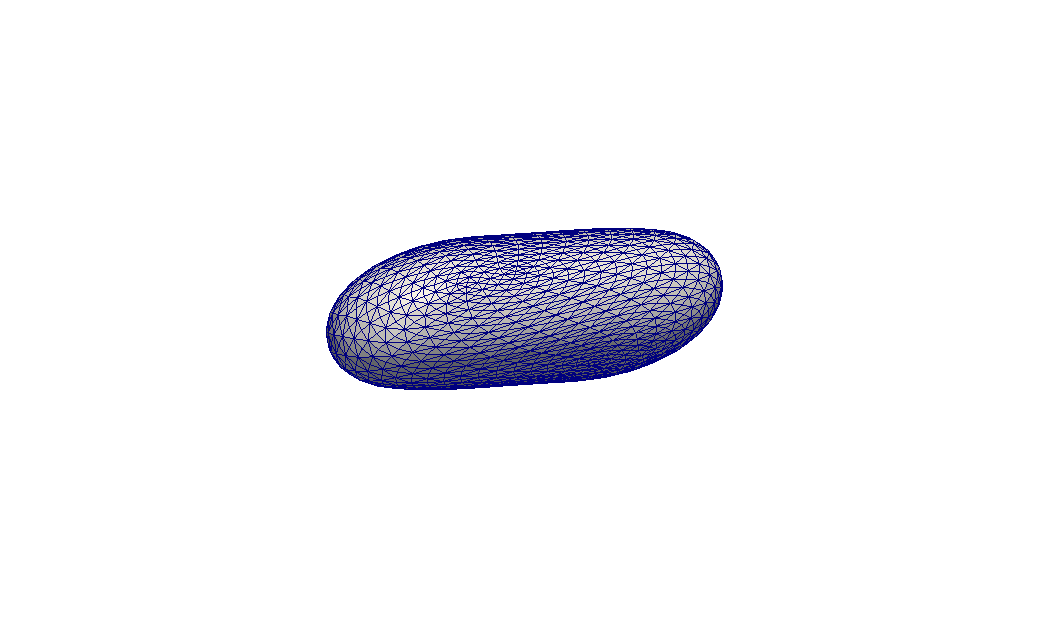} 
&
\includegraphics[angle=-0,width=\localwidth]{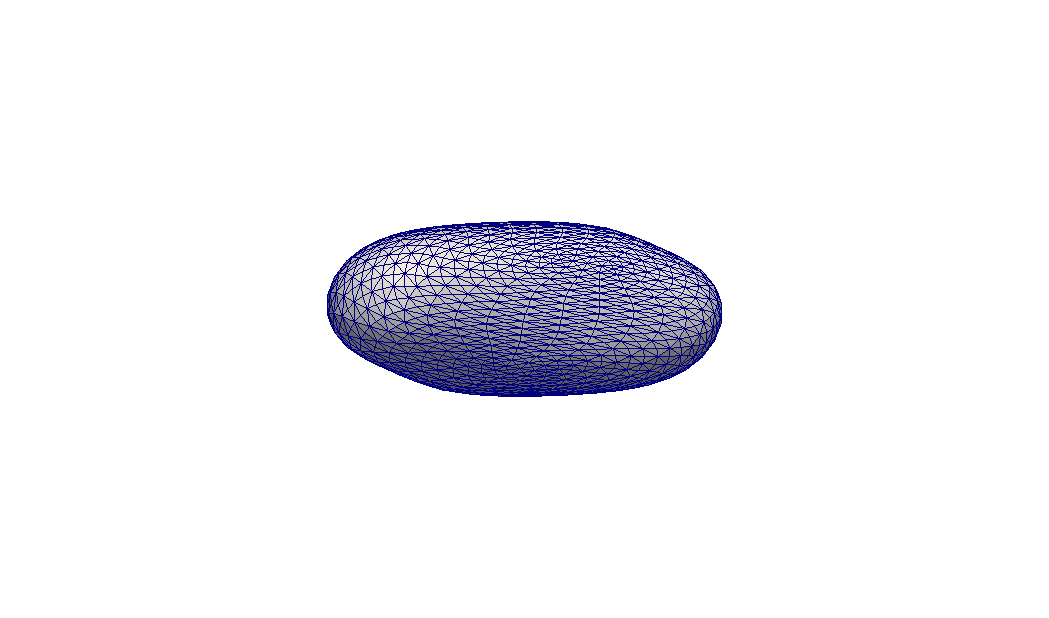}
\\ 
\includegraphics[angle=-0,width=\localwidth]{figures/3dshear8_r0_m1_m3_t0} 
&
\includegraphics[angle=-0,width=\localwidth]{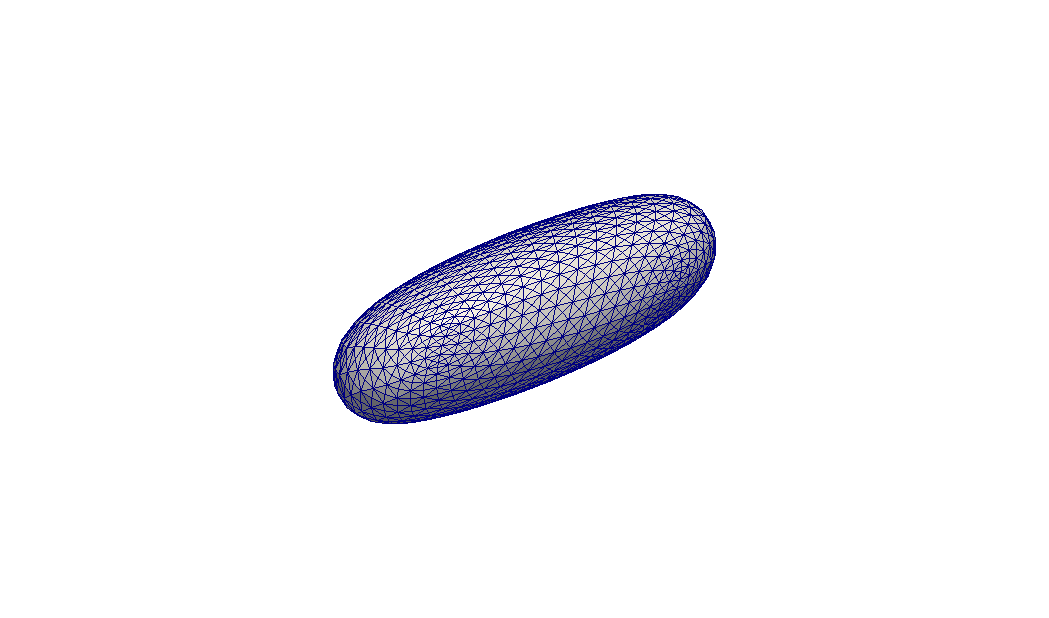} 
&
\includegraphics[angle=-0,width=\localwidth]{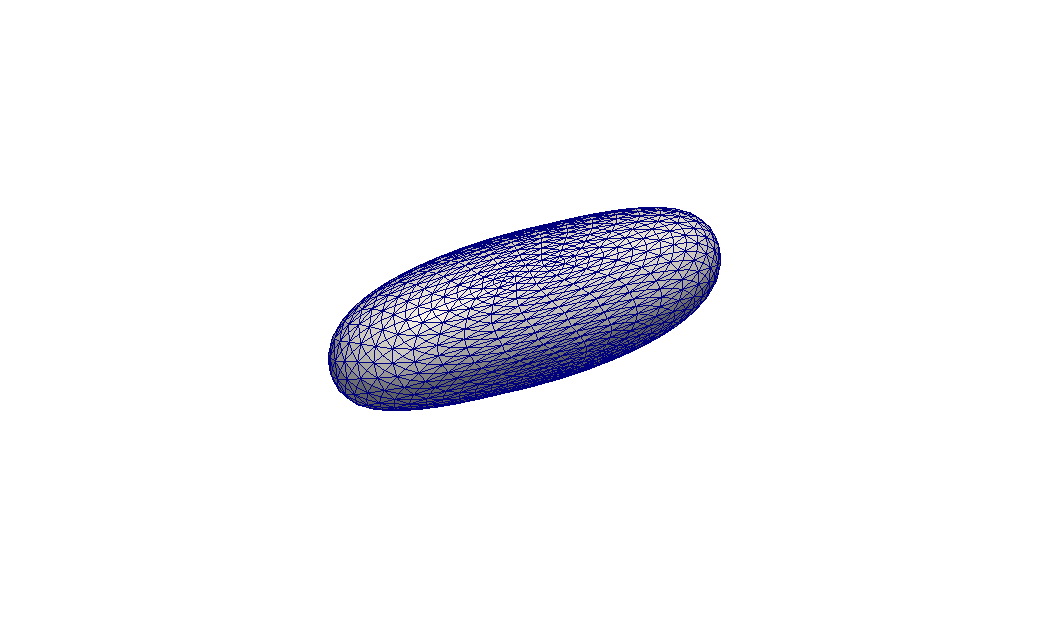} 
&
\includegraphics[angle=-0,width=\localwidth]{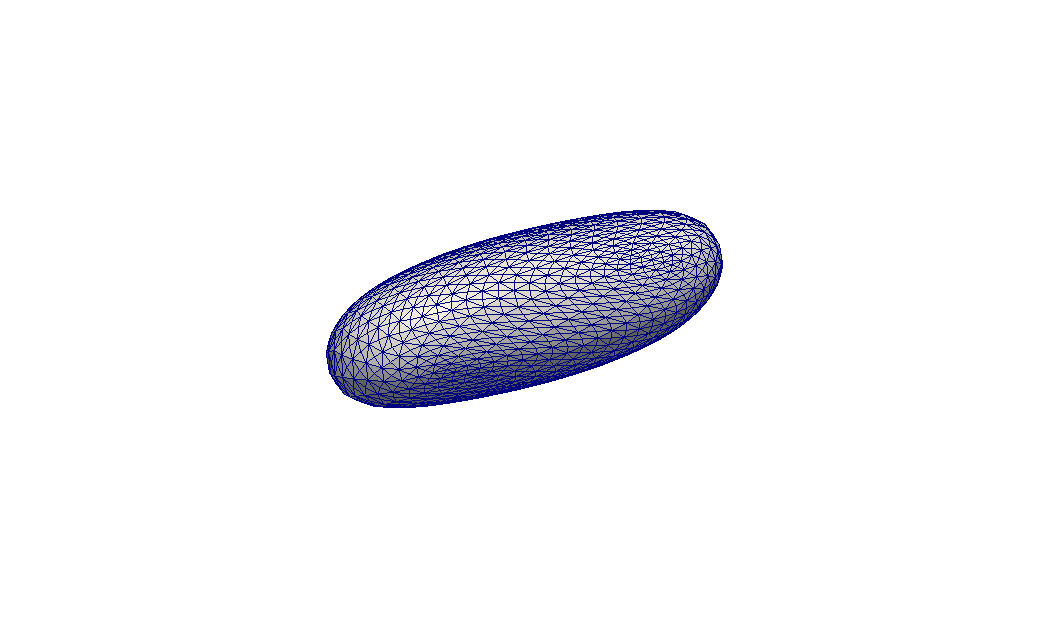} 
\end{tabular}
\caption{Top to bottom: $\mu_\Gamma^* = 3$, $1$, $0.1$ 
for $1/\alpha^*=10$ at times $t=0,\ 3,\ 6,\ 9$.}
\label{fig:sub6b} 
\end{subfigure}
\caption{(Color online)
Phase diagram for $\Lambda = 1$ for 
the domain $\overline\Omega = [-3,3]^3$, starting with a biconcave shape
with $\mathcal{V}_r = 0.8$ and the shortest axis in the $x_1$-direction.
The three big circles in the phase diagram correspond to the simulations in
(b).
}
\label{fig:phase_muGamma}
\end{figure*}

\begin{figure}
\center
\includegraphics[angle=-90,width=0.15\textwidth]{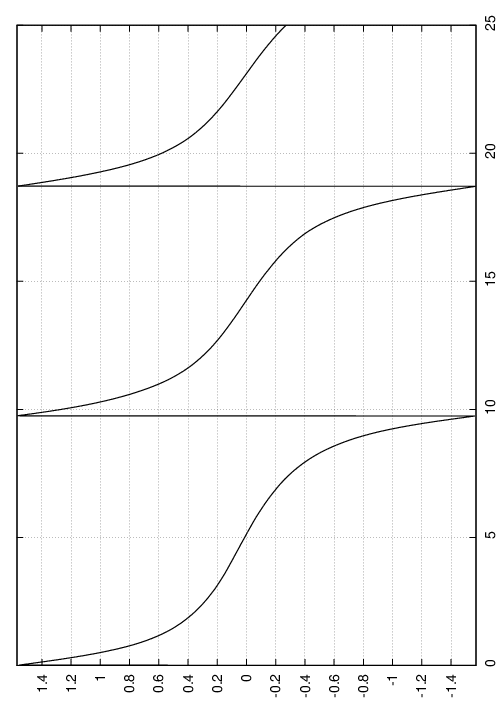}
\includegraphics[angle=-90,width=0.15\textwidth]{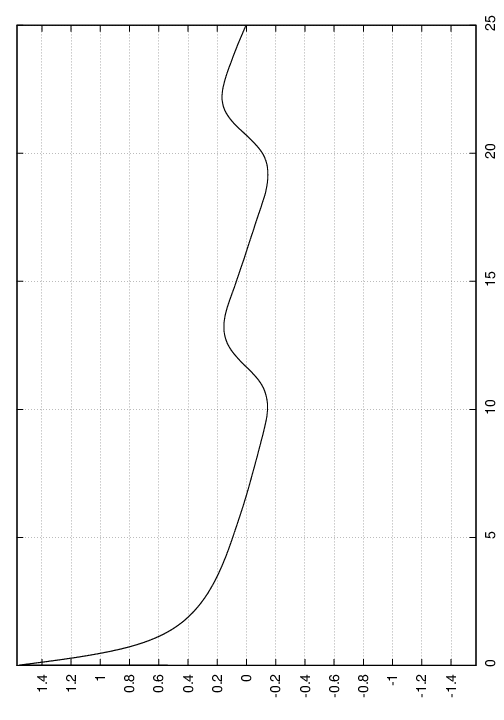}
\includegraphics[angle=-90,width=0.15\textwidth]{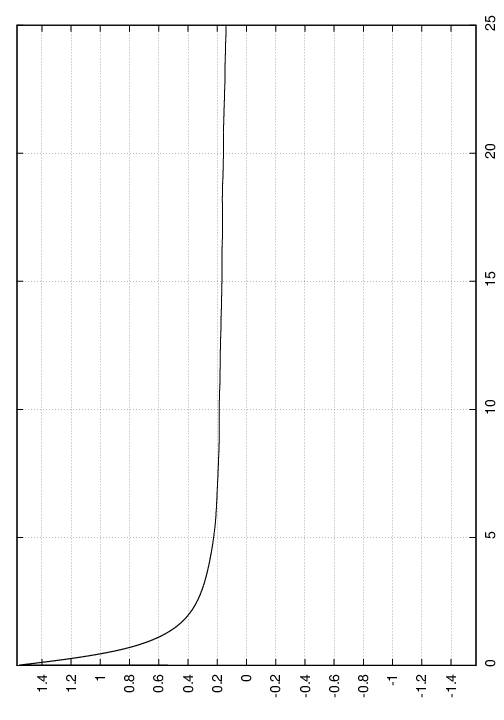}
\caption{The inclination angle $\theta$ for the computations in
Figure~\ref{fig:sub6b}. They correspond to the motions TU, TR
and TT, respectively.}
\label{fig:theta2}
\end{figure}

We consider the effect that surface viscosity has on the TT to TU transition.
To this end, we let $\overline\Omega = [-3,3]^3$, and choose as 
initial shape of the vesicle a biconcave shape with reduced volume $\mathcal{V}_r = 0.8$
and $\mathcal{A}(0) = 4\,\pi$, so that $S=1$. We let $\Lambda = 1$.
In Figure~\ref{fig:sub6a} we present a phase diagram with the axes 
labelled in terms of the
non-dimensional values $\Ca = \frac{1}{\alpha^*}$ and
$\mathcal{M} = \mu_\Gamma^*$, recall (\ref{eq:Spann}) for $S=1$.
The evolutions for $\alpha^* = 0.1$, and either 
$\mu_\Gamma^* = 3$, $\mu_\Gamma^* = 1$, or $\mu_\Gamma^* = 0.1$,
are visualized in Figure~\ref{fig:sub6b},
where we observe the motions TU, TR and TT, respectively.
We stress that the tumbling occurs for a viscosity contrast of
$\Lambda = 1$, and so is only due to the chosen high surface viscosity
$\mu_\Gamma^*$.
The fact that vesicles undergo a transition from steady tank treading
to unsteady tumbling motion has been observed earlier by
\cite{NoguchiG05a}, where, however, the authors used a particle-based 
mesoscopic model to analyze the fluid vesicle dynamics. 
A plot of the inclination angle $\theta$ for the simulations in
Figure~\ref{fig:sub6b} can be seen in Figure~\ref{fig:theta2}.

\subsection{Effect of spontaneous curvature}

\begin{figure}
\renewcommand\localwidth{0.95\textwidth}
\begin{subfigure}[b]{0.23\textwidth}
\includegraphics[angle=-0,width=\localwidth]{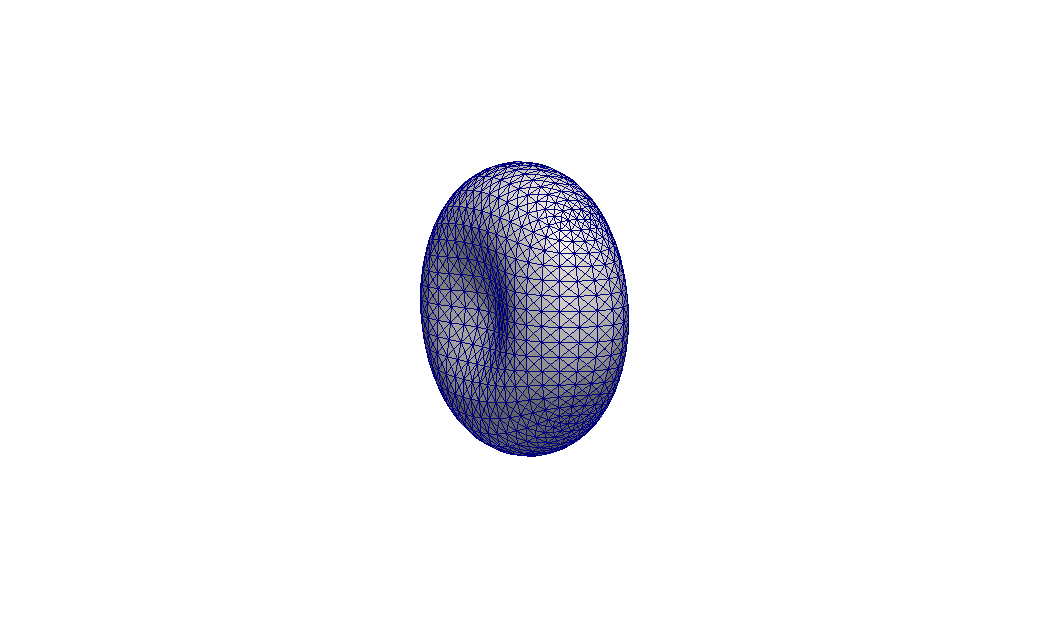}
\caption{$\spont = -5$}
\end{subfigure}
\begin{subfigure}[b]{0.23\textwidth}
\includegraphics[angle=-0,width=\localwidth]{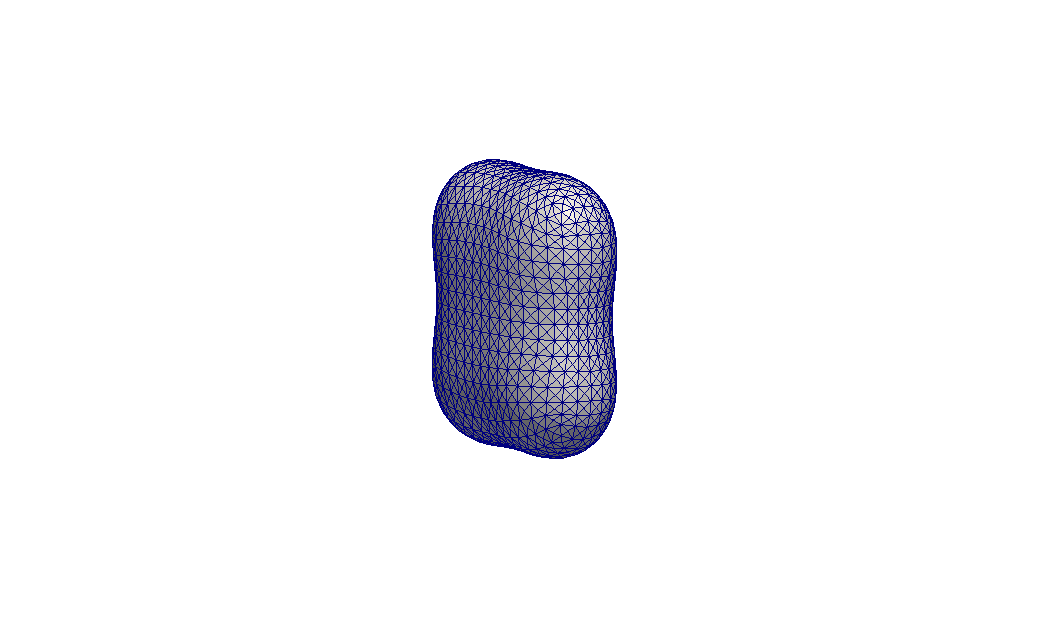}
\caption{$\spont = 5$}
\end{subfigure}
\caption{(Color online)
The vesicles for $\spont^* = \pm5$ 
at time $t=0$.}
\label{fig:initialnew}
\end{figure}

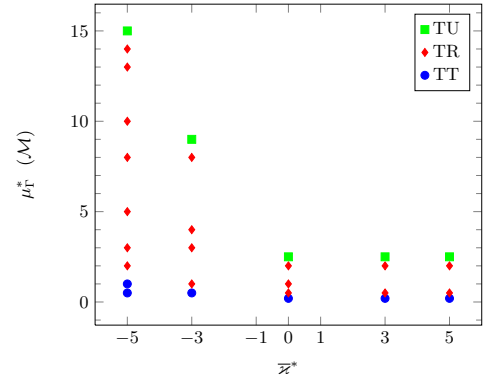
\begin{figure}
\begin{tikzpicture}[scale = \scalefactor]
\begin{axis}[%
xlabel=$\spont^*$,
ylabel=$\mu_\Gamma^*~~(\mathcal{M})$, 
legend pos=north east,%
xtick={-5,-3,-1,0,1,3,5},%
minor tick num=4,%
scatter/classes={%
TU={mark=square*,green},%
TRc={mark=diamond*,red},%
TT={mark=*,blue}}]
\legend{TU, \TRc, TT}
\addplot[scatter,only marks,%
scatter src=explicit symbolic]%
table[meta=motion] {
spont    mu    motion
-5       0.5     TT
-5       1.0     TT
-5       2      TRc
-5       3      TRc
-5       5      TRc
-5       8      TRc
-5       10     TRc
-5       13     TRc
-5       14     TRc
-5       15     TU 

-3       0.5     TT
-3       1      TRc
-3       3      TRc
-3       4      TRc
-3       8      TRc
-3       9      TU

0       0.2     TT
0       0.5     TRc
0        1      TRc
0        2      TRc
0       2.5     TU

3       0.2     TT
3       0.5     TRc
3        2      TRc
3       2.5     TU

5       0.2     TT
5       0.5     TRc
5        2      TRc
5       2.5     TU

};
\end{axis}
\end{tikzpicture}
\caption{(Color online)
Phase diagram for $\Ca = 1/\alpha^* = 10$ for 
the domain $\overline\Omega = [-3,3]^3$, starting with biconcave shapes
with $\mathcal{V}_r = 0.8$ and the shortest axis in the $x_1$-direction.}
\label{fig:phase_muGamma_spont}
\end{figure}

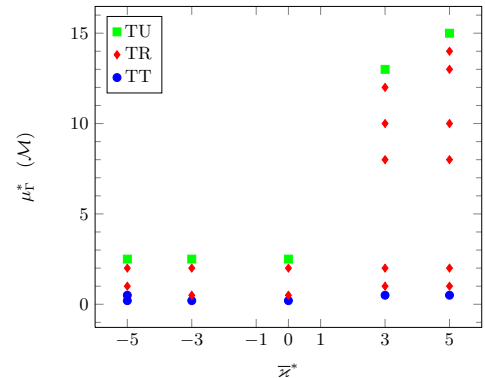
\begin{figure}
\begin{tikzpicture}[scale = \scalefactor]
\begin{axis}[%
xlabel=$\spont^*$,
ylabel=$\mu_\Gamma^*~~(\mathcal{M})$, 
legend pos=north west,%
xtick={-5,-3,-1,0,1,3,5},%
minor tick num=4,%
scatter/classes={%
TU={mark=square*,green},%
TRc={mark=diamond*,red},%
TT={mark=*,blue}}]
\legend{TU, \TRc, TT}
\addplot[scatter,only marks,%
scatter src=explicit symbolic]%
table[meta=motion] {
spont    mu    motion
-5       0.2    TT
-5       0.5    TT
-5       1      TRc
-5       2      TRc
-5       2.5    TU

-3      0.2     TT 
-3      0.5     TRc
-3       2      TRc
-3       2.5    TU

0       0.2     TT
0       0.5     TRc
0        2      TRc
0        2.5    TU

3       0.5    TT
3       1      TRc
3       2      TRc
3       8      TRc
3       10     TRc
3       12     TRc
3       13     TU

5       0.5    TT
5       1      TRc
5       2      TRc
5       8      TRc
5       10     TRc
5       13     TRc
5       14     TRc
5       15     TU 
};
\end{axis}
\end{tikzpicture}
\caption{Phase diagram for $\Ca = 1/\alpha^* = 10$ for 
the domain $\overline\Omega = [-3,3]^3$, starting with biconcave shapes
with $\mathcal{V}_r = 0.8$ and the shortest axis in the $x_2$-direction.}
\label{fig:phase_muGamma_spont_rotated}
\end{figure}

Here the initial shapes
of the vesicles, for a reduced volume of $\mathcal{V}_r = 0.8$ and surface area
$\mathcal{A}(0) = 4\,\pi$, so that $S=1$, 
were chosen to be numerical approximations of local 
minimizers for the curvature energy $\int_\Gamma (\varkappa-\spont^*)^2 \ds$. 
These discrete local minimizers were obtained with 
the help of the gradient flow scheme from \cite{willmore},
and for the choices $\spont^*=\pm5$ they are displayed in 
Figure~\ref{fig:initialnew}.
For $\Ca = 1/\alpha^*=10$ 
we show a phase diagram of $\mathcal{M} = \mu_\Gamma^*$ 
versus $\spont^*$ in Figure~\ref{fig:phase_muGamma_spont},
where the initial vesicles are aligned such that their shortest axis is in 
the $x_1$-direction.
Similarly, in Figure~\ref{fig:phase_muGamma_spont_rotated}
we show a phase diagram of $\mathcal{M} = \mu_\Gamma^*$ 
versus $\spont^*$ when the initial vesicles are aligned such that their 
shortest axis is in the $x_2$-direction.
\begin{table}
\center
\begin{tabular}{r|c|c|c}
$\mu_\Gamma^*$ & $\spont^* = -5$ & $\spont^* = 0$ & $\spont^* = 5$ \\ \hline
0.05& 0.179 & 0.178 & 0.194 \\ 
0.1 & 0.169 & 0.158 & 0.179 \\
0.2 & 0.161 & 0.116 & 0.138 \\
\end{tabular}
\caption{Some inclination angles $\theta$ for the TT motions in 
Figure~\ref{fig:phase_muGamma_spont}.}
\label{tab:theta_muGamma_spont}
\end{table}
\begin{table}
\center
\begin{tabular}{r|c|c|c}
$\mu_\Gamma^*$ & $\spont^* = -5$ & $\spont^* = 0$ & $\spont^* = 5$ \\ \hline
0.05& 0.180 & 0.156 & 0.162 \\ 
0.1 & 0.160 & 0.132 & 0.156 \\
0.2 & 0.118 & 0.084 & 0.143 \\
\end{tabular}
\caption{Some inclination angles $\theta$ for the TT motions in 
Figure~\ref{fig:phase_muGamma_spont_rotated}.}
\label{tab:theta_muGamma_spont_rotated}
\end{table}
The results in Figures~\ref{fig:phase_muGamma_spont}
and \ref{fig:phase_muGamma_spont_rotated}
indicate that the values of the surface
viscosity, at which the transitions between TT, TR and TU take place,
strongly depend on the spontaneous curvature as well as on the orientation
of the initial vesicle. As in \cite{NoguchiG05a}, where the case $\spont^*=0$
was studied, we also observe that the inclination angle in
the tank treading motion decreases as $\mu_\Gamma^*$ increases,
see Tables~\ref{tab:theta_muGamma_spont} and 
\ref{tab:theta_muGamma_spont_rotated}.

\subsection{Shearing for a torus}
\begin{figure*}
\center
\includegraphics[angle=-0,width=\localwidth]{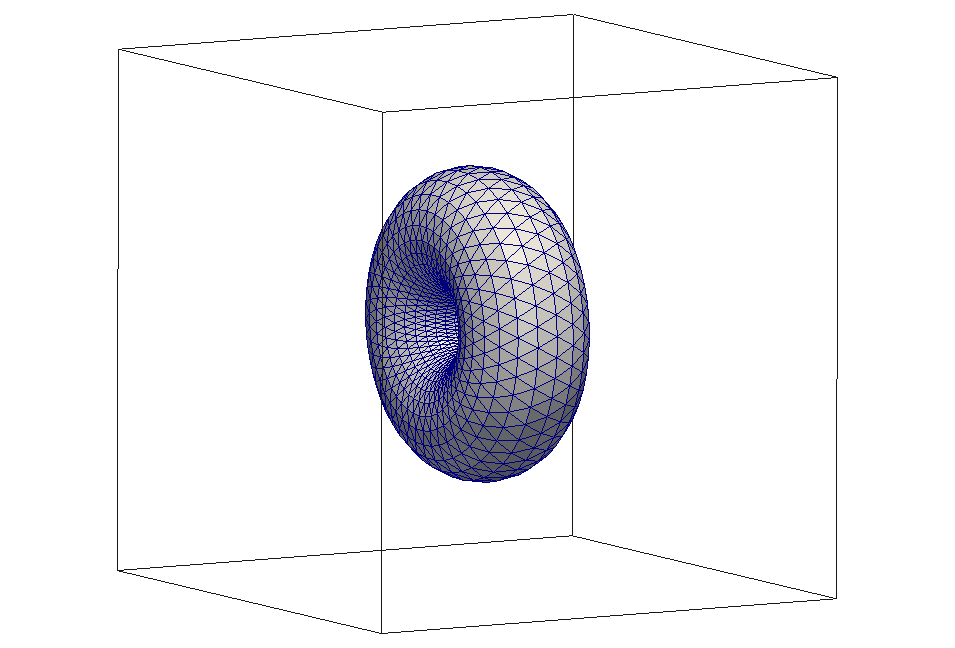} 
\includegraphics[angle=-0,width=\localwidth]{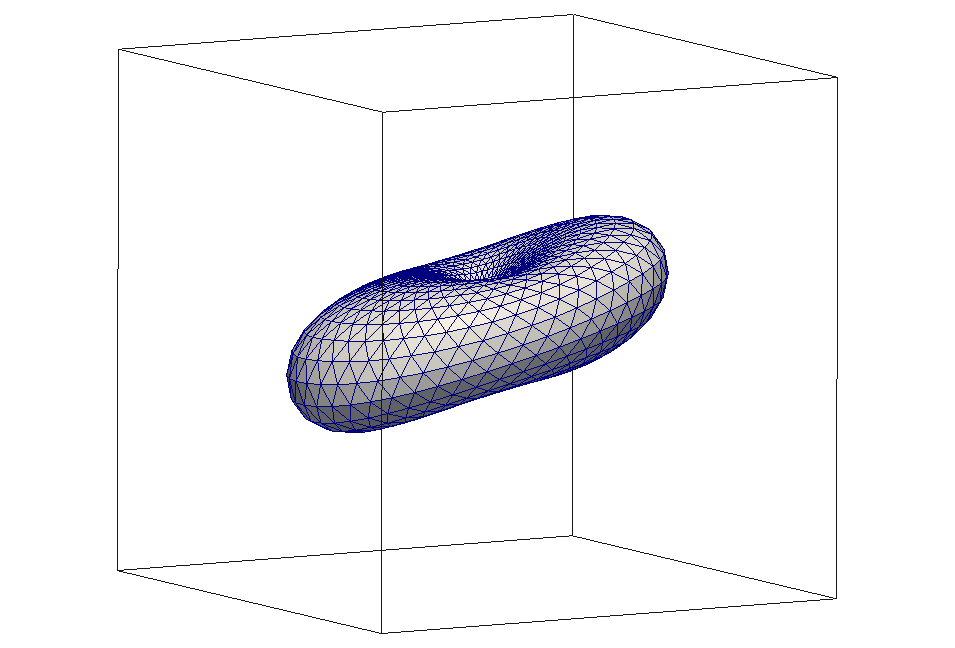} 
\includegraphics[angle=-0,width=\localwidth]{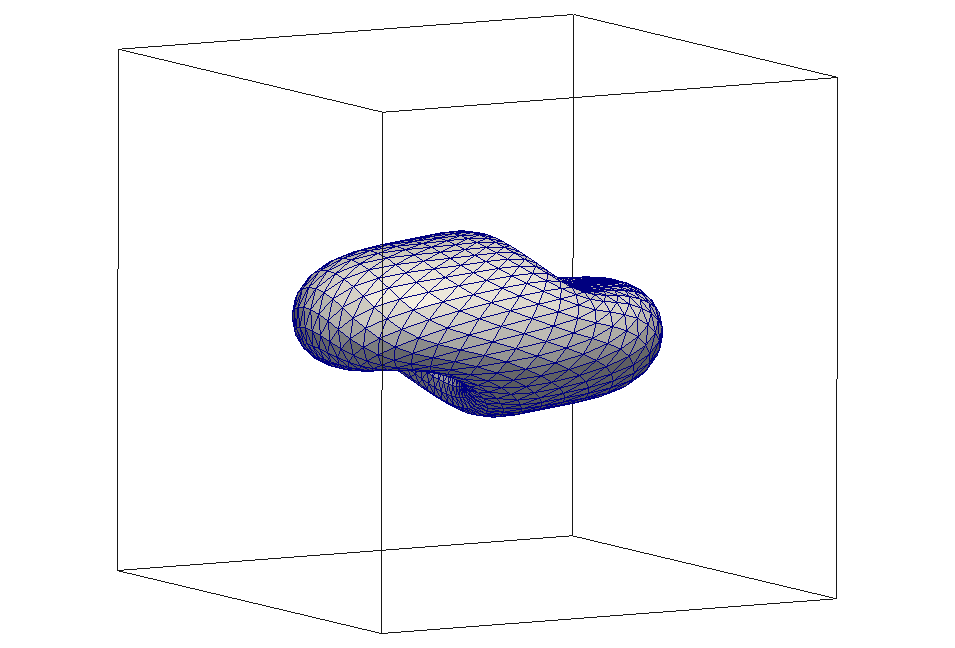} 
\includegraphics[angle=-0,width=\localwidth]{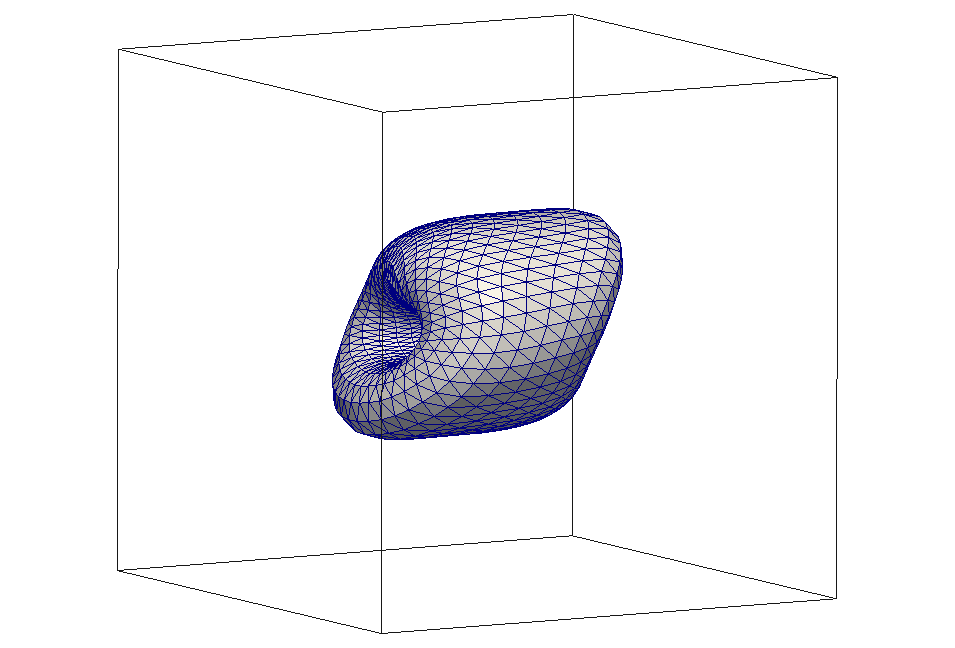}
\\
\includegraphics[angle=-0,width=\localwidth]{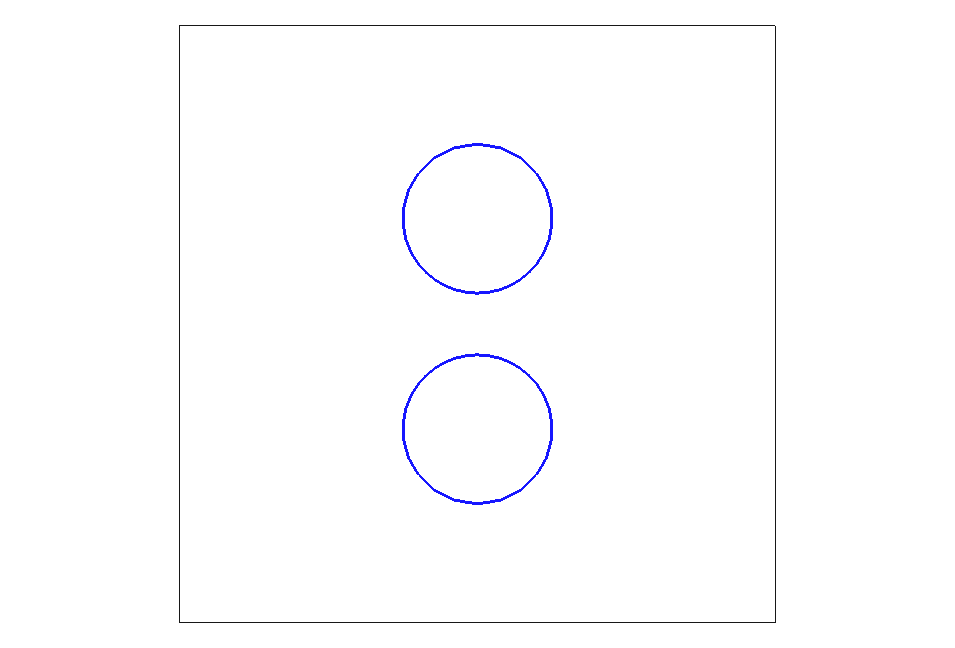} 
\includegraphics[angle=-0,width=\localwidth]{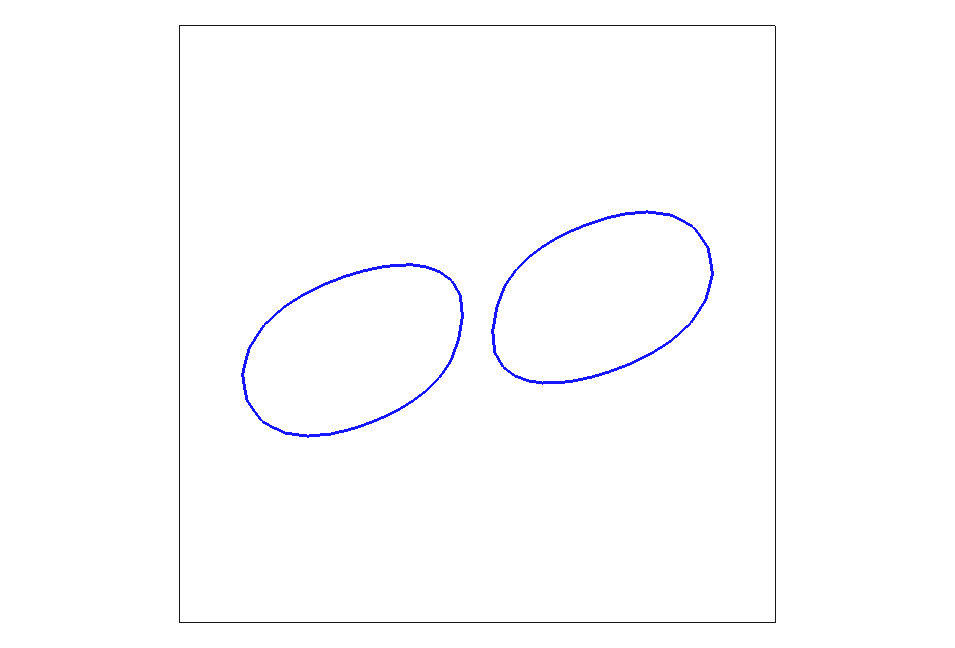} 
\includegraphics[angle=-0,width=\localwidth]{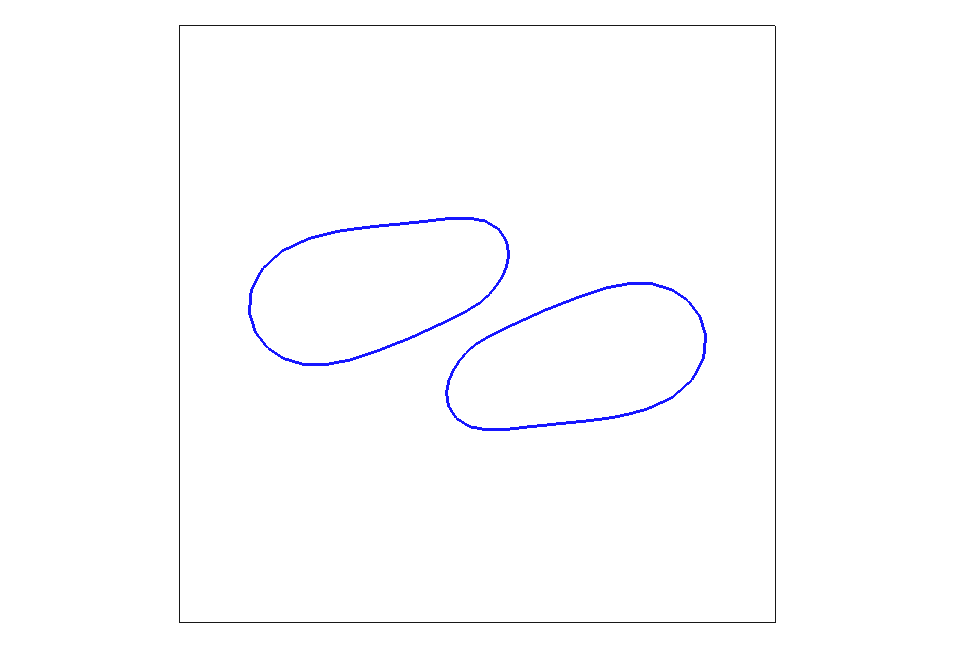} 
\includegraphics[angle=-0,width=\localwidth]{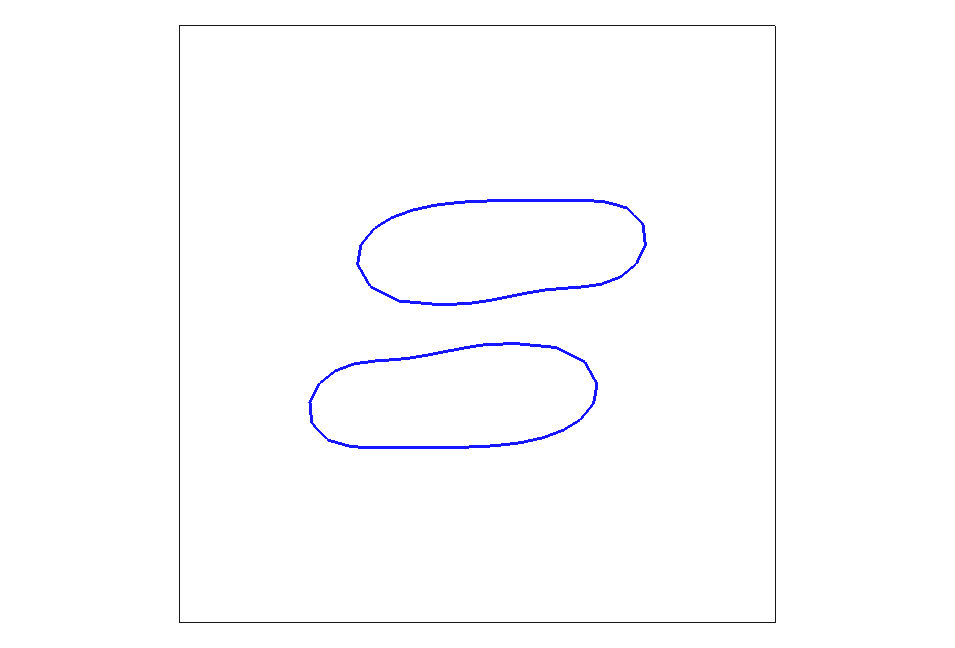} 
\caption{(Color online)
Shear flow for a torus with $\Lambda= \mu_\Gamma^* = 1$.
The plots show the interface $\Gamma^h$ within $\overline\Omega$, as well as
cuts through the $x_1$-$x_3$ plane, 
at times $t=0,\ 2.5,\ 5,\ 7.5$. The interface at $t=10$ is very close to
the plot at $t=2.5$.
}
\label{fig:clifford_rho0_mu1_muGamma1}
\end{figure*}

Here we use as the initial shape a Clifford torus, 
that is aligned with the $x_2$-$x_3$ plane,
with reduced volume 
$\mathcal{V}_r = 0.71$ and $\mathcal{A}(0) = 13.88$, so that $S = 1.05$.
We let $\Lambda = \mu_\Gamma^*=1$, 
$\alpha^*=0.05$ and use the domain $\overline\Omega=[-2,2]^3$.
See Figure~\ref{fig:clifford_rho0_mu1_muGamma1}, where the torus appears to
tumble whilst undergoing strong deformations.

\begin{figure*}
\center
\includegraphics[angle=-0,width=\localwidth]{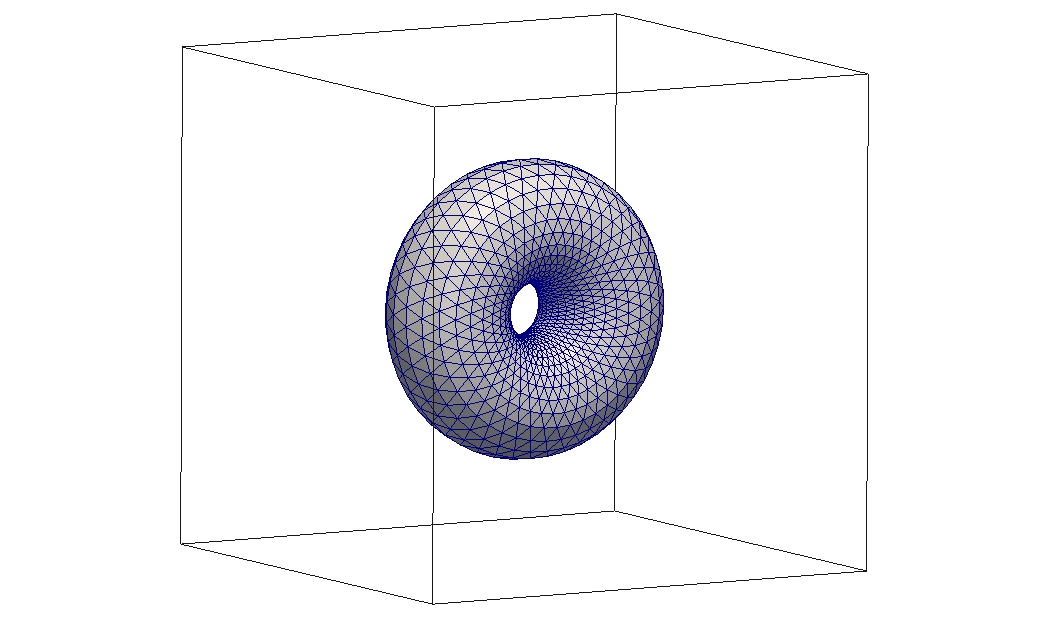} 
\includegraphics[angle=-0,width=\localwidth]{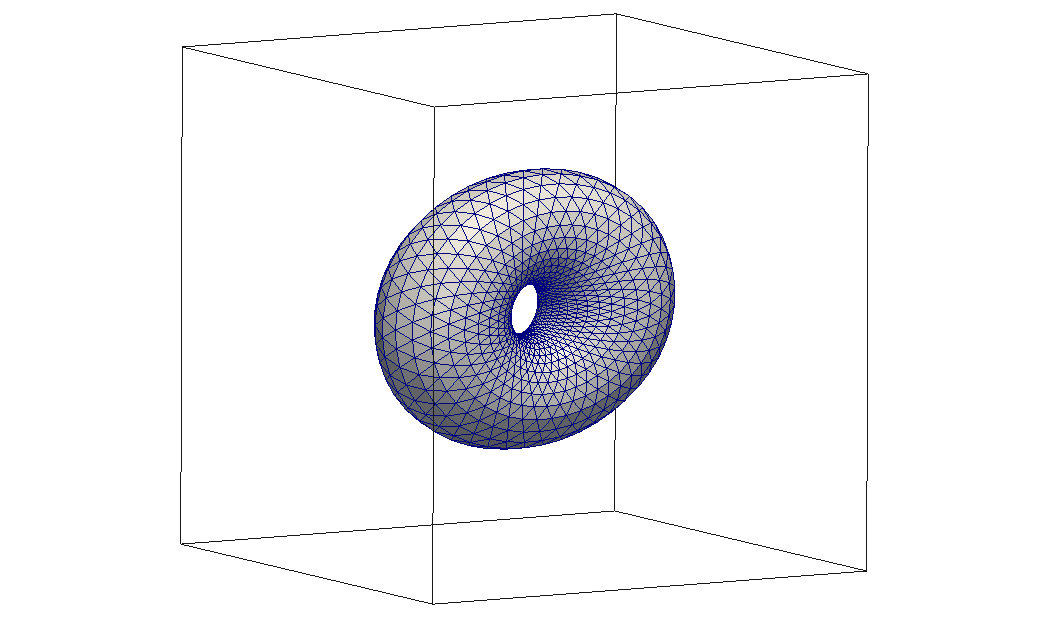} 
\includegraphics[angle=-0,width=\localwidth]{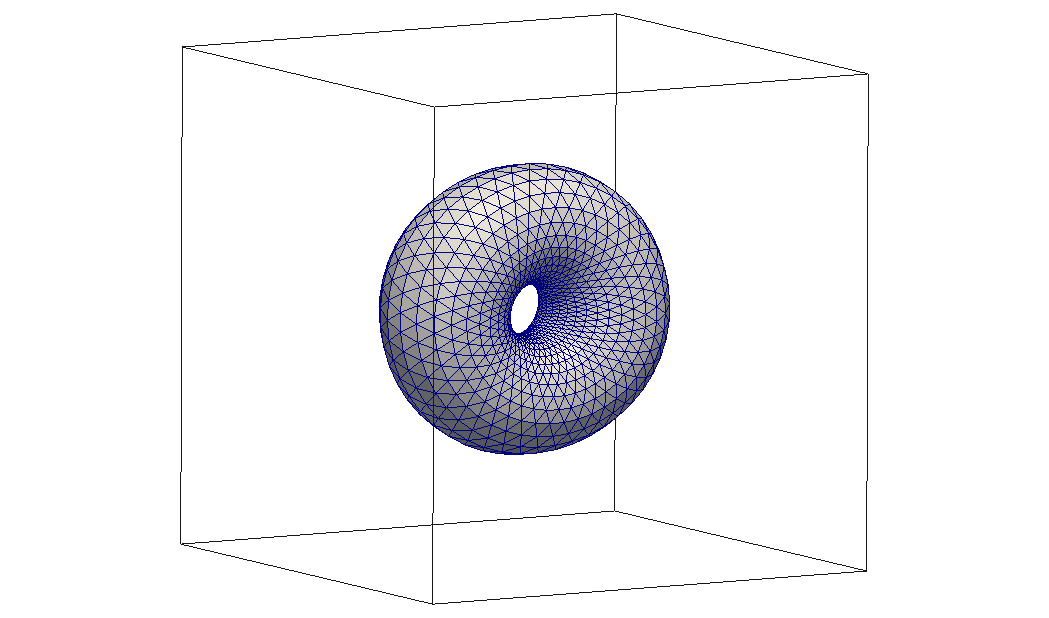} 
\includegraphics[angle=-0,width=\localwidth]{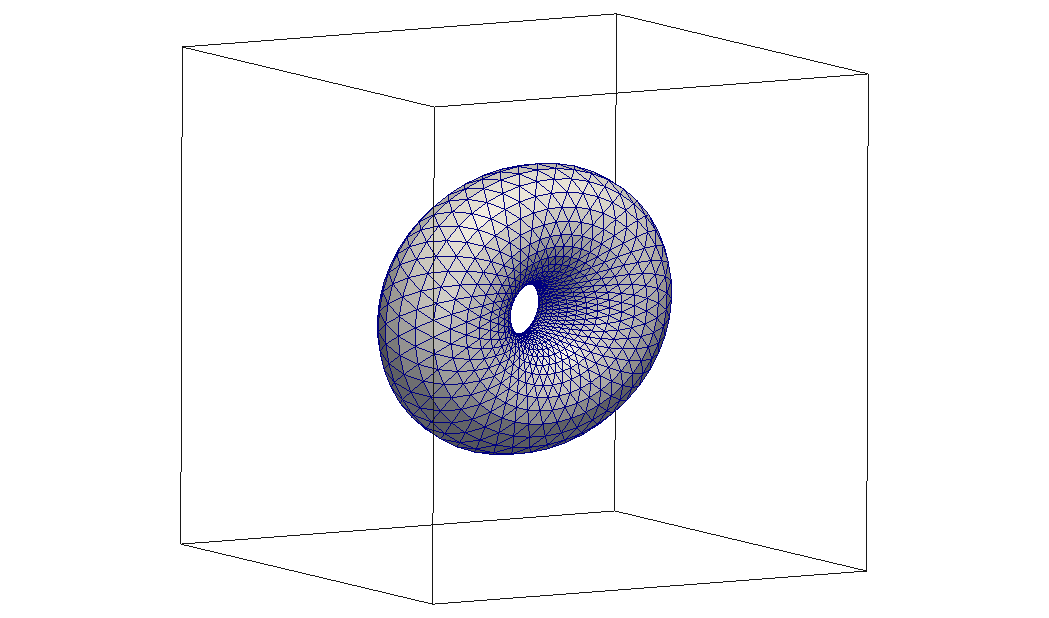} \\
\includegraphics[angle=-0,width=\localwidth]{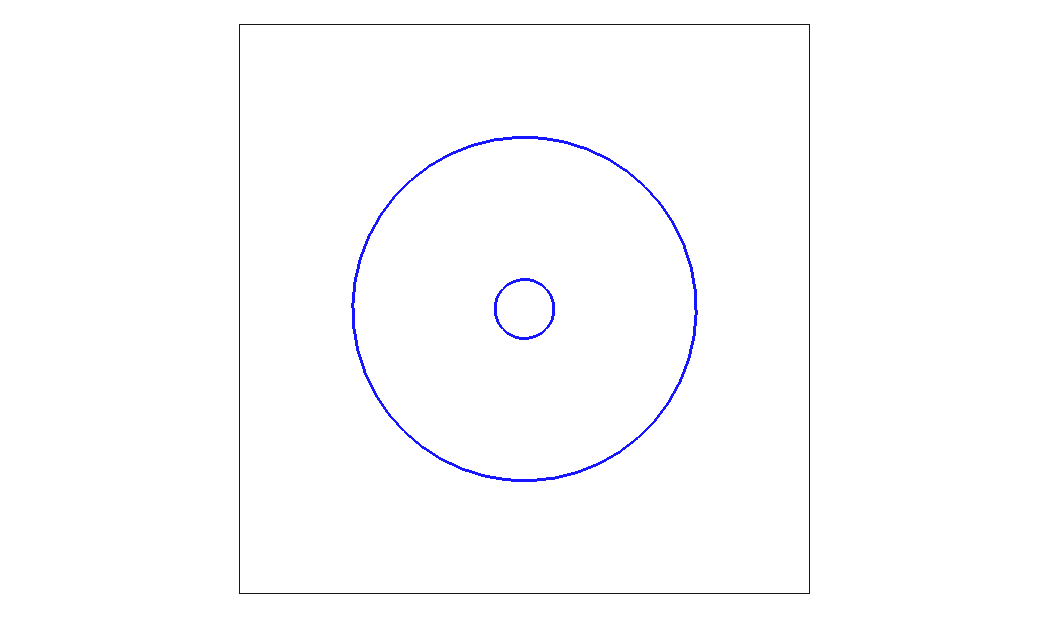} 
\includegraphics[angle=-0,width=\localwidth]{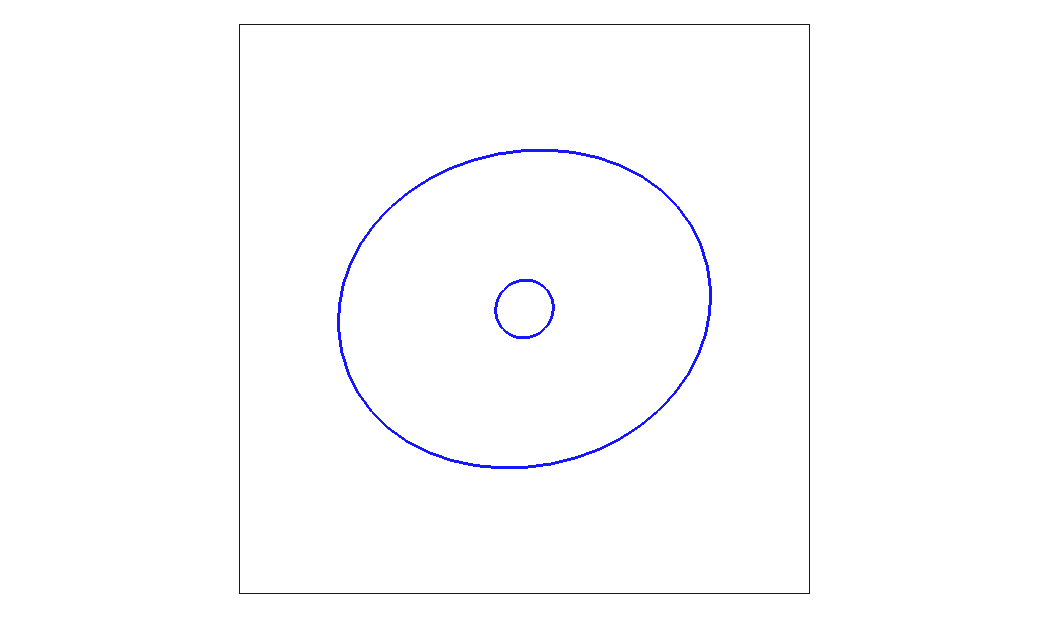} 
\includegraphics[angle=-0,width=\localwidth]{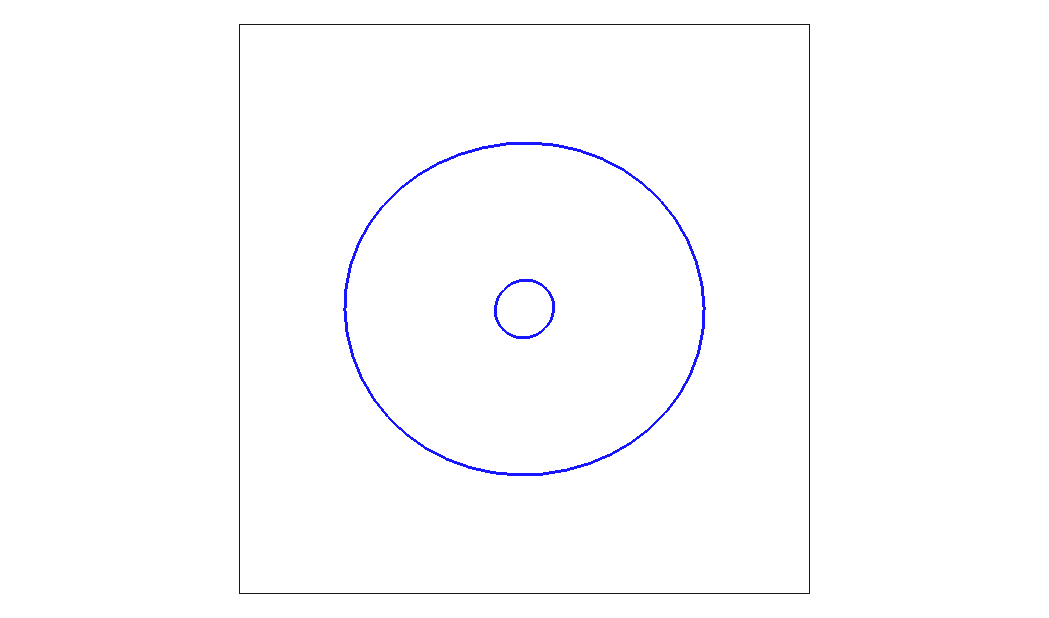} 
\includegraphics[angle=-0,width=\localwidth]{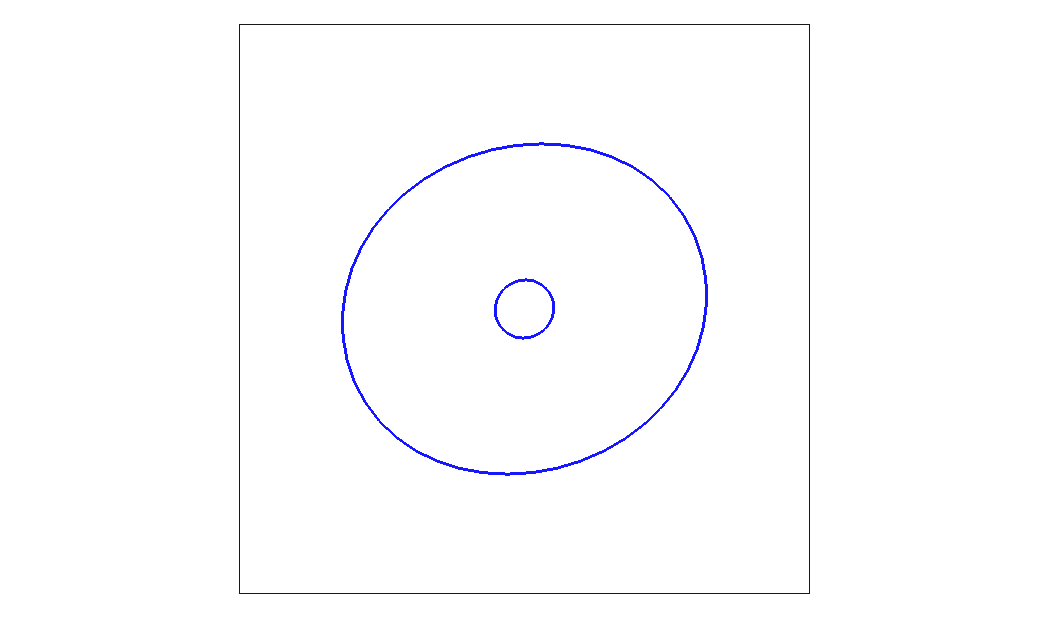} 
\caption{(Color online)
Shear flow for a torus with $\Lambda =1$, $\mu_\Gamma^* = 10$.
The plots show the interface $\Gamma^h$ within $\overline\Omega$, as well as
cuts through the $x_1$-$x_3$ plane, 
at times $t=0,\ 2.5,\ 5,\ 7.5$.
}
\label{fig:clifford2a1_rho0_mu1_muGamma10}
\end{figure*}

\begin{figure*}
\center
\includegraphics[angle=-0,width=\localwidth]{figures/clifford2a1_r0_m1_m10_t0} 
\includegraphics[angle=-0,width=\localwidth]{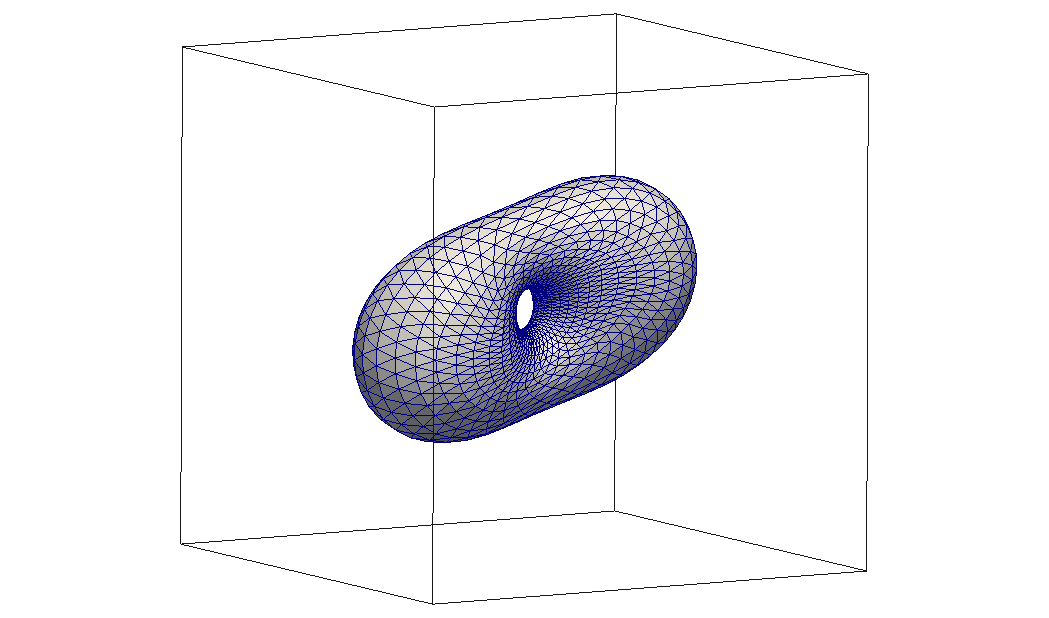} 
\includegraphics[angle=-0,width=\localwidth]{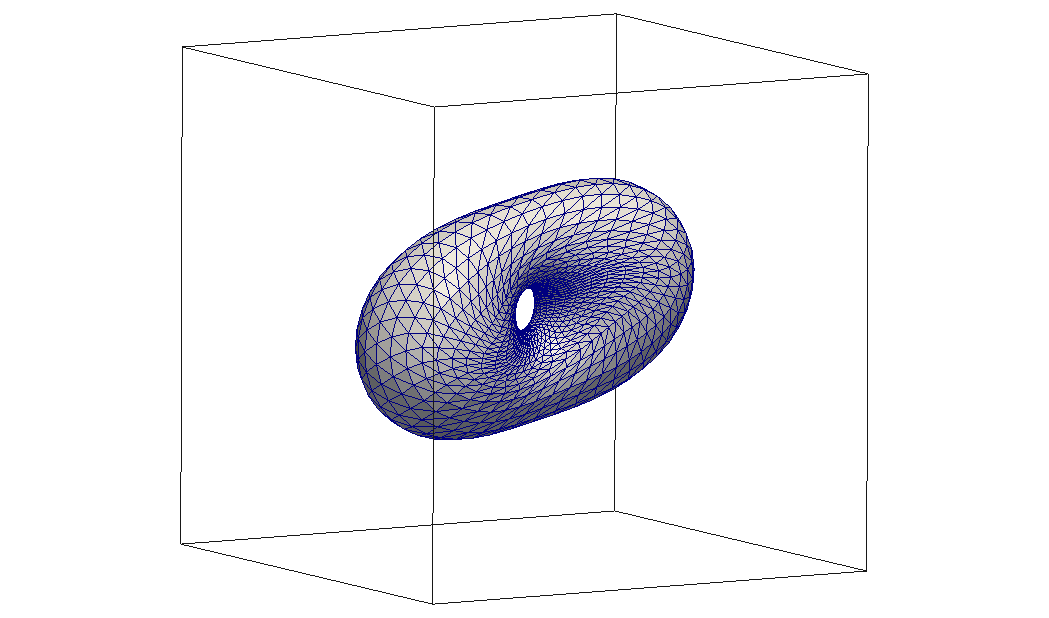} 
\includegraphics[angle=-0,width=\localwidth]{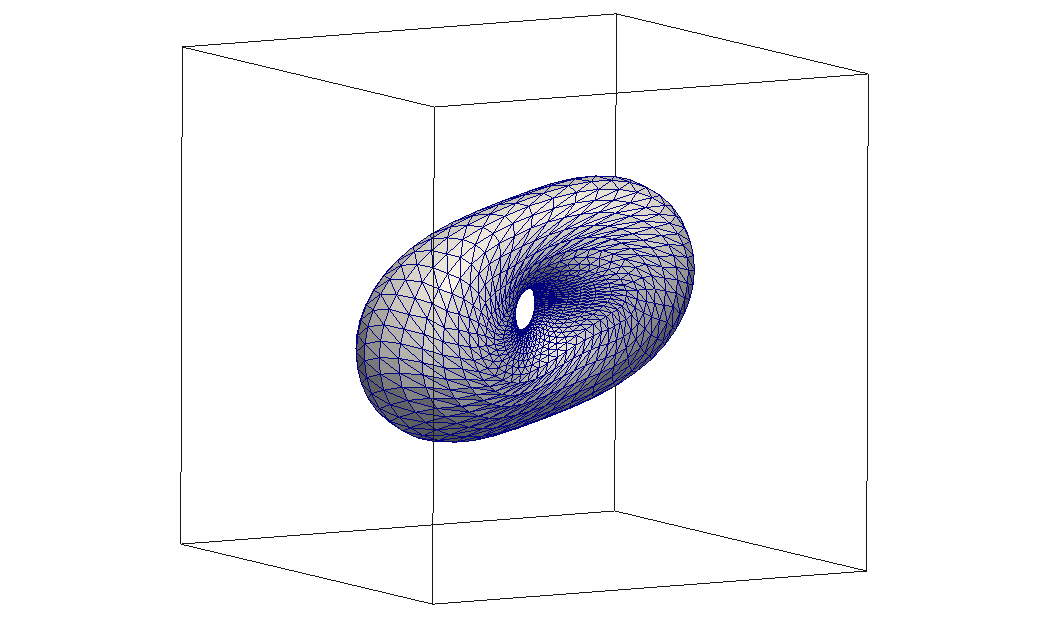} \\
\includegraphics[angle=-0,width=\localwidth]{figures/clifford2a1_r0_m1_m10_t0cut} 
\includegraphics[angle=-0,width=\localwidth]{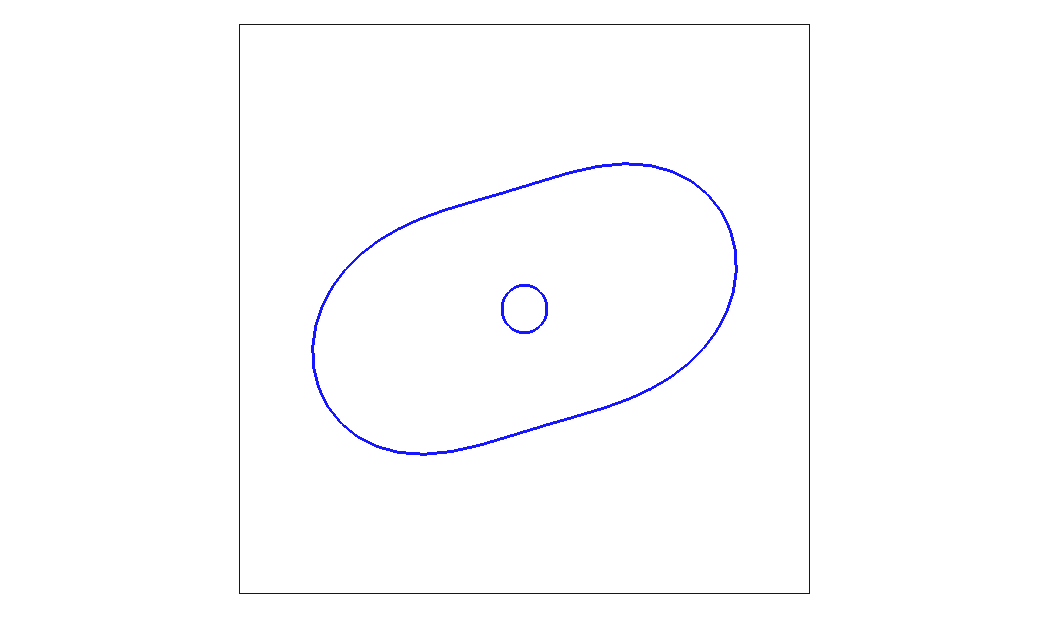} 
\includegraphics[angle=-0,width=\localwidth]{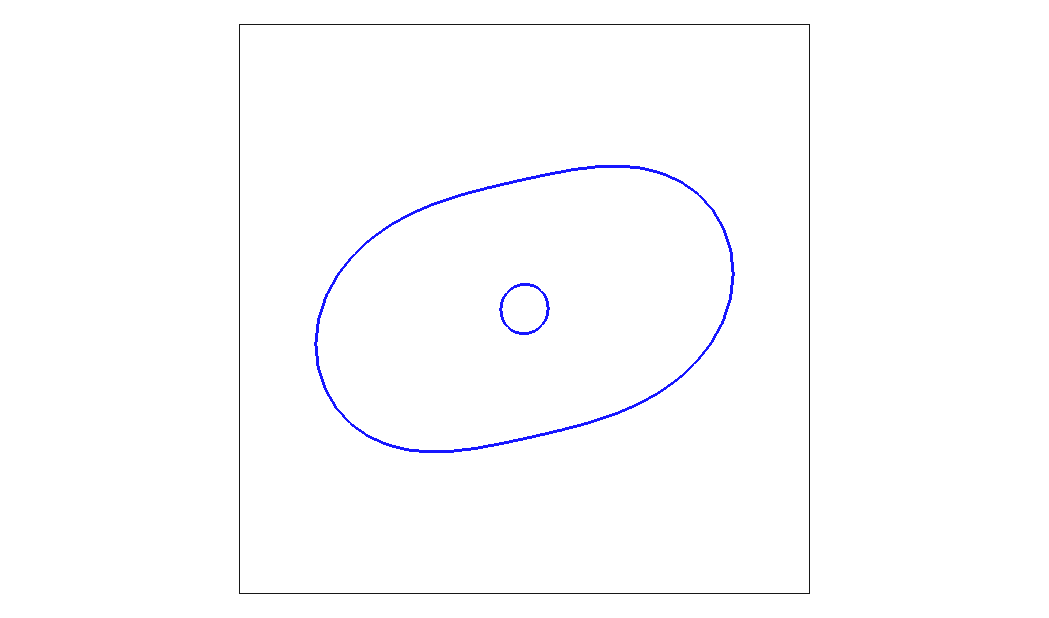} 
\includegraphics[angle=-0,width=\localwidth]{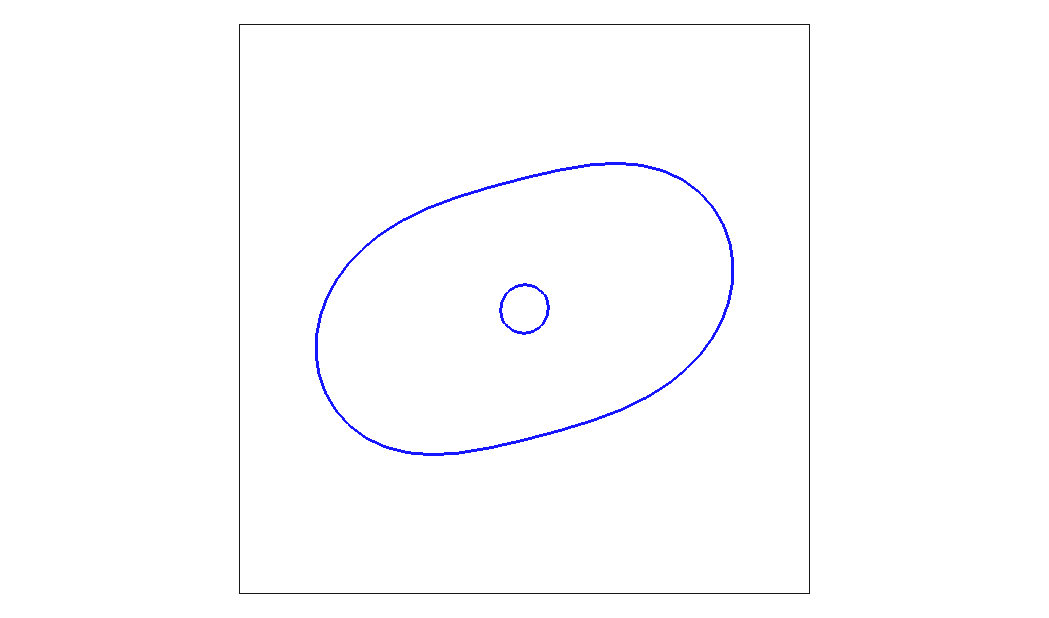} 
\caption{(Color online)
Shear flow for a torus with $\Lambda =1$, $\mu_\Gamma^* = 0$.
The plots show the interface $\Gamma^h$ within $\overline\Omega$, as well as
cuts through the $x_1$-$x_3$ plane, at times $t=0,\ 2.5,\ 5,\ 7.5$.
}
\label{fig:clifford2a1_rho0_mu1_muGamma0}
\end{figure*}

\begin{figure}
\center
\mbox{
\includegraphics[angle=-90,width=\localwidth]{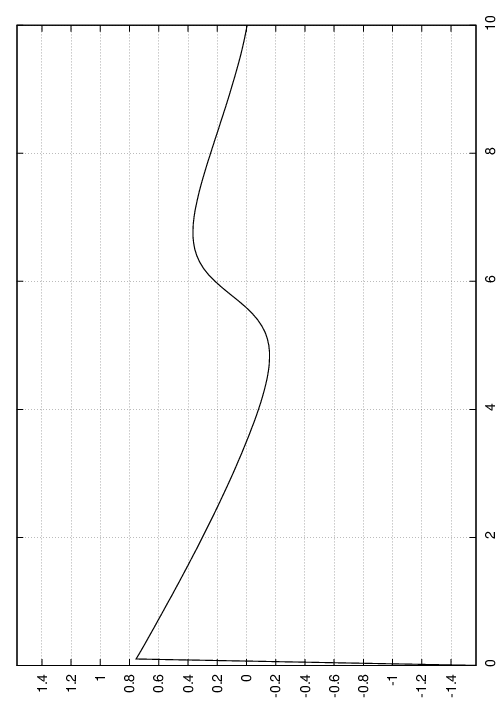} 
\includegraphics[angle=-90,width=\localwidth]{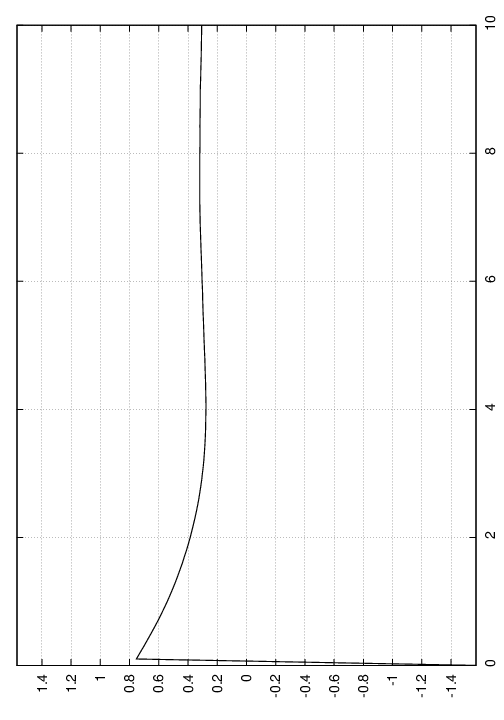} 
}
\caption{The inclination angle $\theta$ for the simulations in 
Figures~\ref{fig:clifford2a1_rho0_mu1_muGamma10} and
\ref{fig:clifford2a1_rho0_mu1_muGamma0}.}
\label{fig:theta_clifford}
\end{figure}

\begin{figure}
\center
\mbox{
\includegraphics[angle=-0,width=\localwidth]{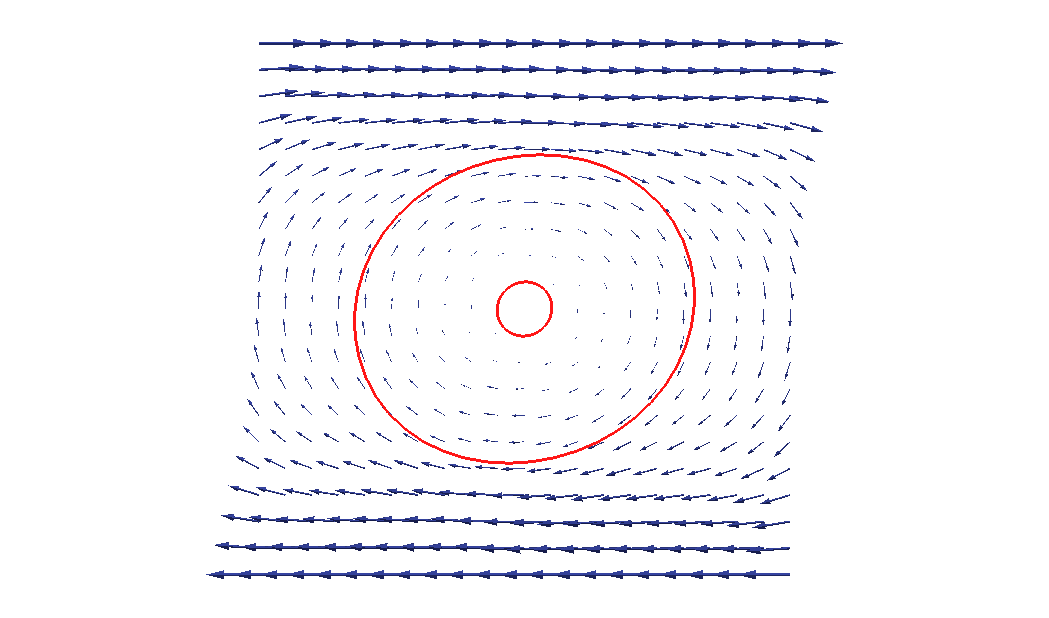}
\includegraphics[angle=-0,width=\localwidth]{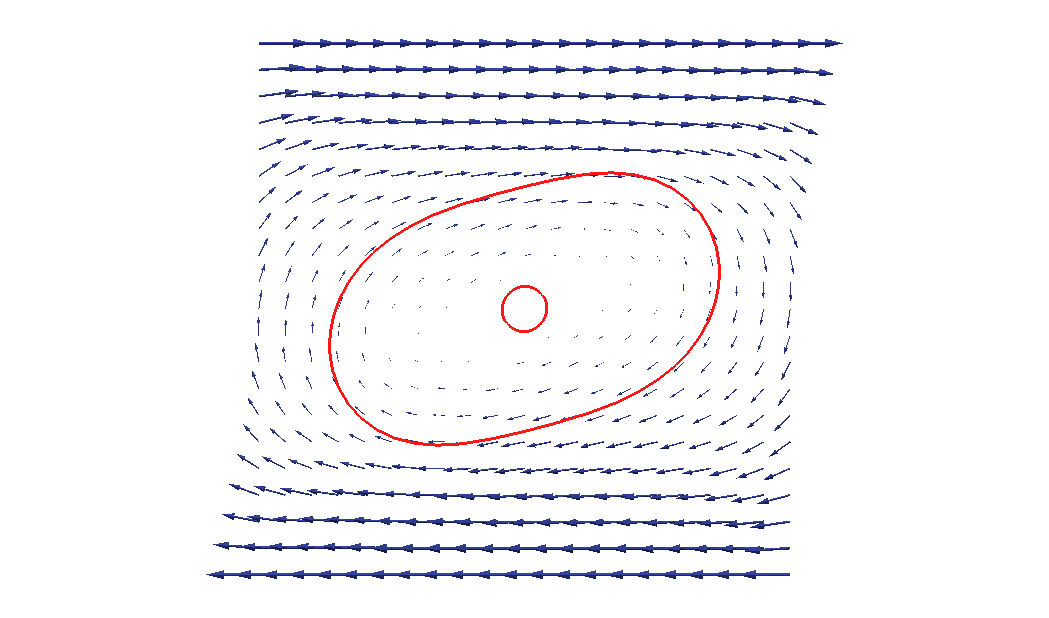}
}
\caption{(Color online)
The flow at time $t=7.5$ in the $x_1$-$x_3$ plane for the simulations
in Figures~\ref{fig:clifford2a1_rho0_mu1_muGamma10} and
\ref{fig:clifford2a1_rho0_mu1_muGamma0}.}
\label{fig:flow_clifford}
\end{figure}

Repeating the experiment for an initial torus aligned with the shear flow
direction, and setting $\alpha^* = 1$ and $\mu_\Gamma^* =10$, 
leads to the results shown in 
Figure~\ref{fig:clifford2a1_rho0_mu1_muGamma10}. This shows a TR motion.
Setting $\mu_\Gamma^* = 0$, on the other hand, leads to TT, as shown in 
Figure~\ref{fig:clifford2a1_rho0_mu1_muGamma0}.
A plot of the inclination angle $\theta$ for the simulations in
Figures~\ref{fig:clifford2a1_rho0_mu1_muGamma10} and
\ref{fig:clifford2a1_rho0_mu1_muGamma0} 
can be seen in Figure~\ref{fig:theta_clifford}, while we visualize the flow in
the $x_1$-$x_3$ plane in Figure~\ref{fig:flow_clifford}.

\subsection{Effect of area difference elasticity}
\begin{figure*}
\center
\mbox{
\hspace{-4mm}
\includegraphics[angle=-0,width=\localwidth]{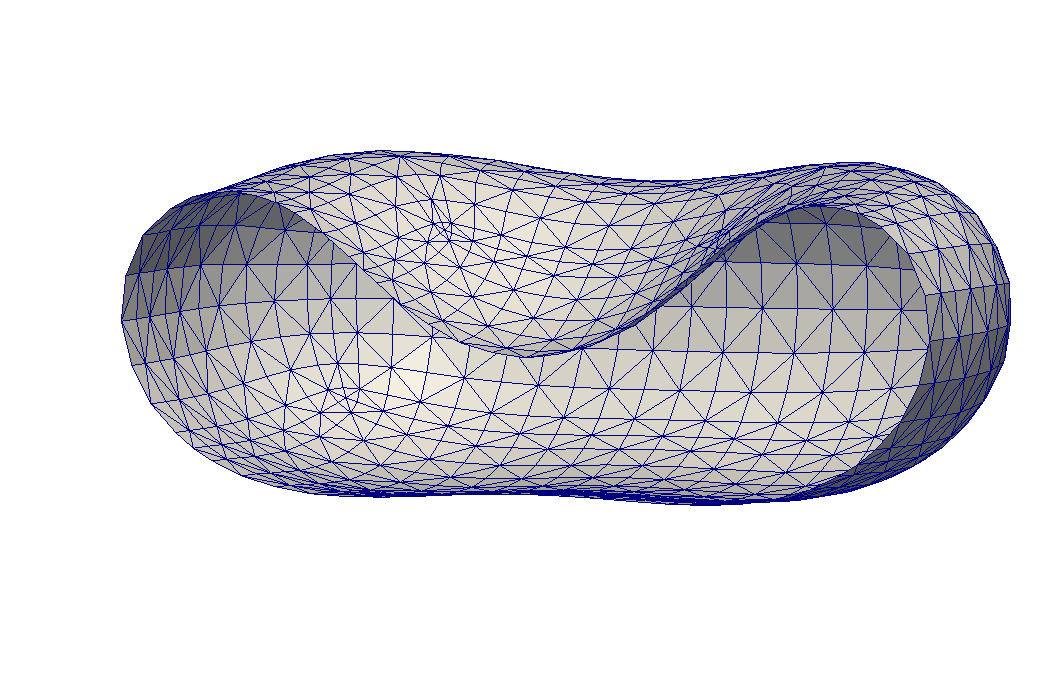} 
\includegraphics[angle=-0,width=\localwidth]{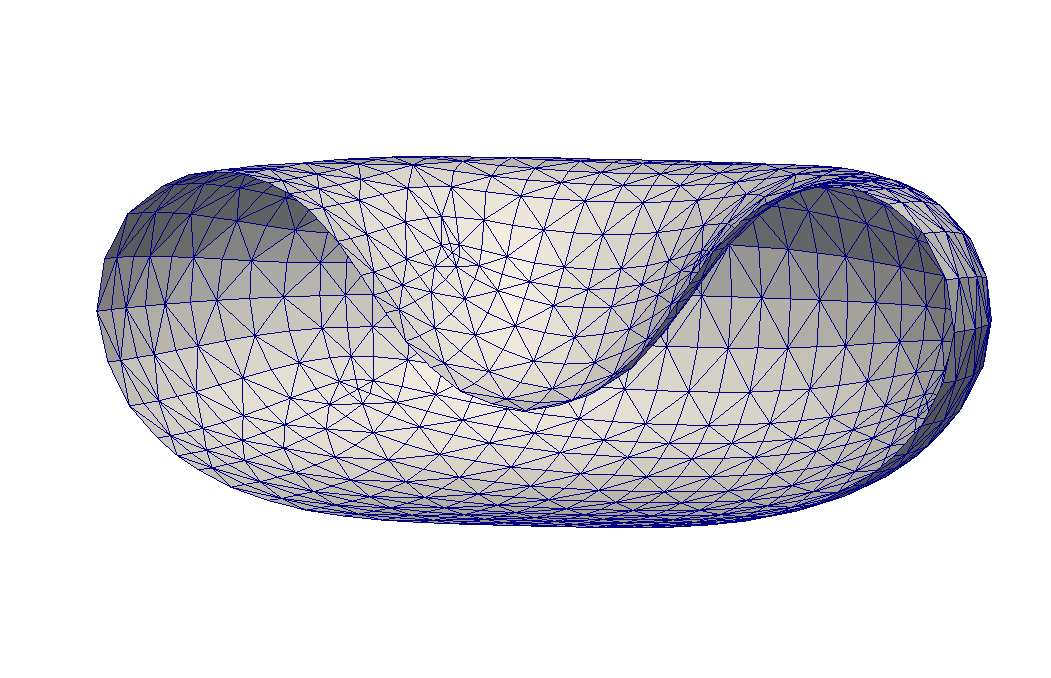} 
\includegraphics[angle=-0,width=\localwidth]{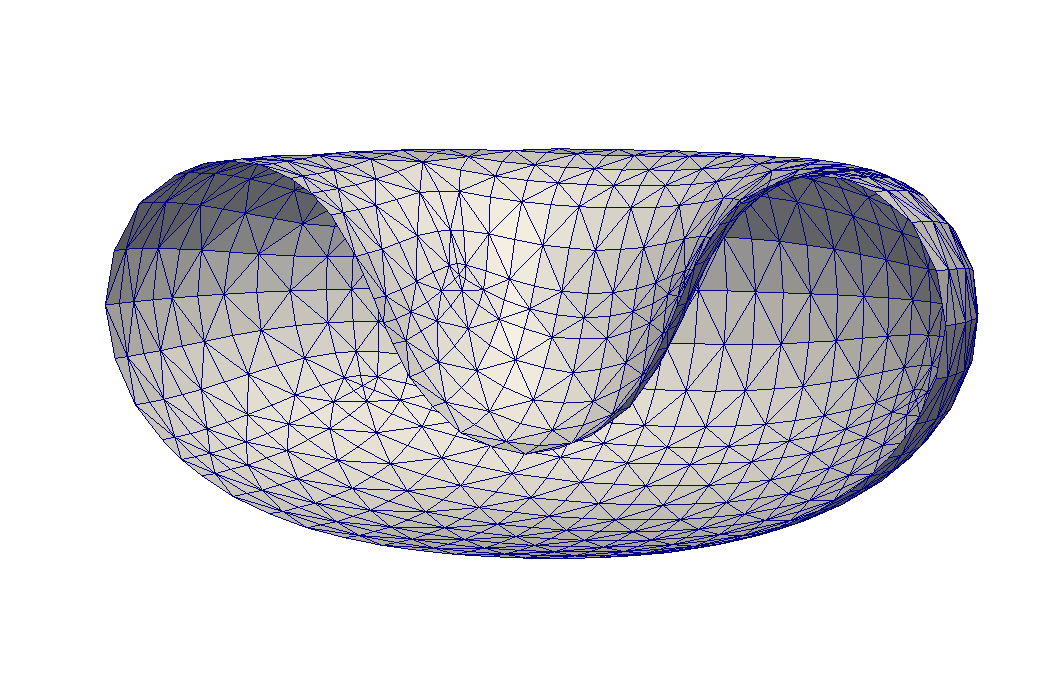} 
\includegraphics[angle=-0,width=\localwidth]{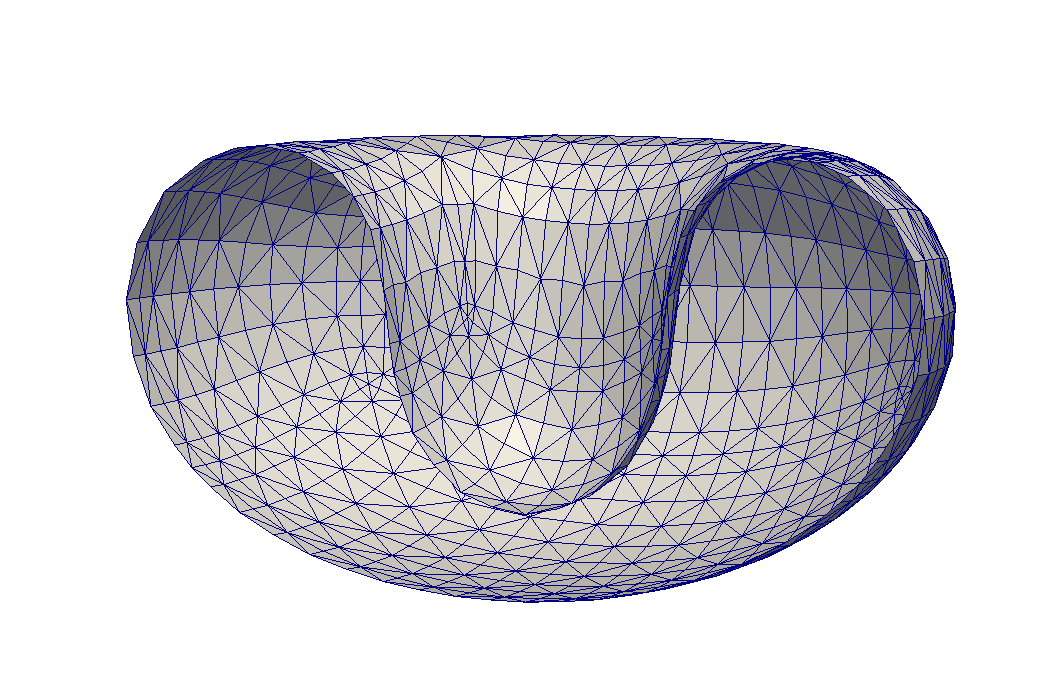} 
}
\mbox{
\hspace{-4mm}
\includegraphics[angle=-0,width=\localwidth]{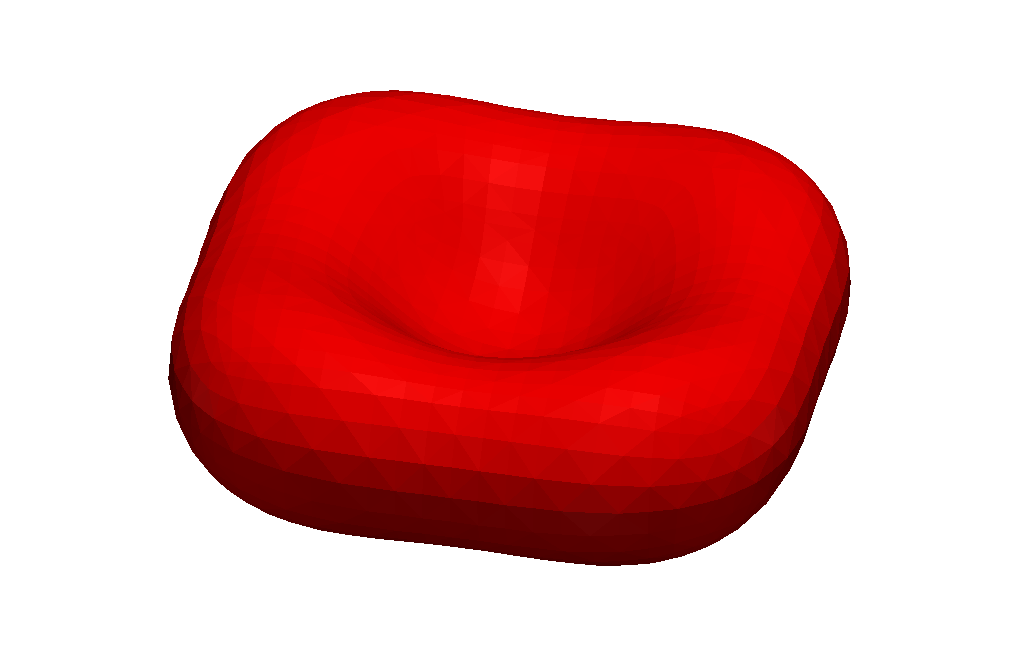} 
\includegraphics[angle=-0,width=\localwidth]{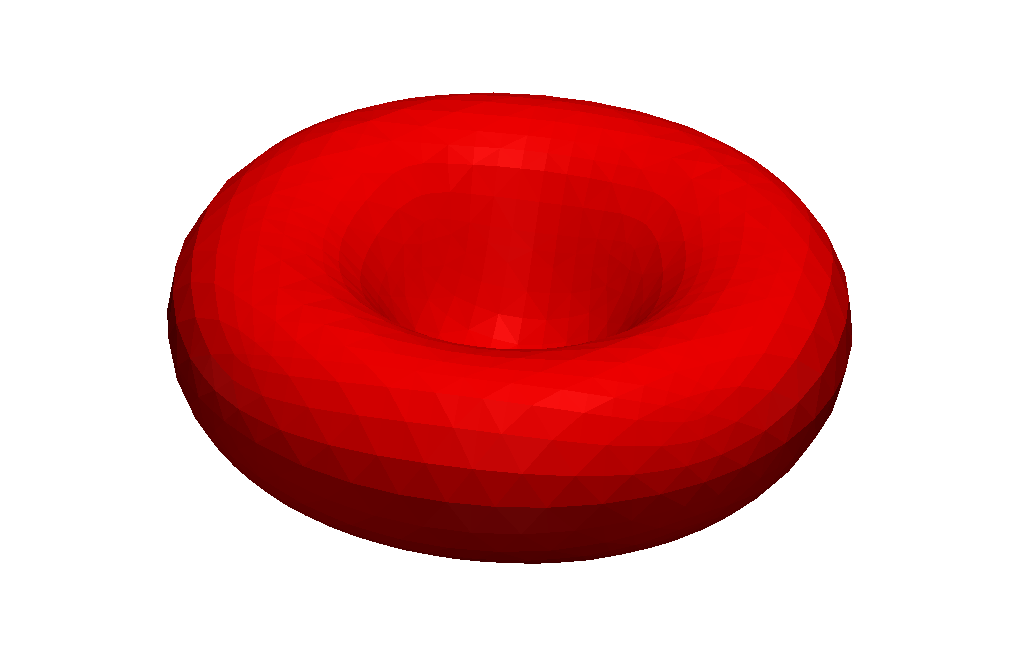} 
\includegraphics[angle=-0,width=\localwidth]{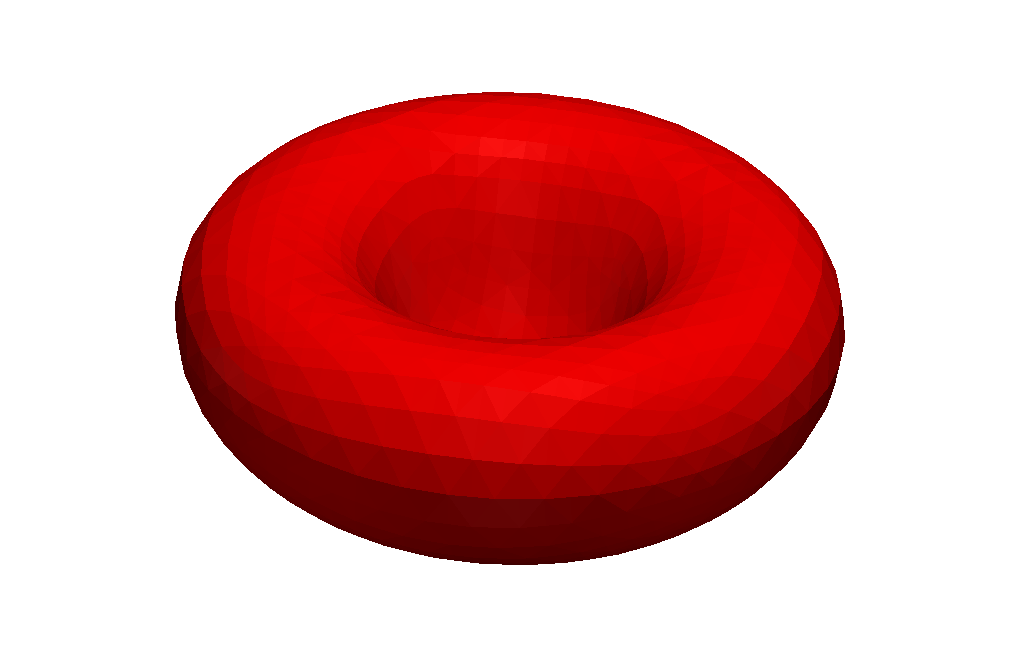} 
\includegraphics[angle=-0,width=\localwidth]{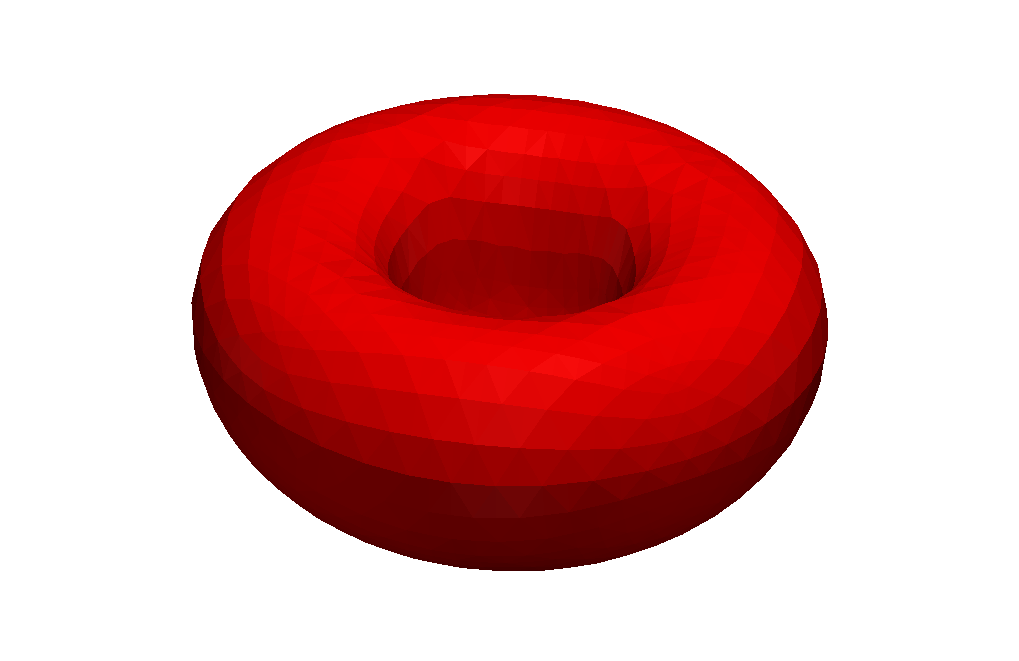} 
}
\caption{(Color online)
Flow for a cup-like stomatocyte shape with $\mathcal{V}_r = 0.65$
for $M_0^* = -48.24$ and $\beta^* = 0.053$.
The plots show the interface $\Gamma^h$ at times $t=0,\ 5,\ 10,\ 20$,
with the top row visualizing the triangulations by showing half the vesicle.
}
\label{fig:stokes_cup2_vr065}
\end{figure*}

\begin{figure*}
\center
\mbox{
\includegraphics[angle=-0,width=\localwidth]{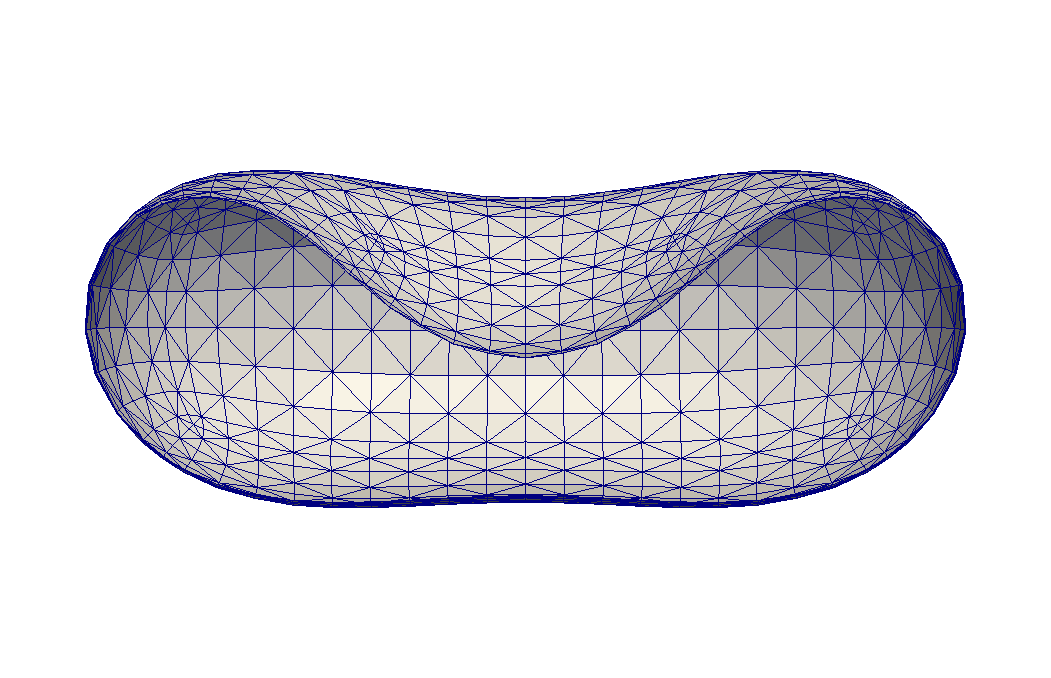} 
\includegraphics[angle=-0,width=\localwidth]{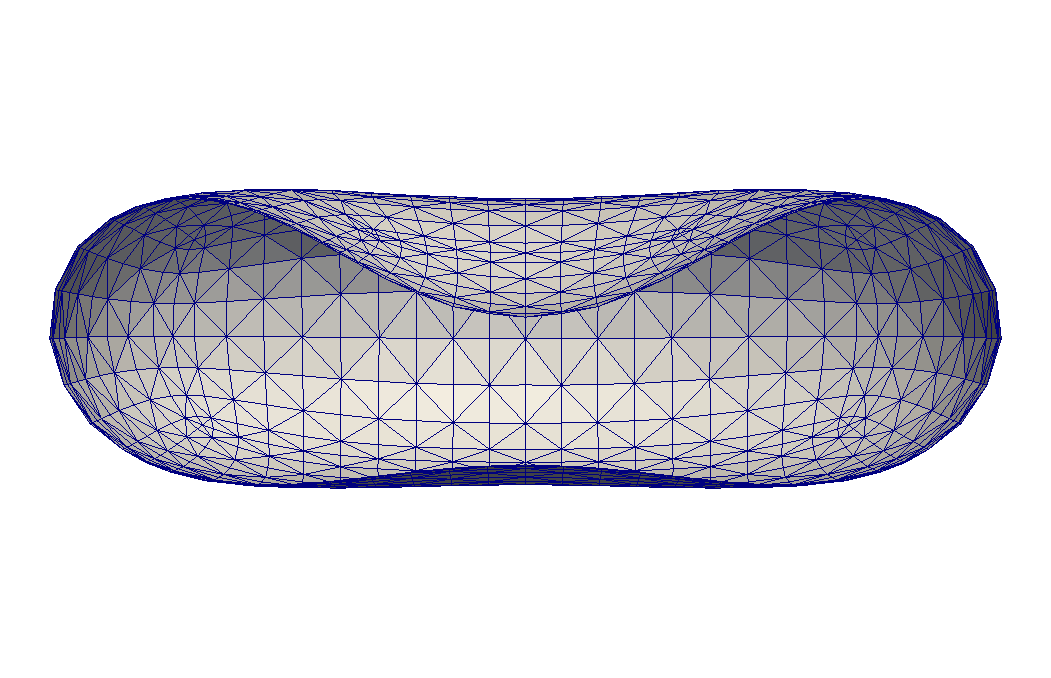} 
\includegraphics[angle=-0,width=\localwidth]{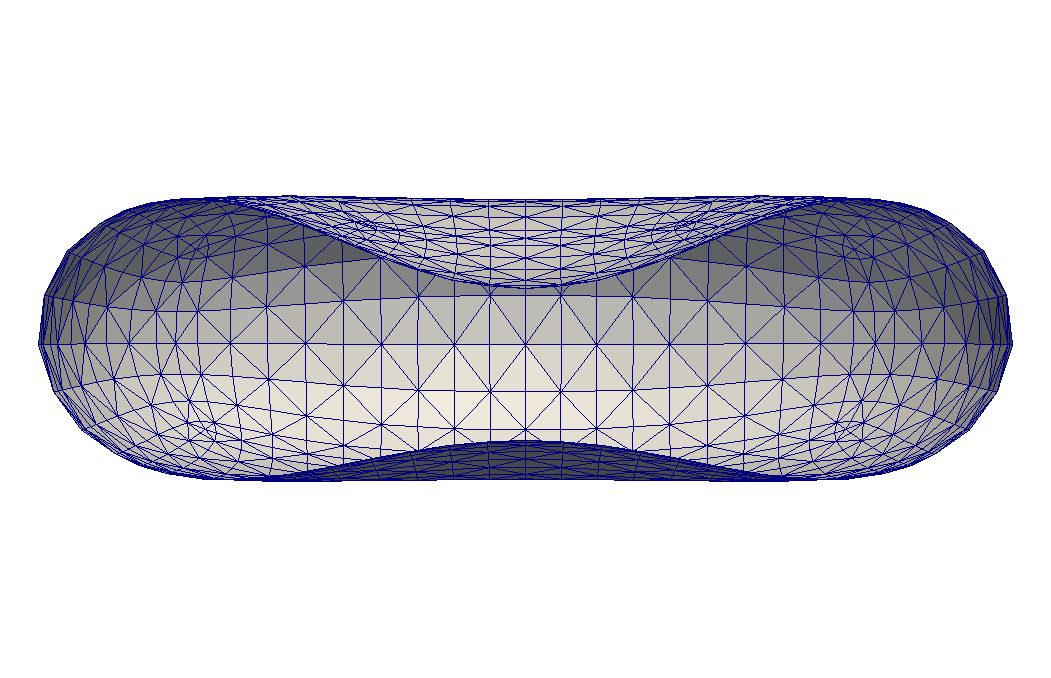} 
\includegraphics[angle=-0,width=\localwidth]{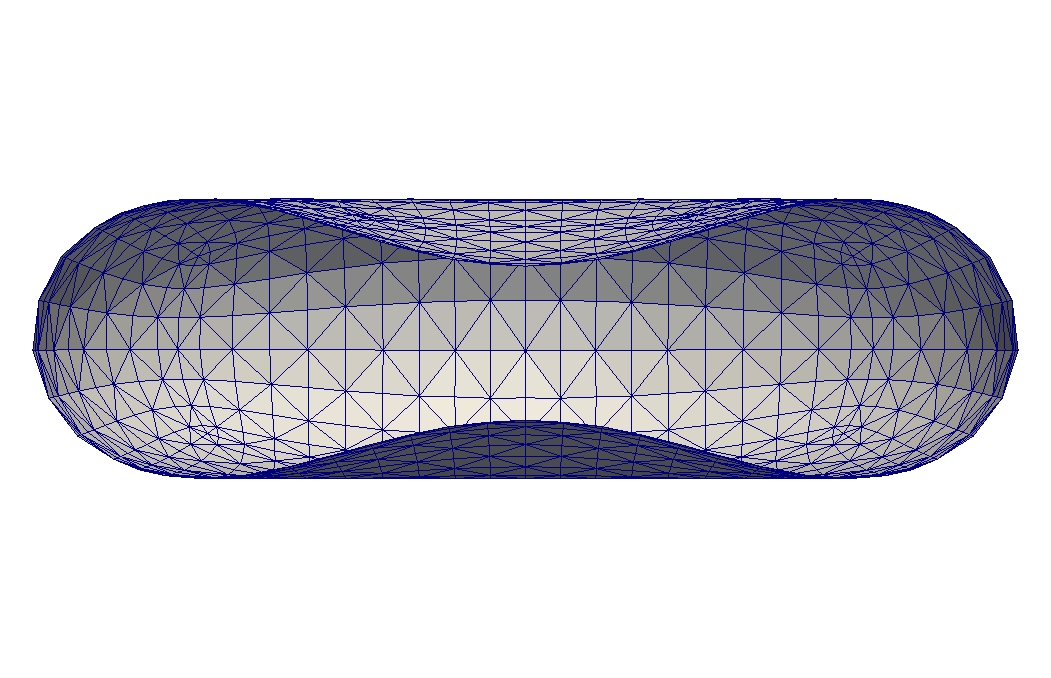} 
} 
\mbox{
\includegraphics[angle=-0,width=\localwidth]{figures/GDs_cup2_vr065_0v2} 
\includegraphics[angle=-0,width=\localwidth]{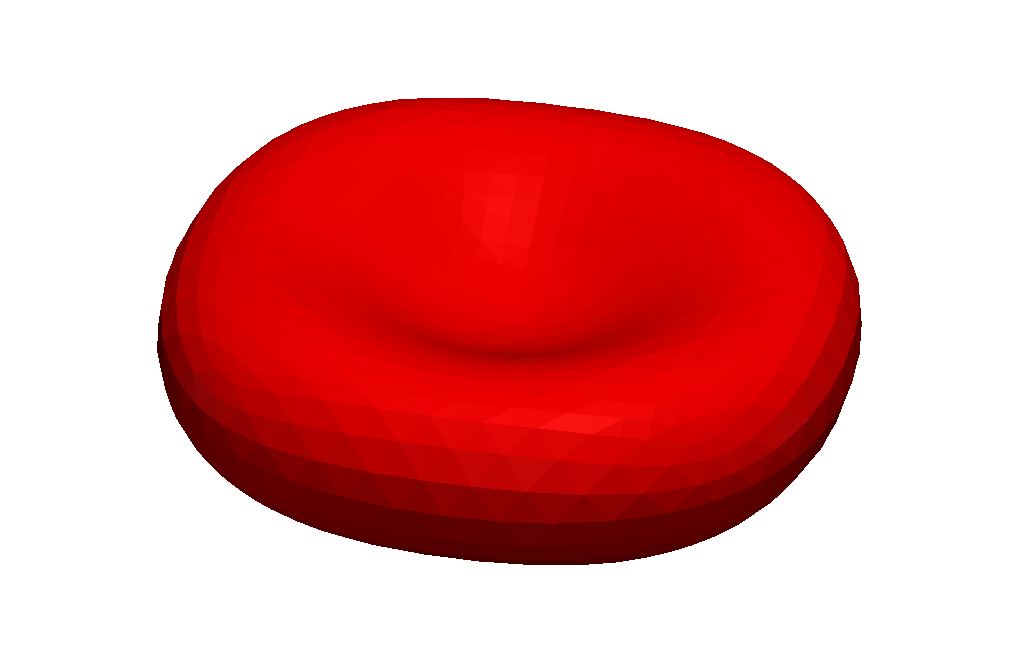} 
\includegraphics[angle=-0,width=\localwidth]{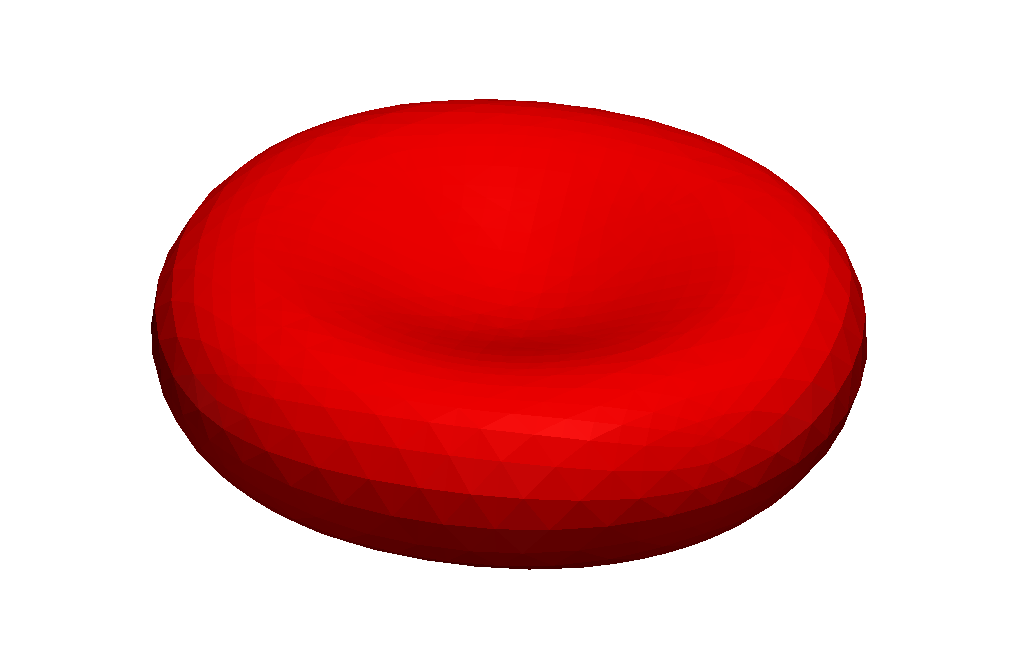} 
\includegraphics[angle=-0,width=\localwidth]{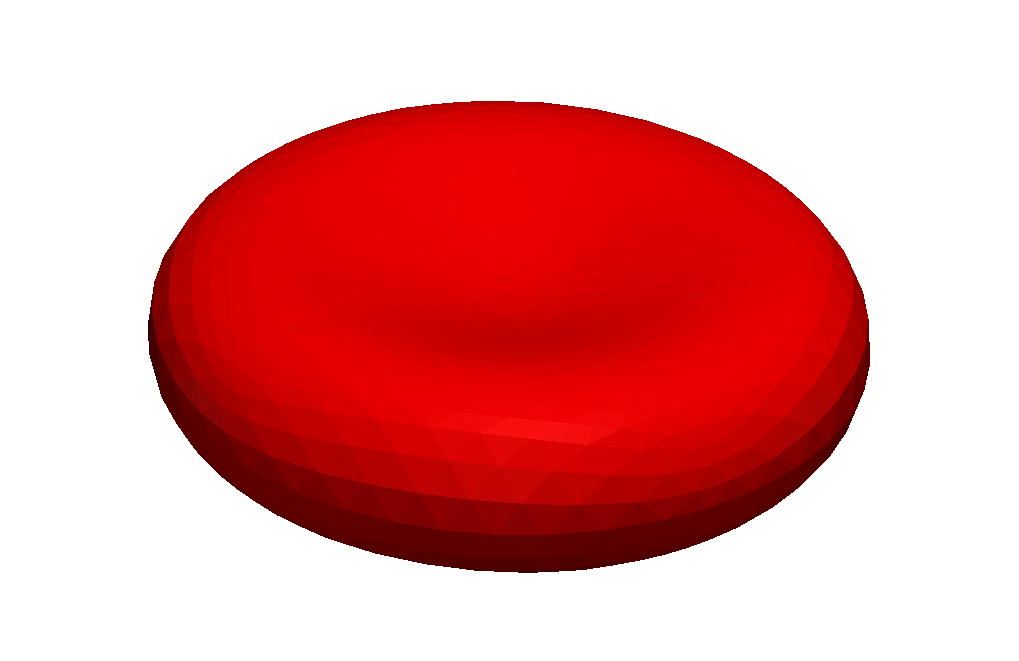} 
} 
\caption{(Color online)
Same as Figure~\ref{fig:stokes_cup2_vr065} with $\beta^*=0$.
}
\label{fig:stokes_cup2_vr065rho0}
\end{figure*}

We consider $\overline\Omega = [-4,4]^3$
and set $\Lambda = \mu_\Gamma^* = \alpha^* = 1$. 
The parameters for $\vec f_\Gamma^*$ are
$\beta^* = 0.053$ and $M_0^* = -48.24$. For the vesicle
we use a cup-like stomatocyte initial shape with $\mathcal{V}_r = 0.65$ and 
$\mathcal{A}(0) = 82.31$, so that $S = 2.56$.
See Figure~\ref{fig:stokes_cup2_vr065} for a numerical simulation.
As a comparison, we show the same simulation with $\beta^* = 0$ in 
Figure~\ref{fig:stokes_cup2_vr065rho0}.

\begin{figure}
\center
\mbox{
\psfrag{v(t) / v(0)}{$~\large\mathcal{V}(t)/\mathcal{V}(0)$}
\psfrag{a(t) / a(0)}{$~\large\mathcal{A}(t)/\mathcal{A}(0)$}
\includegraphics[angle=-90,width=0.49\textwidth]{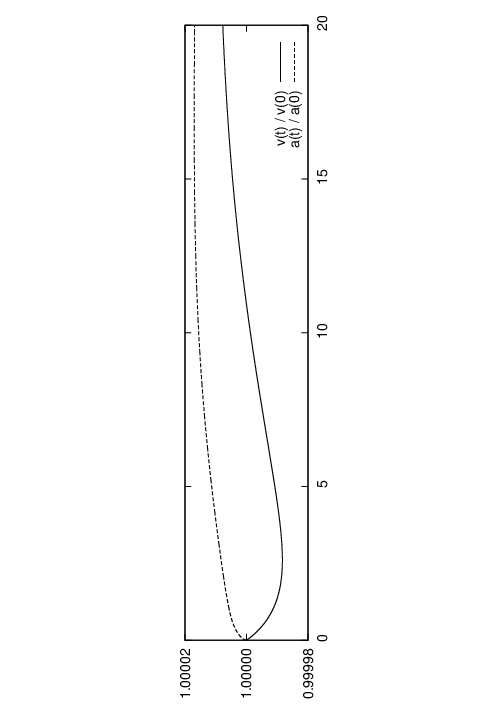}
}
\caption{The evolutions of the relative discrete volume 
$\mathcal{V}(t)/\mathcal{V}(0)$, and the
relative discrete surface area $\mathcal{A}(t)/\mathcal{A}(0)$ over time.}
\label{fig:stokes_cup2_volarea} 
\end{figure}

In Figure~\ref{fig:stokes_cup2_volarea} we show the evolutions of the discrete
volume of the inner phase and the discrete surface area over time. Clearly
these two quantities are preserved almost exactly for our numerical scheme
in this simulation. In fact, a semidiscrete variant of our scheme conserves
these two quantities exactly, and so in practice the fully discrete algorithm
will preserve them well for sufficiently small time step sizes.

\begin{figure*}
\center
\mbox{
\hspace{-4mm}
\includegraphics[angle=-0,width=\localwidth]{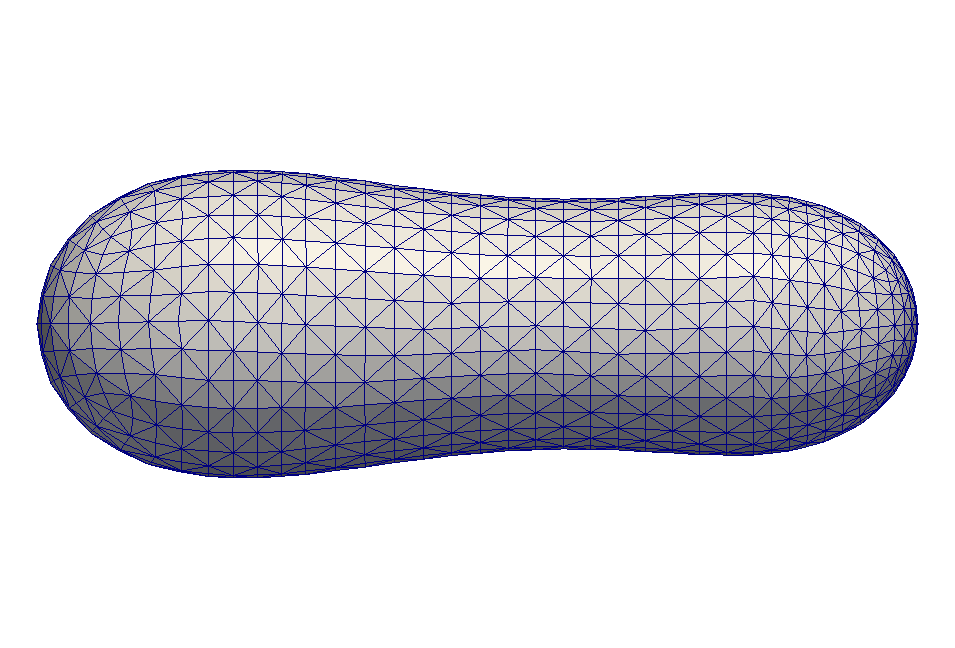} 
\includegraphics[angle=-0,width=\localwidth]{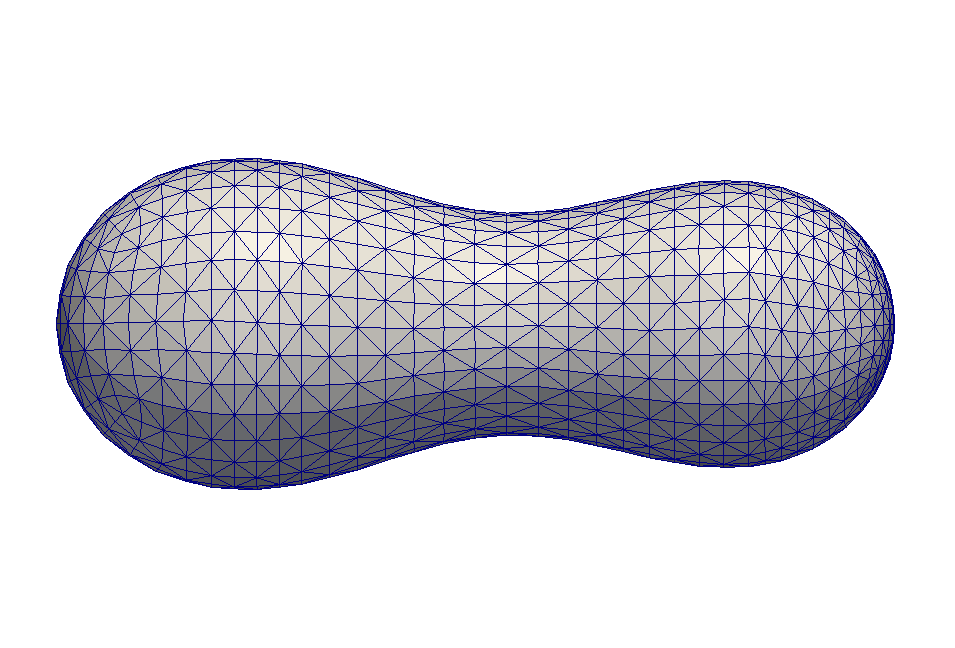} 
\includegraphics[angle=-0,width=\localwidth]{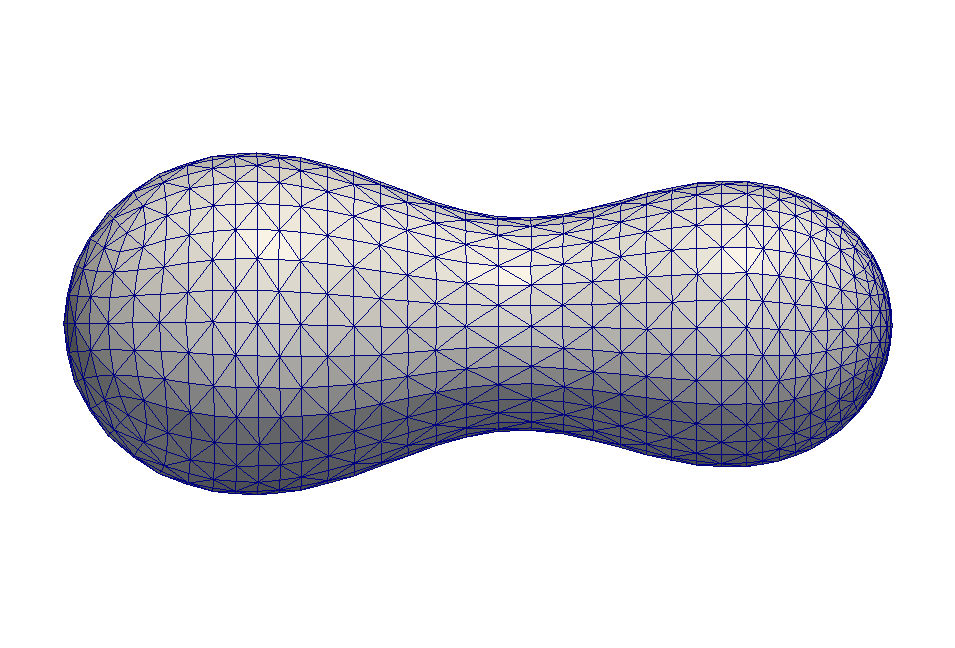} 
\includegraphics[angle=-0,width=\localwidth]{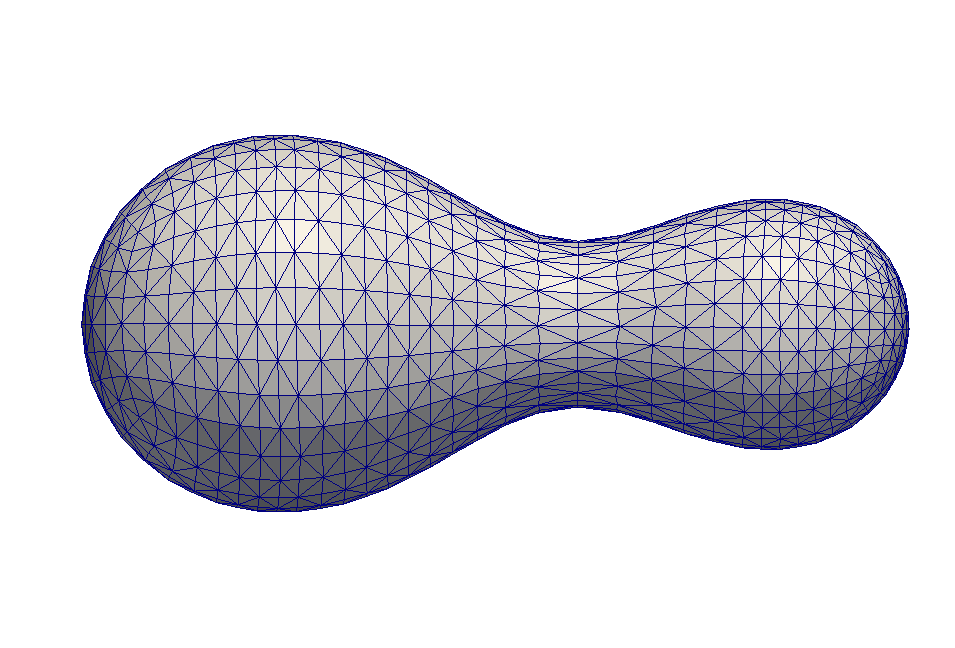} 
}
\mbox{
\hspace{-4mm}
\includegraphics[angle=-0,width=\localwidth]{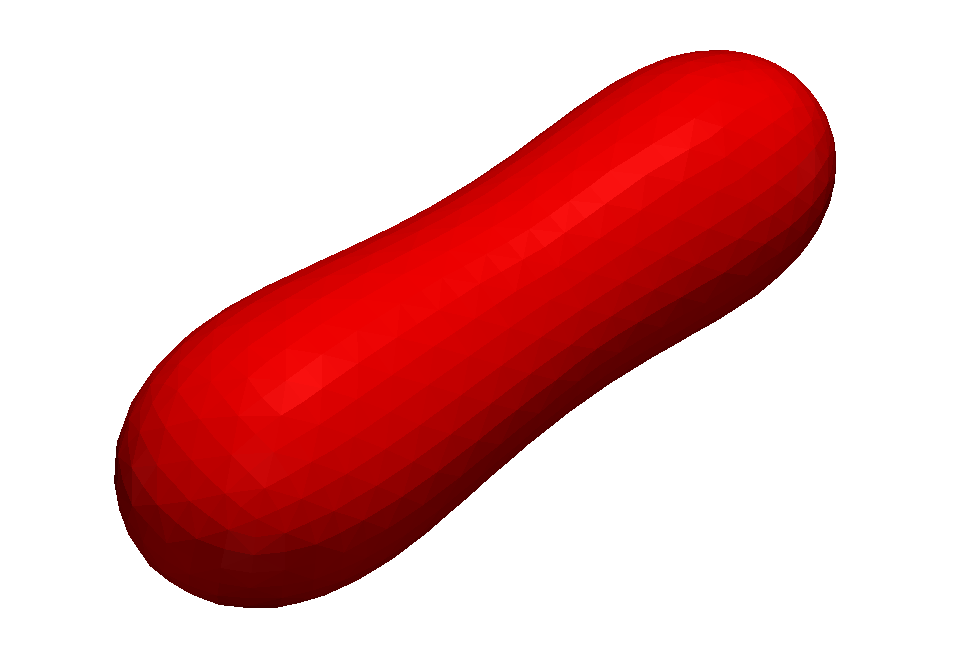} 
\includegraphics[angle=-0,width=\localwidth]{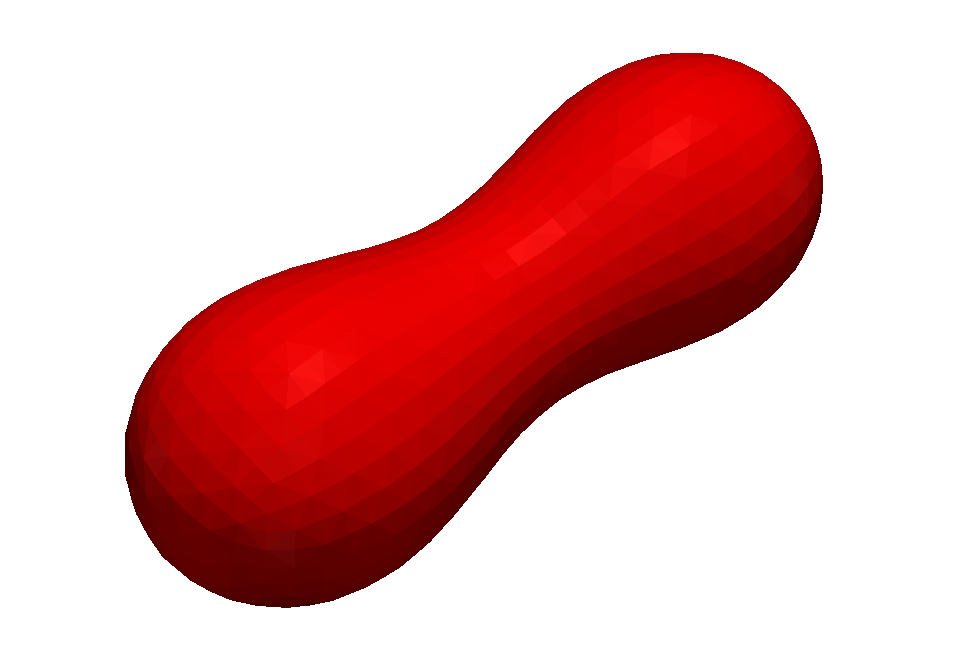} 
\includegraphics[angle=-0,width=\localwidth]{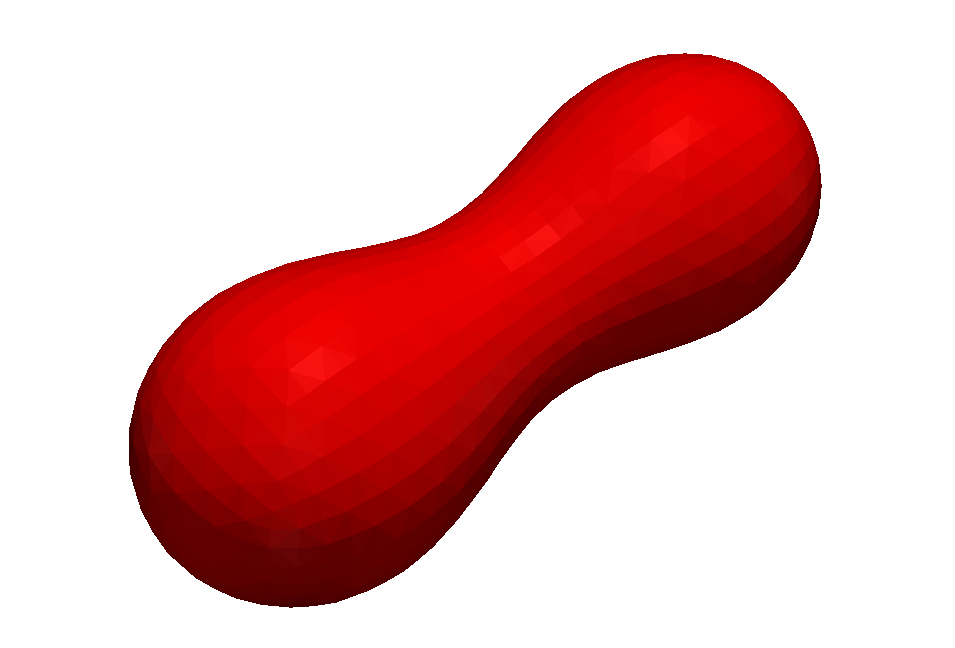} 
\includegraphics[angle=-0,width=\localwidth]{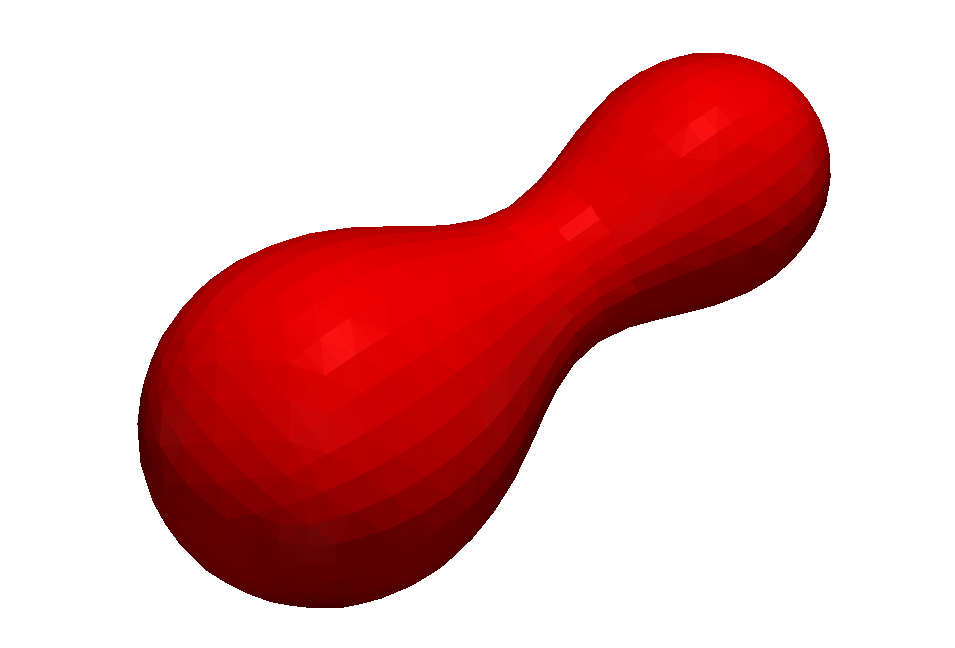} 
}
\caption{(Color online)
Flow for a varying-diameter cigar-like shape with $\mathcal{V}_r = 0.75$ 
for $M_0^* = -33.5$ and $\beta^* = 0.46$.
The plots show the interface $\Gamma^h$ at times $t=0,\ 1,\ 10,\ 50$,
with the top row visualizing the triangulations.
}
\label{fig:stokes_cigar2_vr075M335}
\end{figure*}

\begin{figure*}
\center
\mbox{
\hspace{-4mm}
\includegraphics[angle=-0,width=\localwidth]{figures/GDs_cig2_vr075M335_0cm} 
\includegraphics[angle=-0,width=\localwidth]{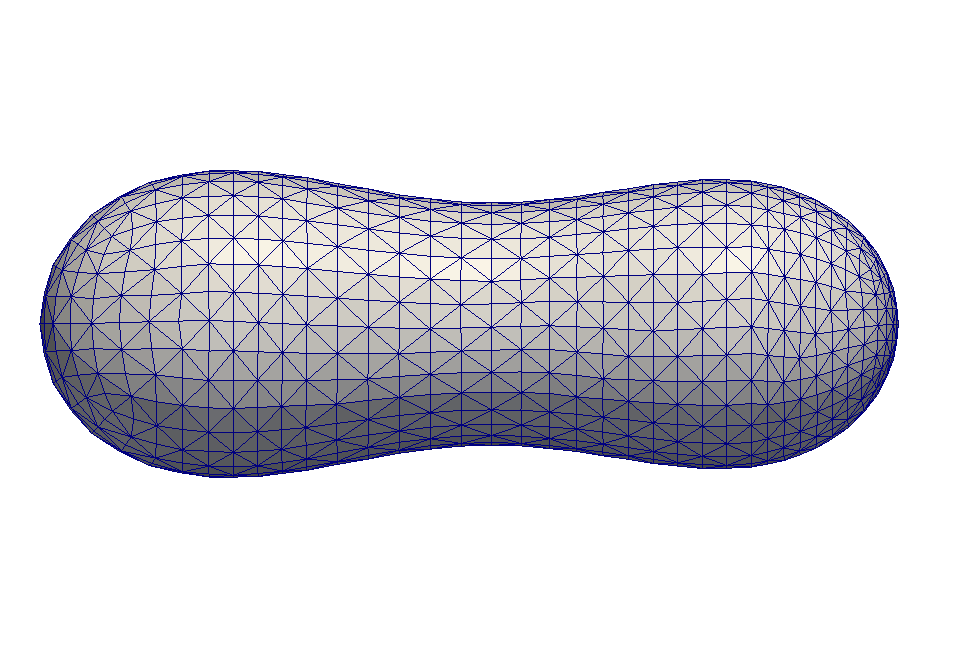} 
\includegraphics[angle=-0,width=\localwidth]{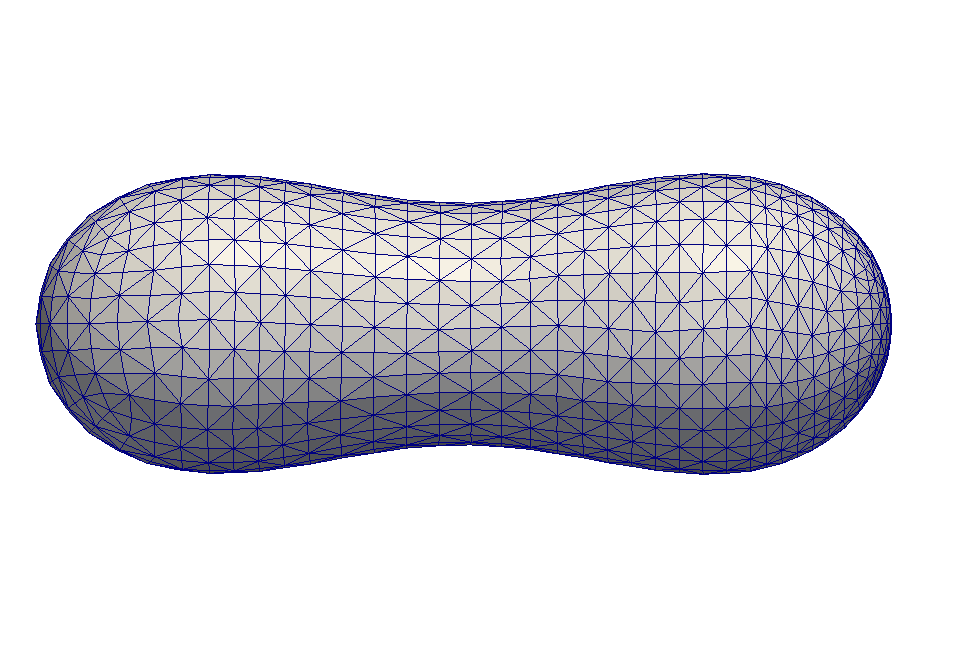} 
\includegraphics[angle=-0,width=\localwidth]{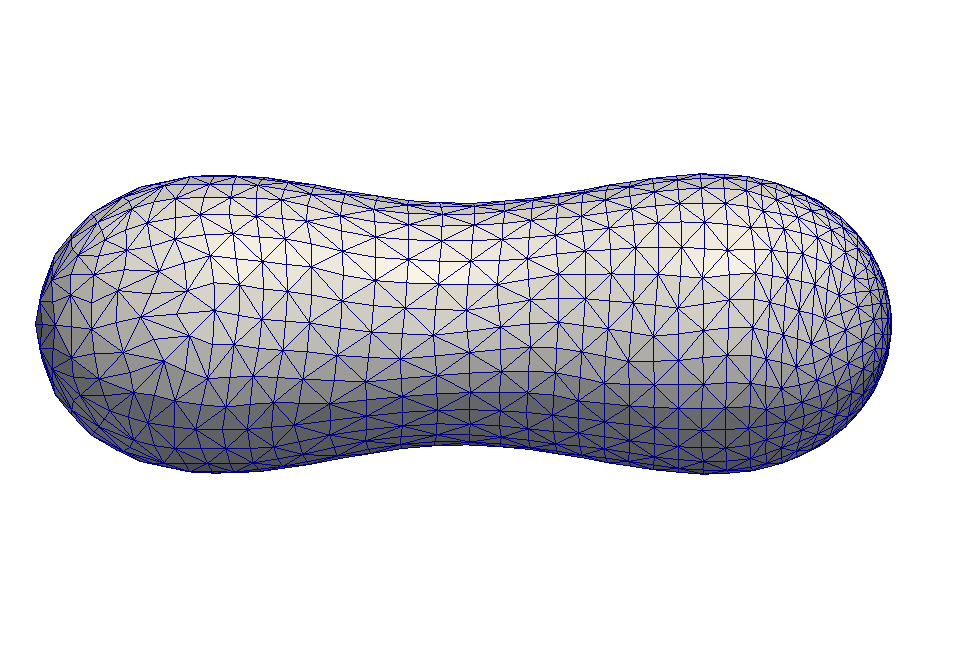} 
}
\mbox{
\hspace{-4mm}
\includegraphics[angle=-0,width=\localwidth]{figures/GDs_cig2_vr075M335_0c} 
\includegraphics[angle=-0,width=\localwidth]{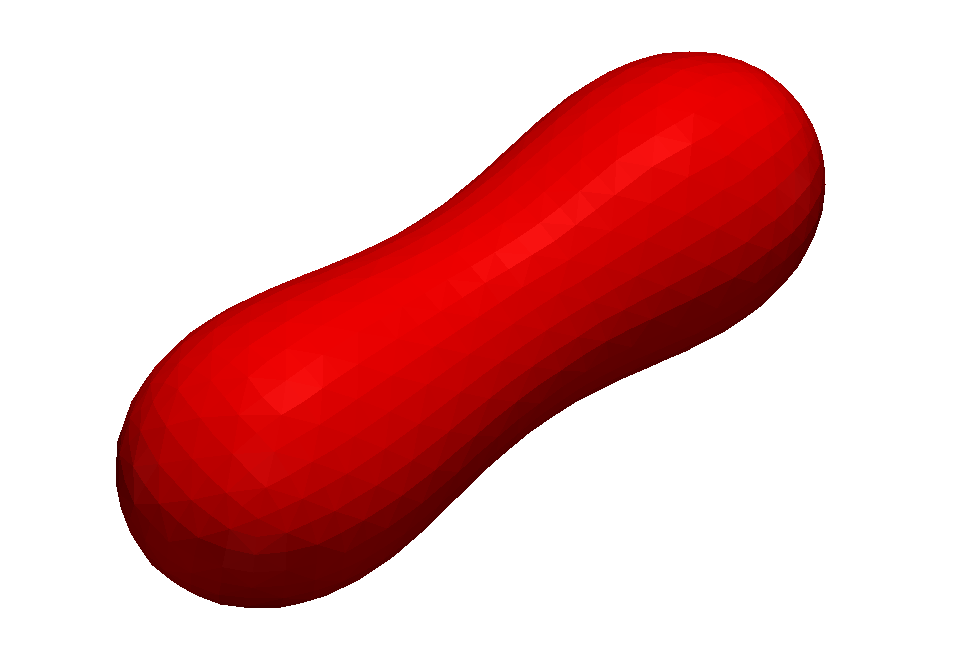} 
\includegraphics[angle=-0,width=\localwidth]{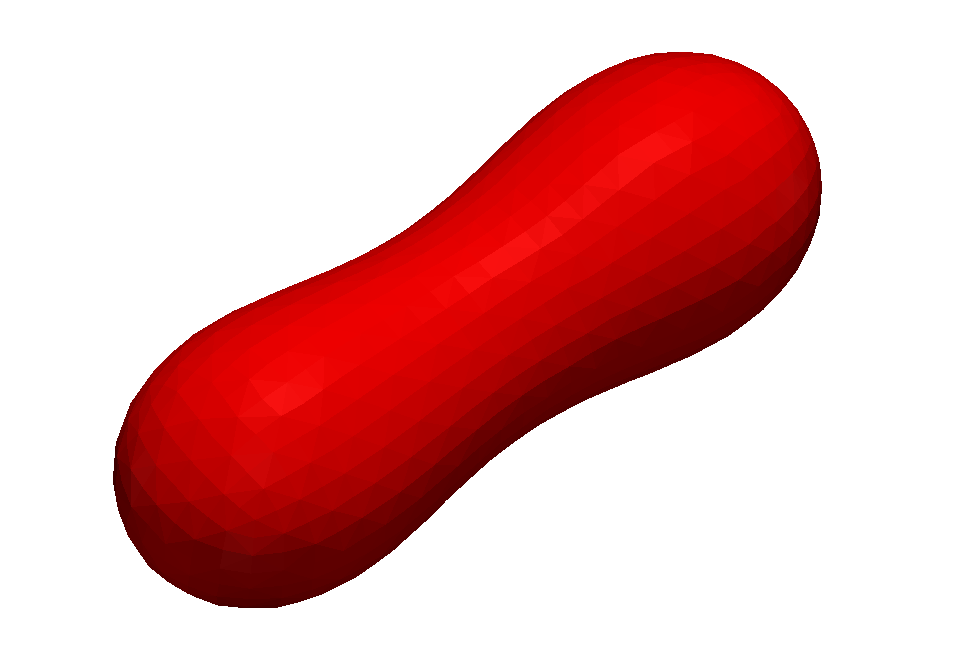} 
\includegraphics[angle=-0,width=\localwidth]{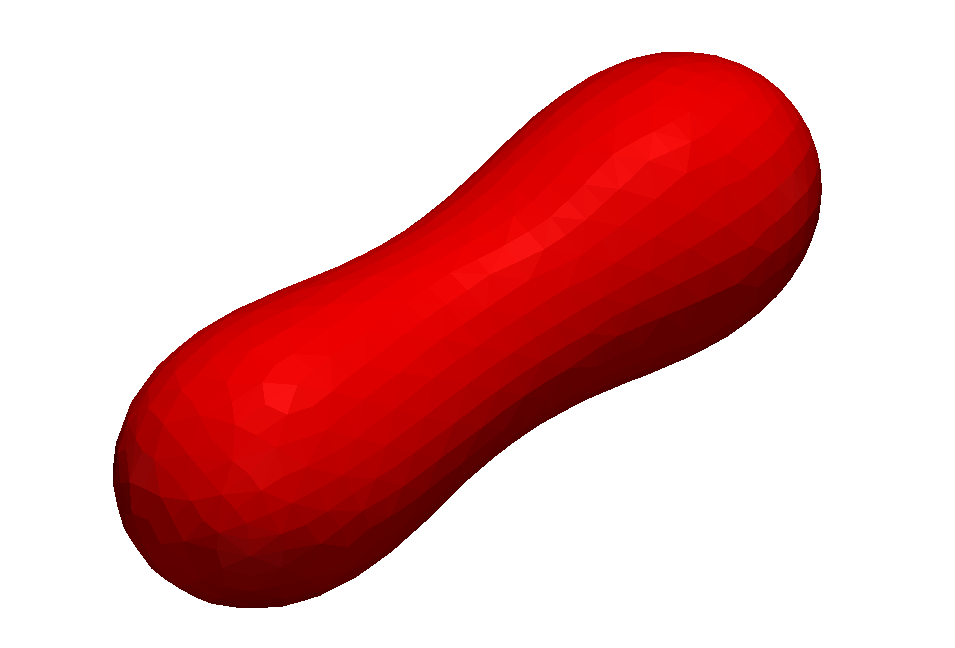} 
}
\caption{(Color online)
Same as Figure~\ref{fig:stokes_cigar2_vr075M335} with $\beta^*=0$.
}
\label{fig:stokes_cigar2_vr075rho0}
\end{figure*}

In our next simulation, we let $\overline\Omega = [-2.5,2.5]^3$ and set
$\Lambda=\alpha^*=1$, as well as $\beta^* = 0.46$ and $M_0^* = -33.5$. 
As initial vesicle we take
a varying-diameter cigar-like shape that has $\mathcal{V}_r = 0.75$ and 
$\mathcal{A}(0) = 9.65$, so that $S = 0.88$.
A simulation can be seen in Figure~\ref{fig:stokes_cigar2_vr075M335}. 
As a comparison, we show the
simulation with $\beta^* = 0$ in Figure~\ref{fig:stokes_cigar2_vr075rho0}.
Similarly to previous studies, where an energy involving area difference 
elasticity terms was minimized, we also observe in our hydrodynamic model
that less symmetric shapes occur when the ADE-energy contributions are 
taken into account.

\subsection{Shearing for budded shape (two arms)}
\begin{figure*}
\center
\includegraphics[angle=-0,width=\localwidth]{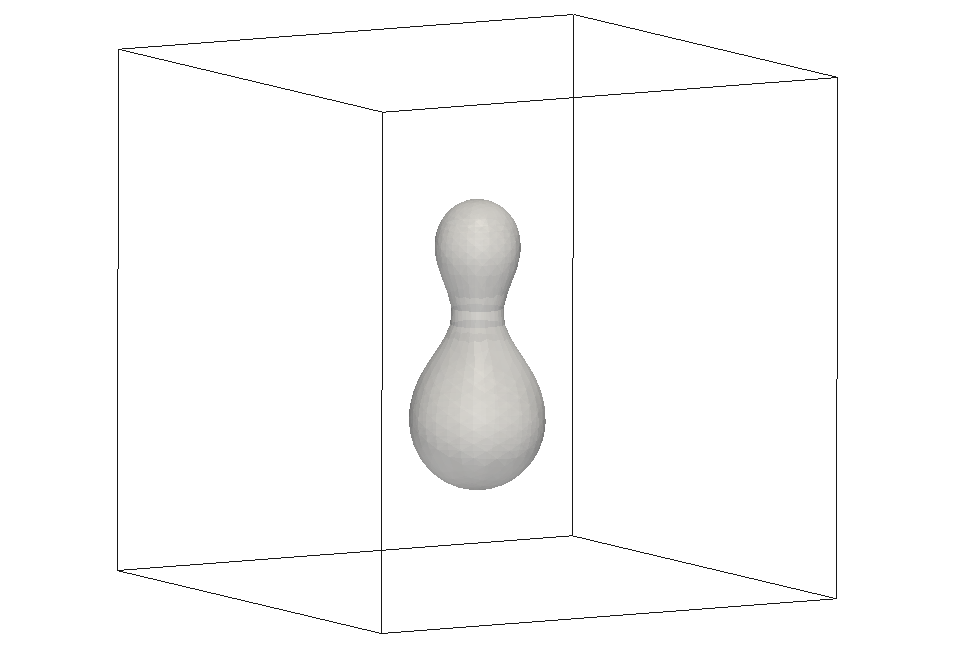} 
\includegraphics[angle=-0,width=\localwidth]{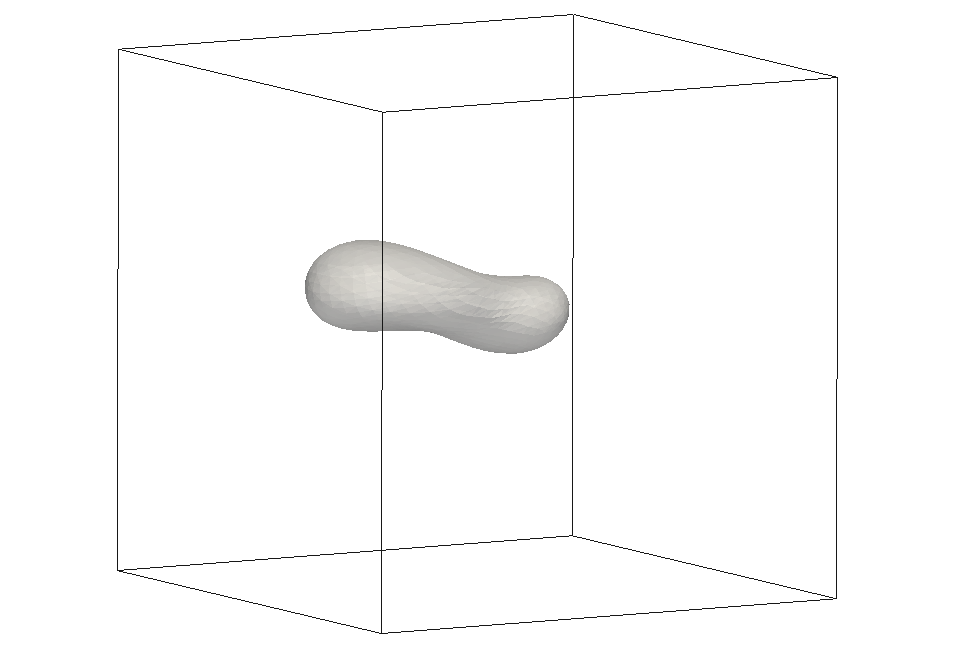} 
\includegraphics[angle=-0,width=\localwidth]{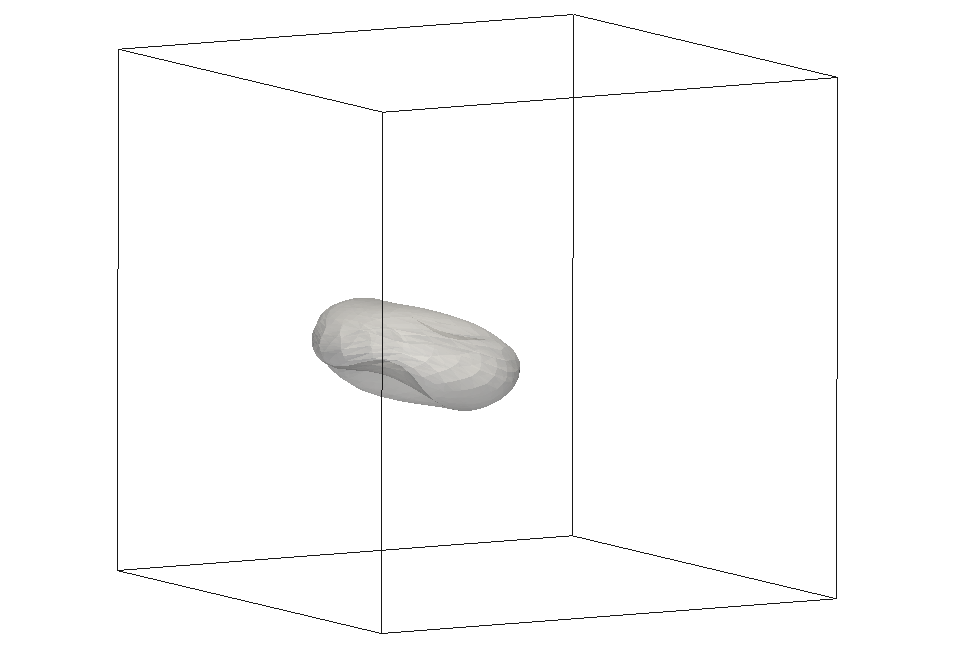} 
\includegraphics[angle=-0,width=\localwidth]{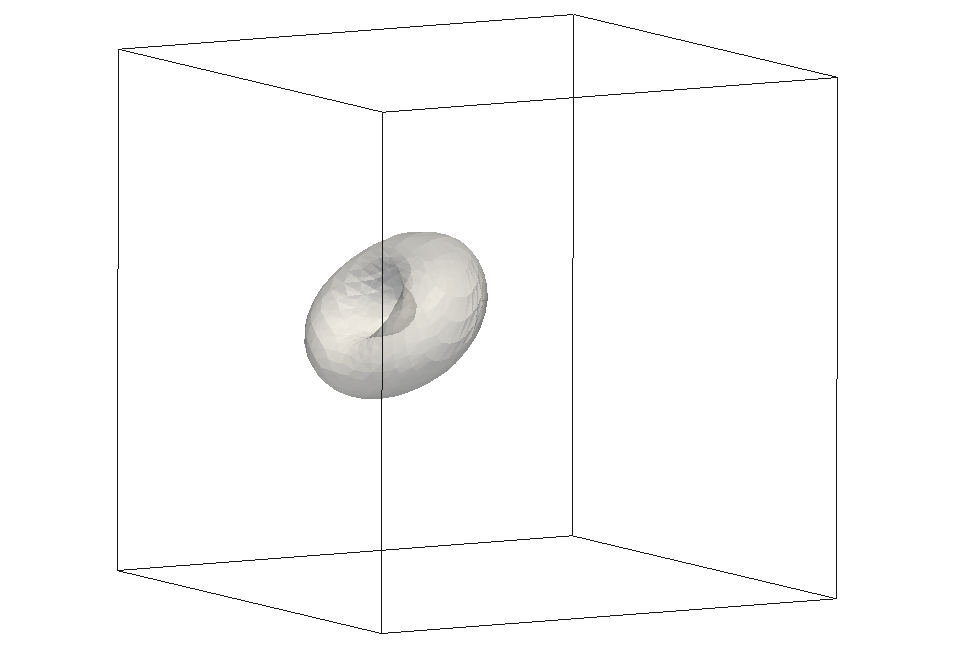}
\\
\includegraphics[angle=-0,width=\localwidth]{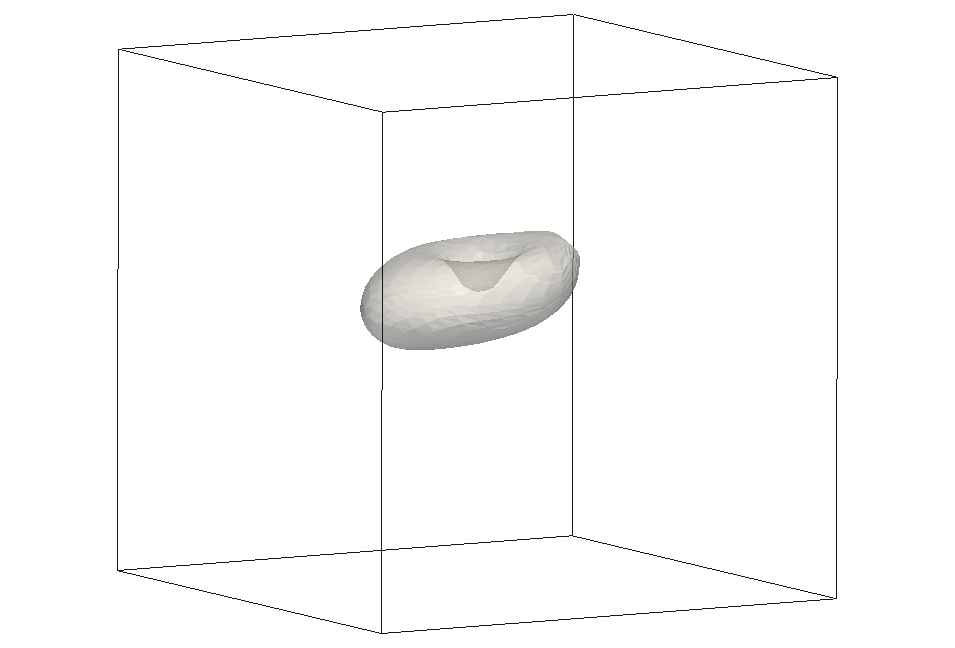}
\includegraphics[angle=-0,width=\localwidth]{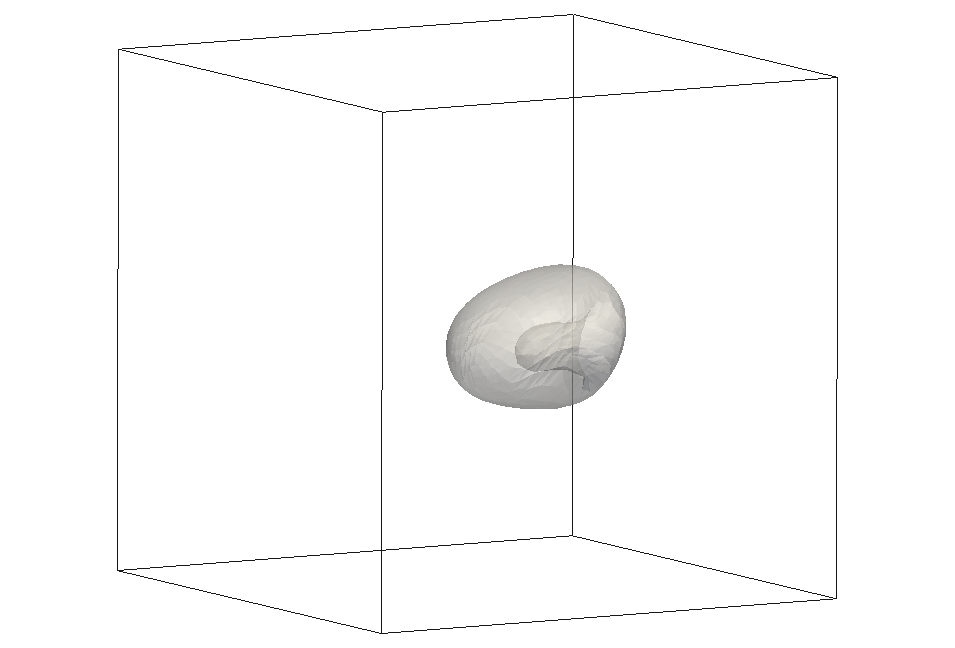} 
\includegraphics[angle=-0,width=\localwidth]{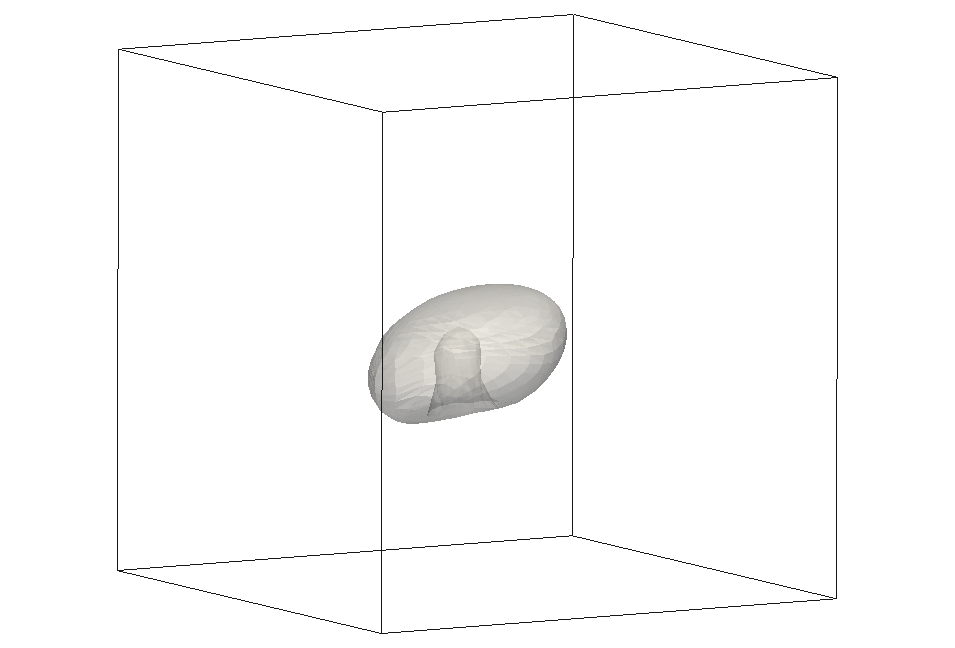} 
\includegraphics[angle=-0,width=\localwidth]{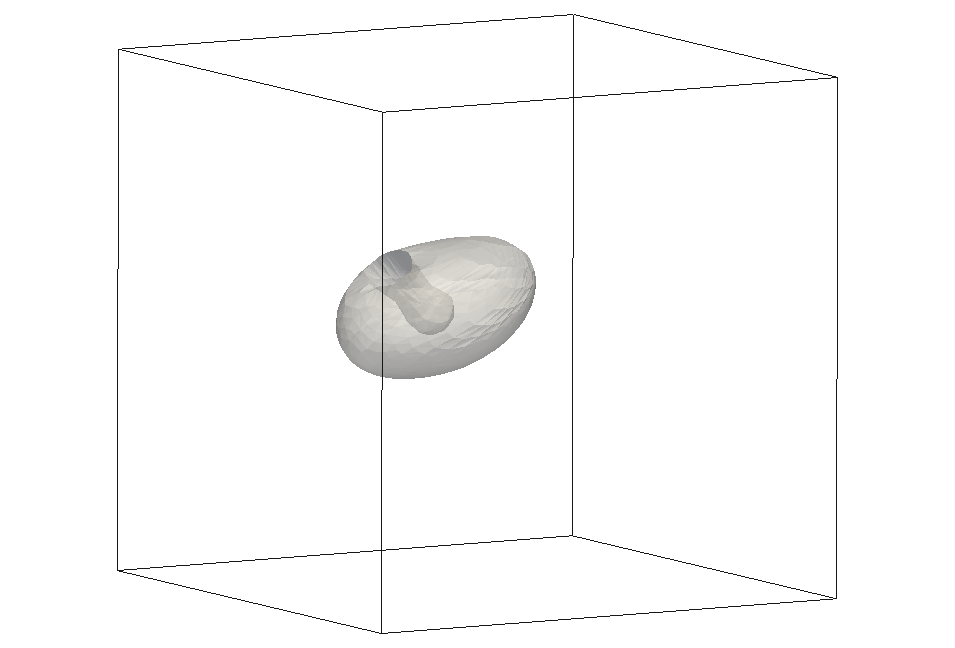}
\\ 
\includegraphics[angle=-0,width=\localwidth]{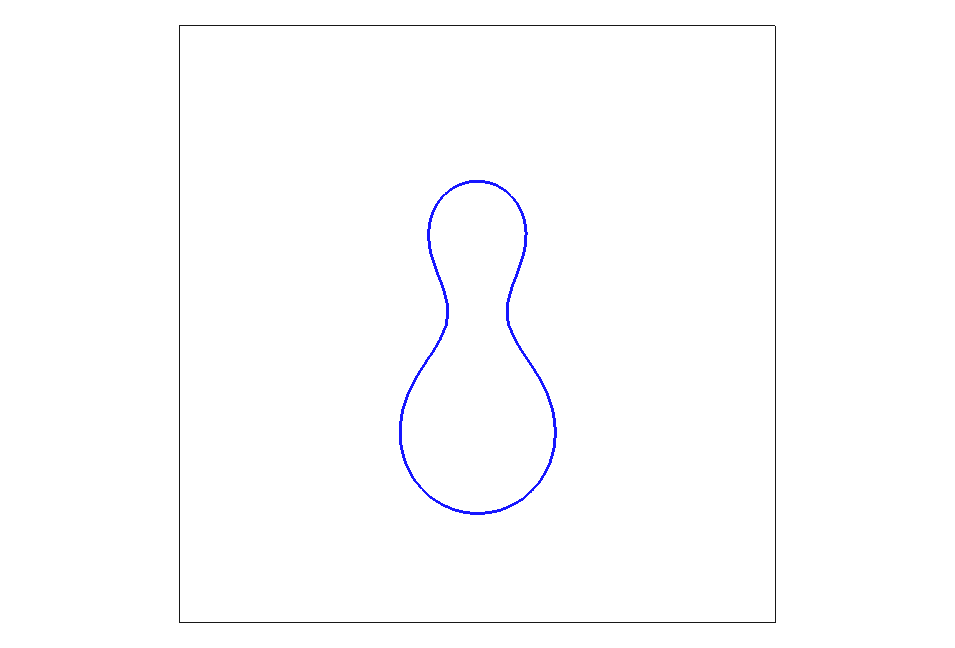} 
\includegraphics[angle=-0,width=\localwidth]{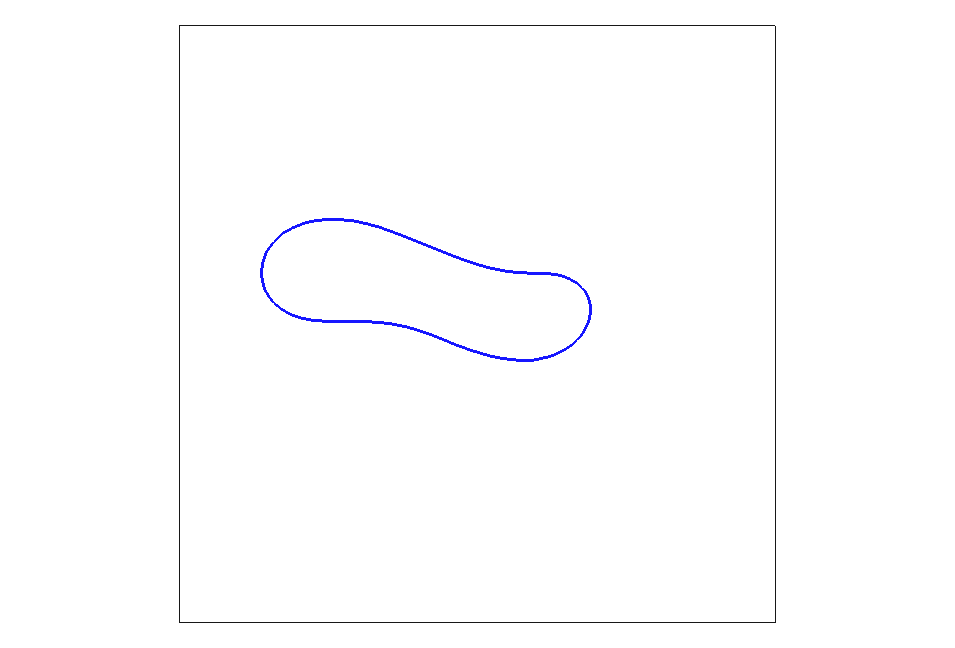} 
\includegraphics[angle=-0,width=\localwidth]{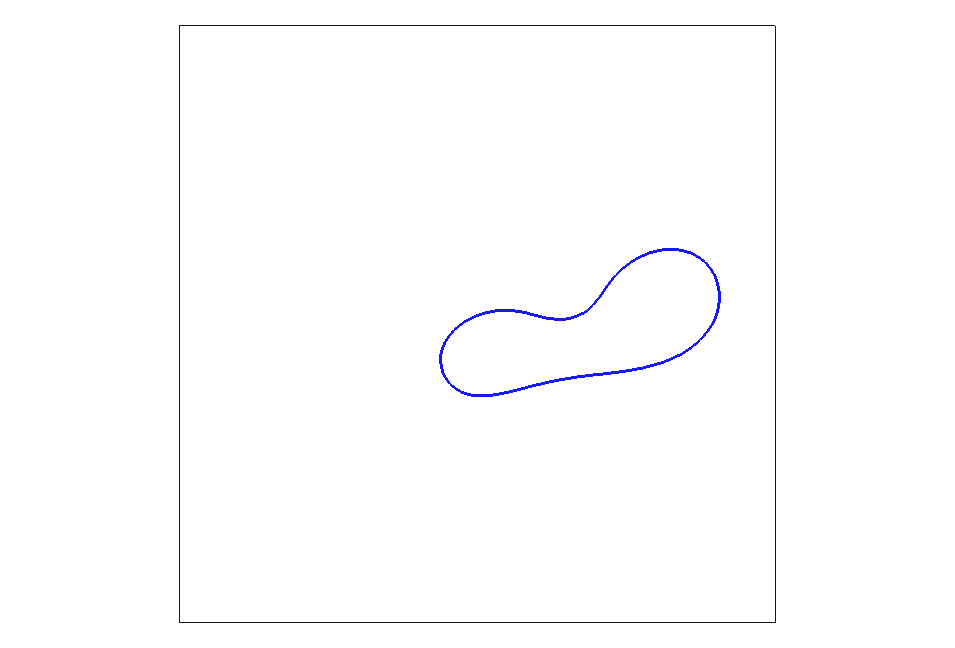} 
\includegraphics[angle=-0,width=\localwidth]{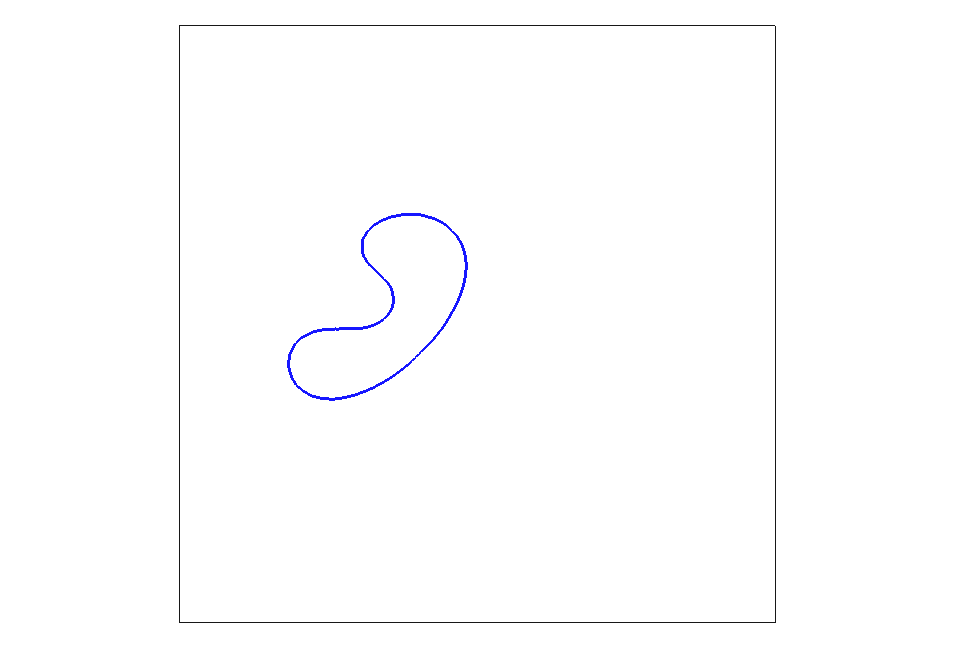} \\
\includegraphics[angle=-0,width=\localwidth]{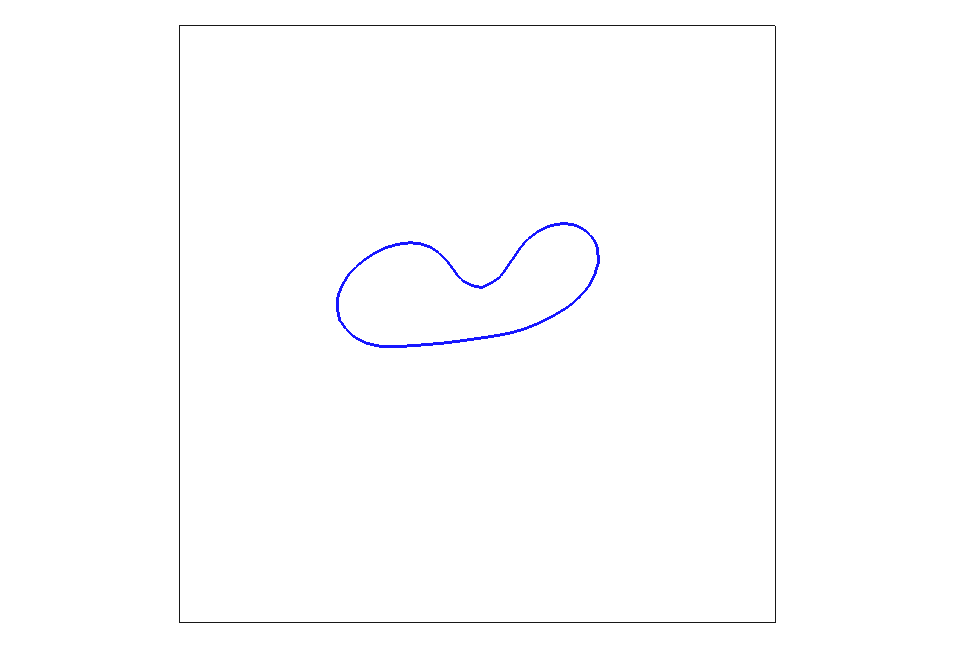} 
\includegraphics[angle=-0,width=\localwidth]{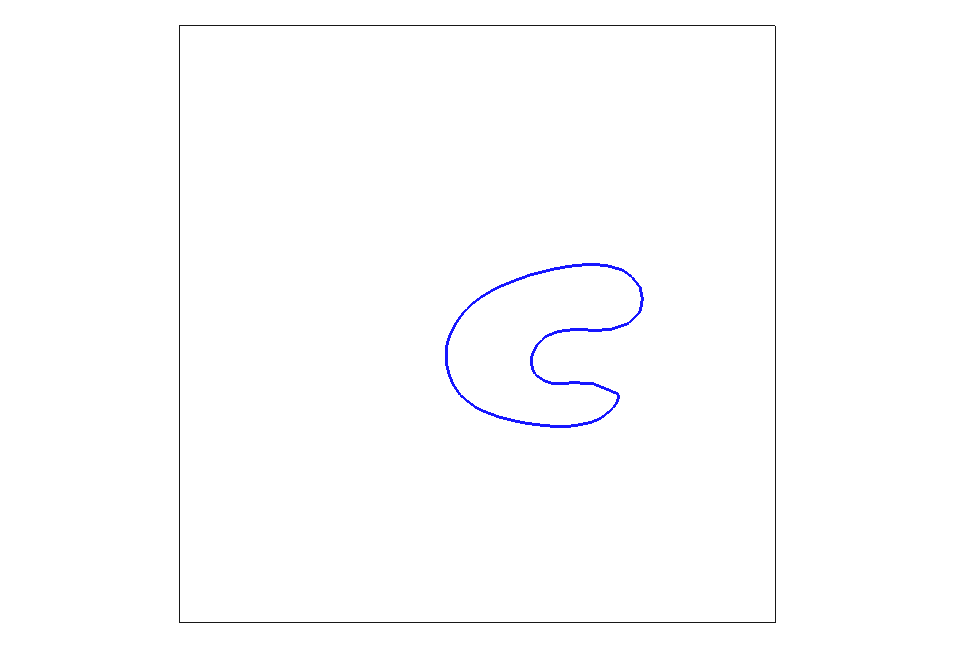} 
\includegraphics[angle=-0,width=\localwidth]{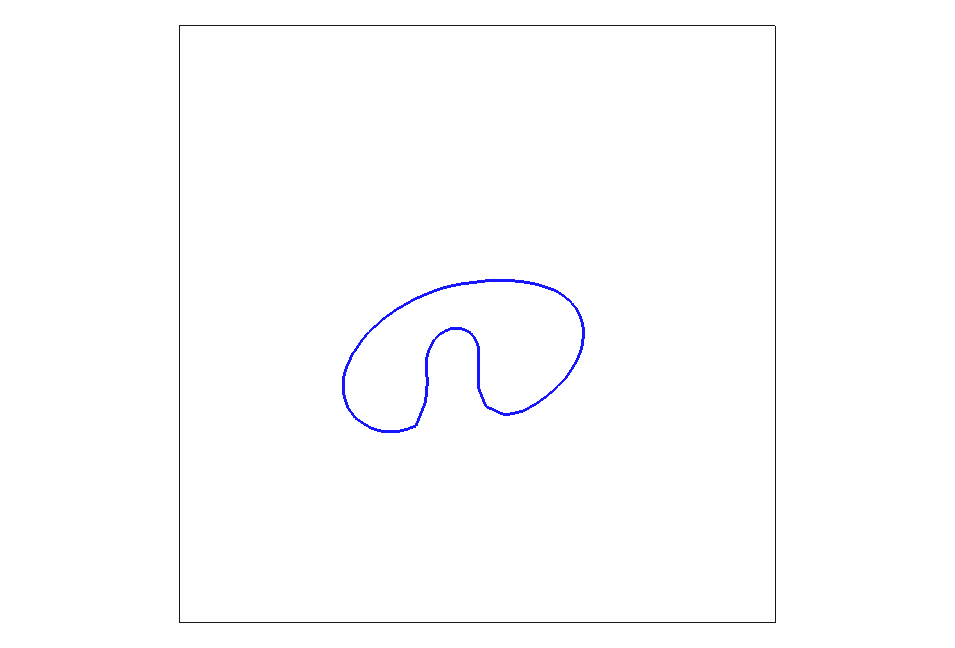} 
\includegraphics[angle=-0,width=\localwidth]{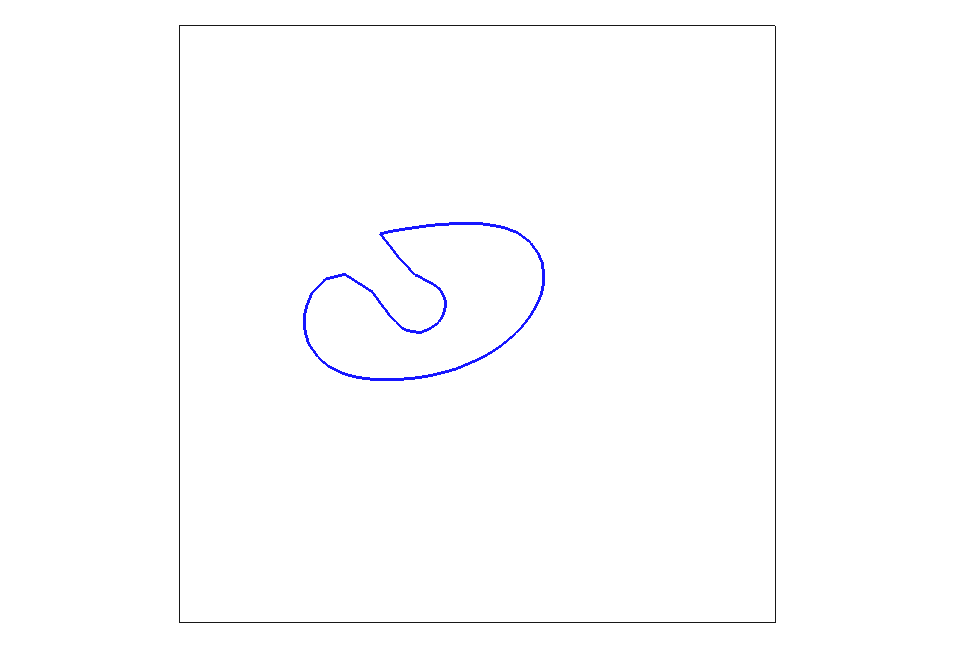} 
\caption{(Color online)
Shear flow for a budding shape with $\Lambda= \mu_\Gamma^* = 1$. Here
$\beta^* = 0.1$ and $M_0^* = -33.5$.
The plots show the interface $\Gamma^h$ within $\overline\Omega$, as well as
cuts through the $x_1$-$x_3$ plane, 
at times $t=0,\ 5,\ 15,\ 17.5,\ 20,\ 25,\ 27.5,\ 32.5$. 
}
\label{fig:budding_rho0_mu1_muGamma1}
\end{figure*}

\begin{figure*}
\center
\includegraphics[angle=-0,width=\localwidth]{figures/budding_r0_m1_m1_t0} 
\includegraphics[angle=-0,width=\localwidth]{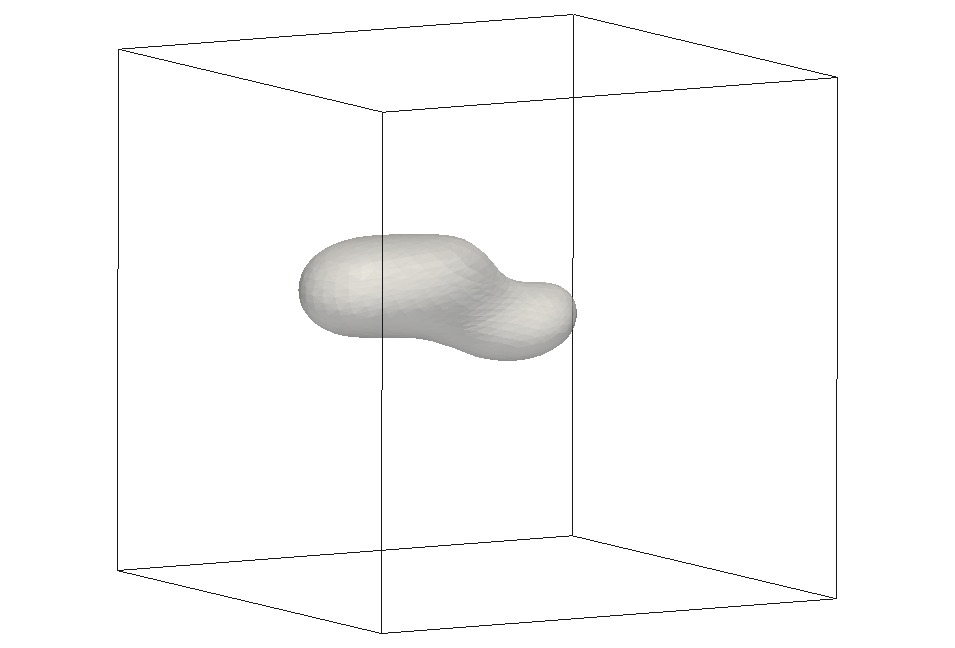} 
\includegraphics[angle=-0,width=\localwidth]{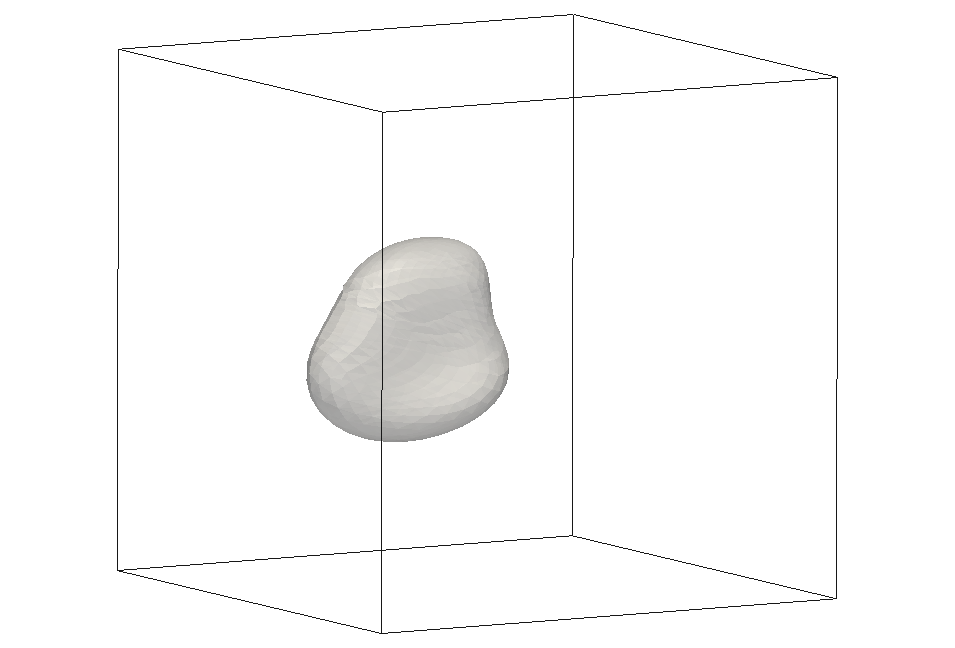} 
\includegraphics[angle=-0,width=\localwidth]{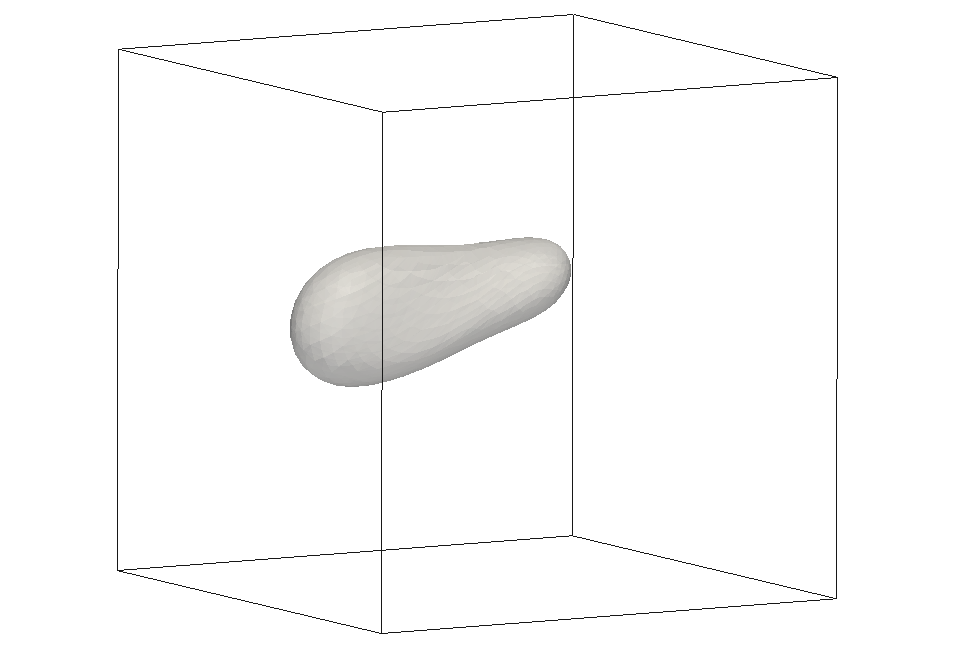}
\\ 
\includegraphics[angle=-0,width=\localwidth]{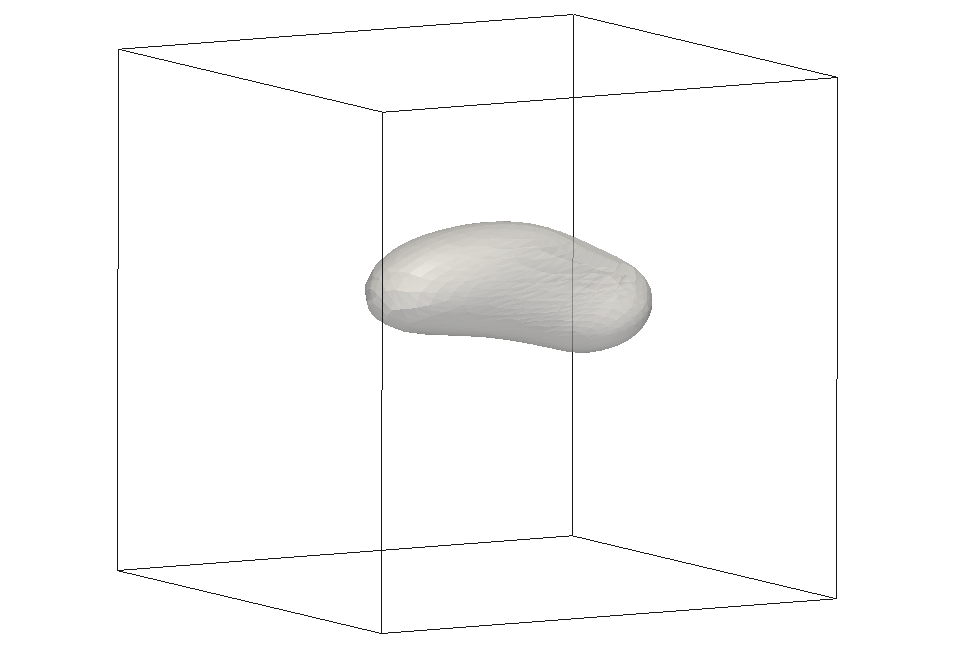} 
\includegraphics[angle=-0,width=\localwidth]{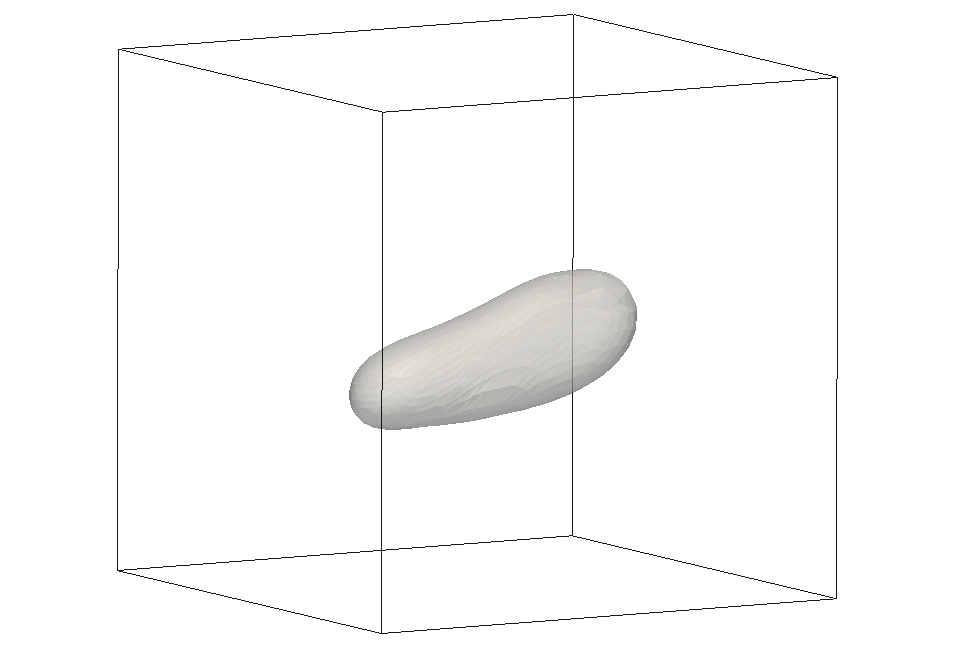} 
\includegraphics[angle=-0,width=\localwidth]{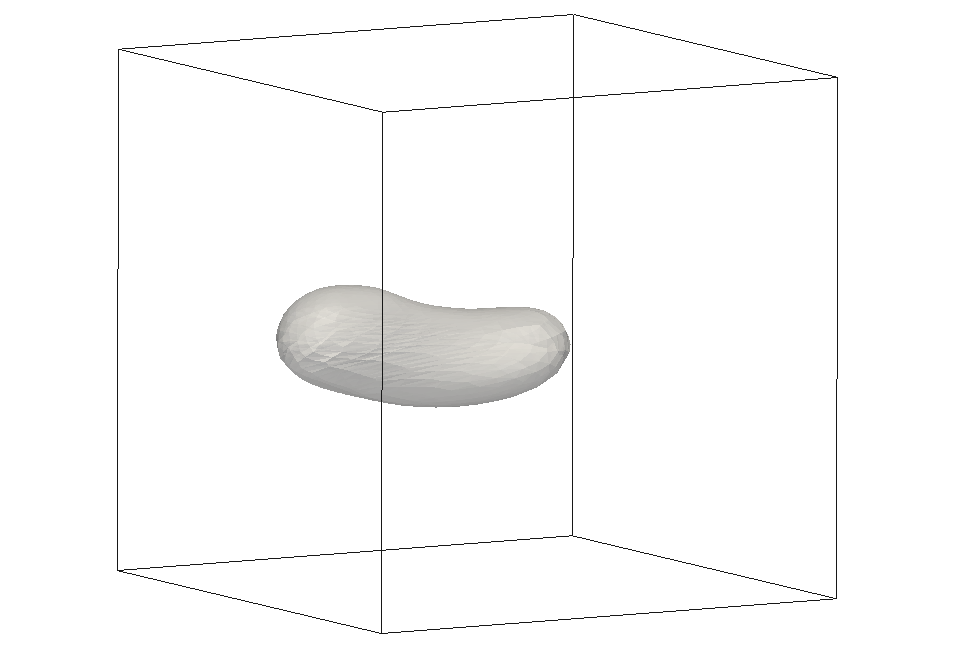} 
\includegraphics[angle=-0,width=\localwidth]{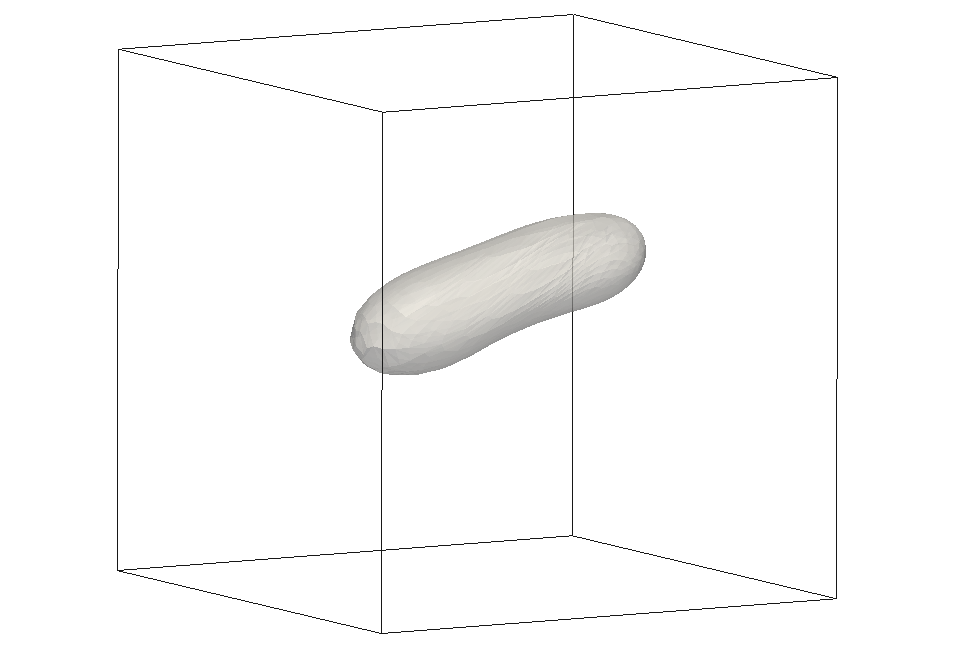} 
\\
\includegraphics[angle=-0,width=\localwidth]{figures/budding_r0_m1_m1_t0cut} 
\includegraphics[angle=-0,width=\localwidth]{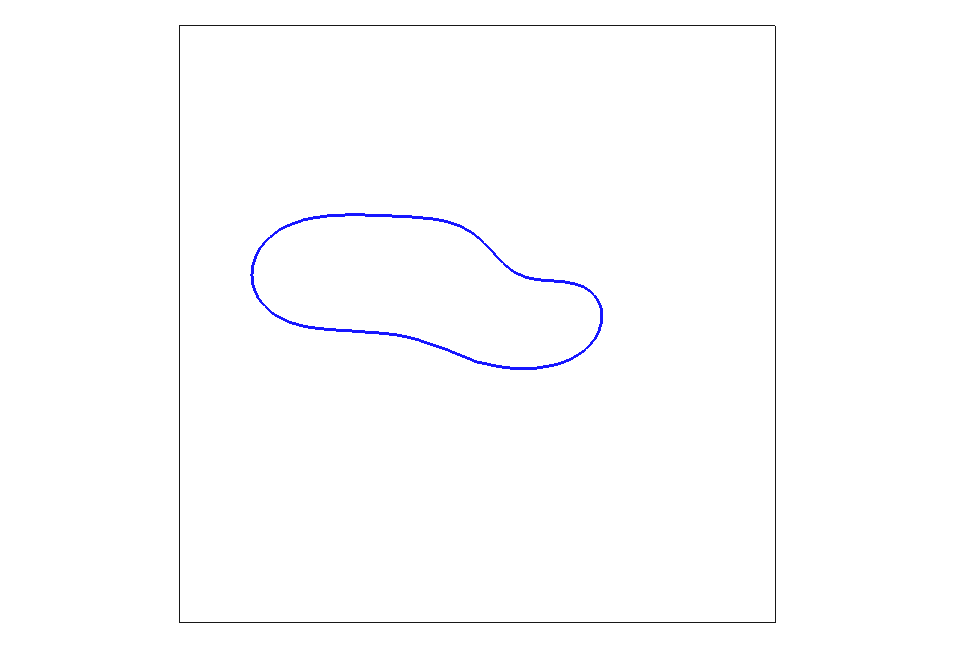} 
\includegraphics[angle=-0,width=\localwidth]{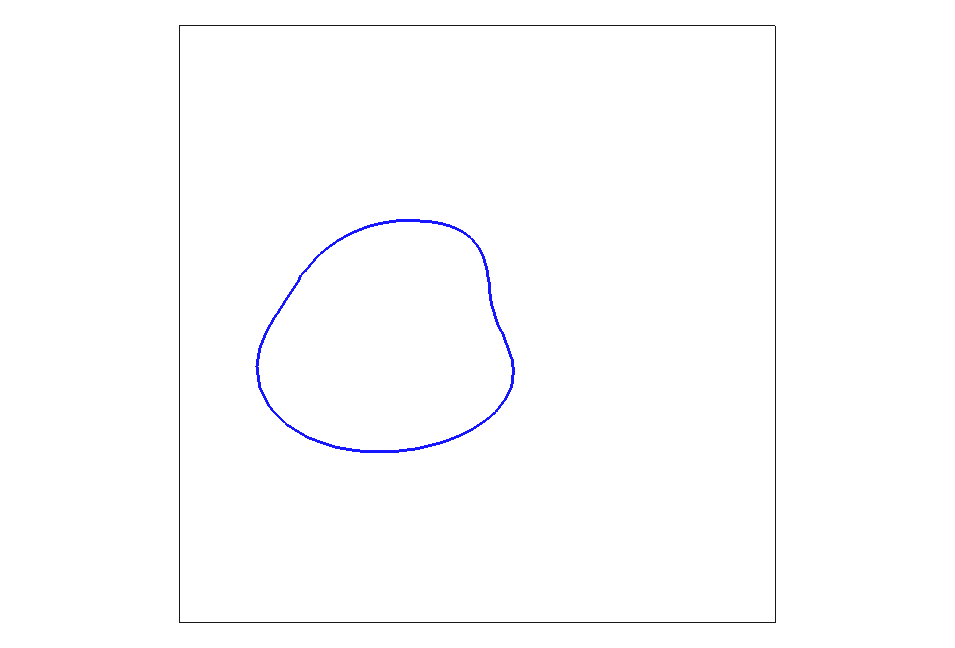} 
\includegraphics[angle=-0,width=\localwidth]{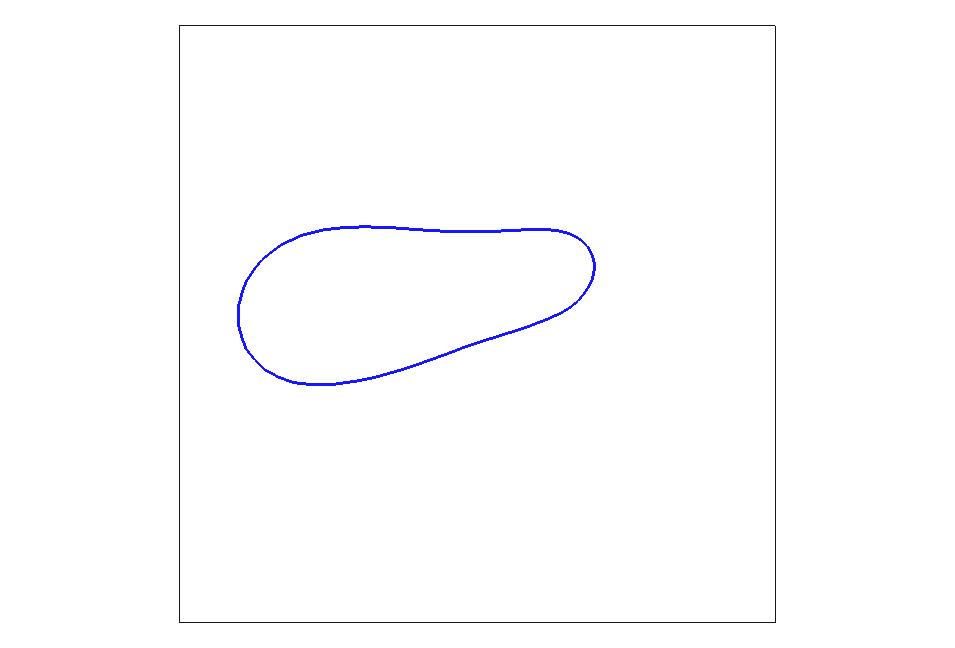} \\
\includegraphics[angle=-0,width=\localwidth]{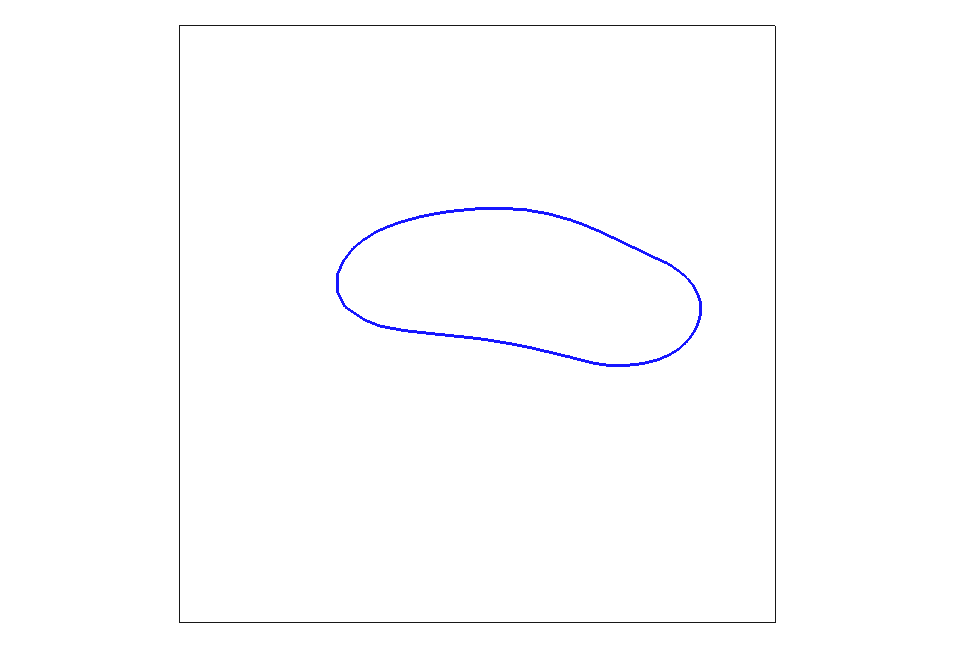} 
\includegraphics[angle=-0,width=\localwidth]{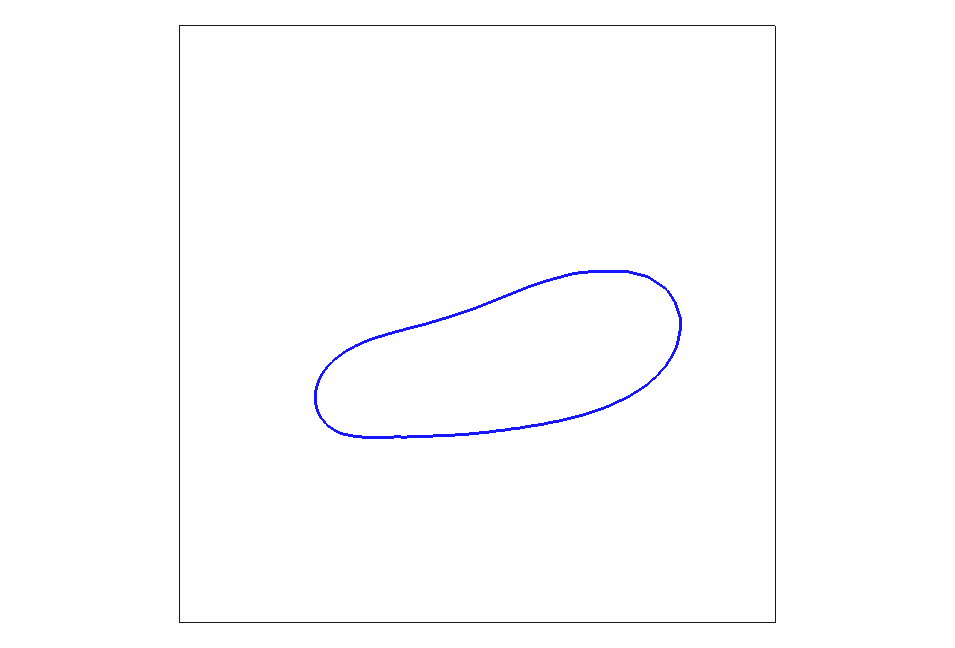} 
\includegraphics[angle=-0,width=\localwidth]{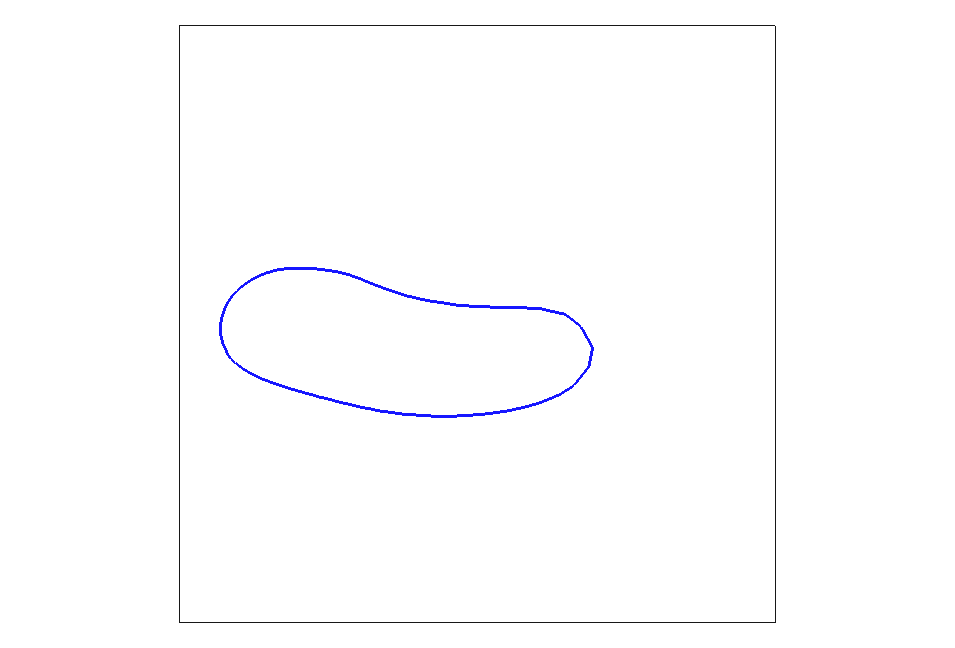} 
\includegraphics[angle=-0,width=\localwidth]{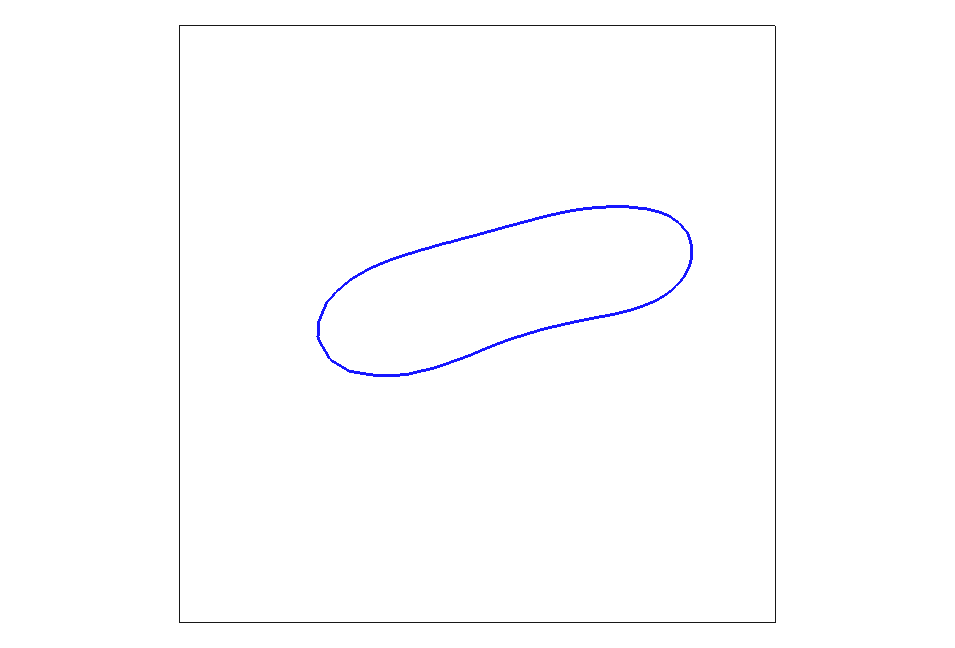} 
\caption{(Color online)
Same as Figure~\ref{fig:budding_rho0_mu1_muGamma1} but with $\beta^* = 0$.
}
\label{fig:budding0_rho0_mu1_muGamma1}
\end{figure*}

\begin{figure}
\center
\includegraphics[angle=-90,width=0.22\textwidth]{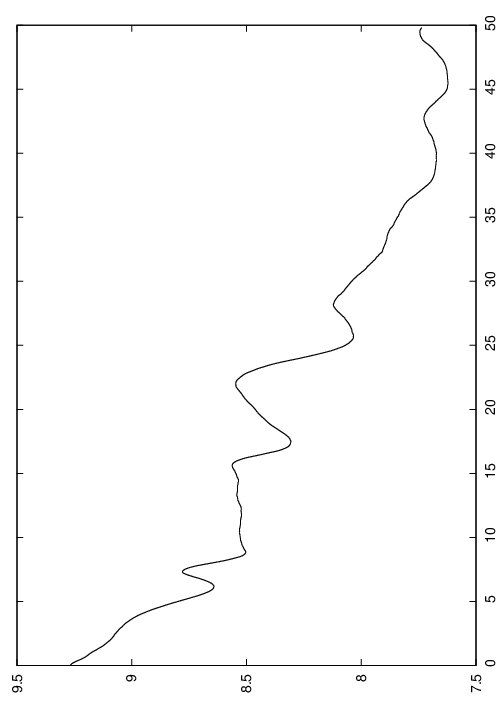}
\includegraphics[angle=-90,width=0.22\textwidth]{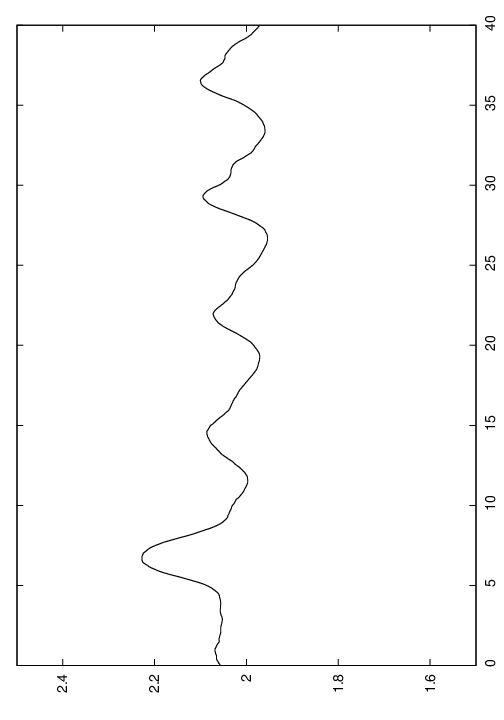}
\caption{The bending energy $\alpha^*\,E^*(\Gamma^h)$ for the computations in
Figures~\ref{fig:budding_rho0_mu1_muGamma1} and
\ref{fig:budding0_rho0_mu1_muGamma1}.}
\label{fig:budding_e}
\end{figure}

We start a scaled variant of the final shape from 
Figure~\ref{fig:stokes_cigar2_vr075M335} in a shear flow experiment
in $\overline\Omega=[-2,2]^3$.
In particular, the initial shape is axisymmetric, with reduced volume 
$\mathcal{V}_r = 0.75$ and $\mathcal{A}(0) = 5.43$, so that $S = 0.66$.
We set $\Lambda = \mu_\Gamma^* = 1$, $\alpha^* = 0.05$.
See Figure~\ref{fig:budding_rho0_mu1_muGamma1} for a run with
$\beta^* = 0.1$ and $M_0^* = -33.5$. We observe that the shape of the vesicle
changes drastically, with part of the surface growing inwards. This is 
similar to the shapes observed in Figure~\ref{fig:stokes_cup2_vr065}, 
where the presence of a lower reduced volume led to cup-like stomatocyte 
shapes.
We repeat the same experiment for $\beta^* = 0$ in 
Figure~\ref{fig:budding0_rho0_mu1_muGamma1}. Now the budding shape loses its
strong nonconvexity completely, as can be clearly seen in the plots of the 
two-dimensional cuts in Figure~\ref{fig:budding0_rho0_mu1_muGamma1}.
Plots of the bending energy $\alpha^*\, E^*(\Gamma^h)$ 
are shown in Figure~\ref{fig:budding_e}, where we recall
that the energy inequality in (\ref{EnI}) does not hold for the 
inhomogeneous boundary conditions employed in the present simulations.

\subsection{Shearing for a seven-arm starfish}
\begin{figure}
\center
\includegraphics[angle=-0,width=0.22\textwidth]{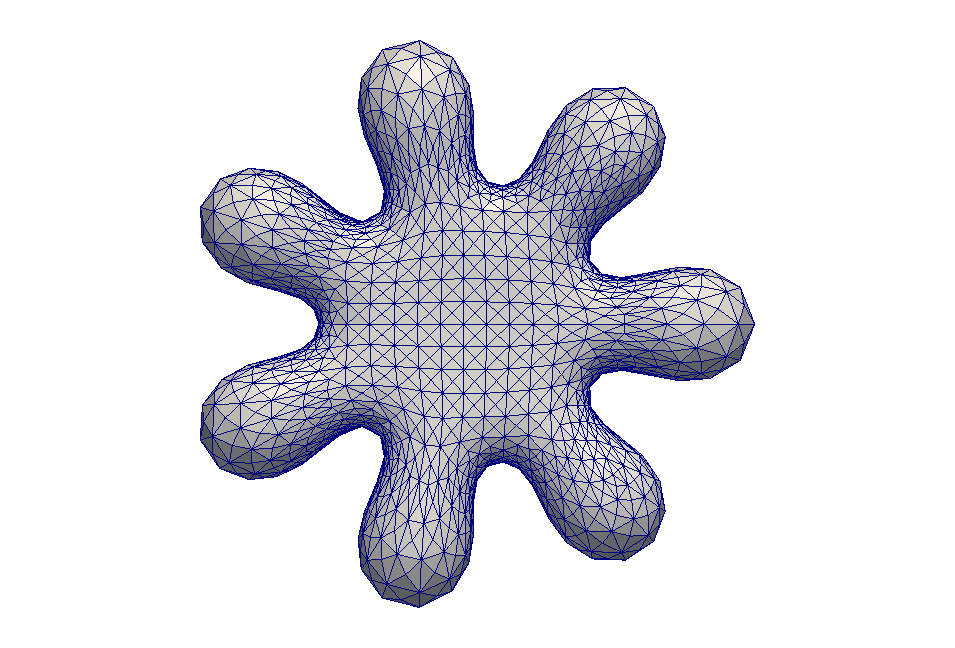}
\includegraphics[angle=-0,width=0.22\textwidth]{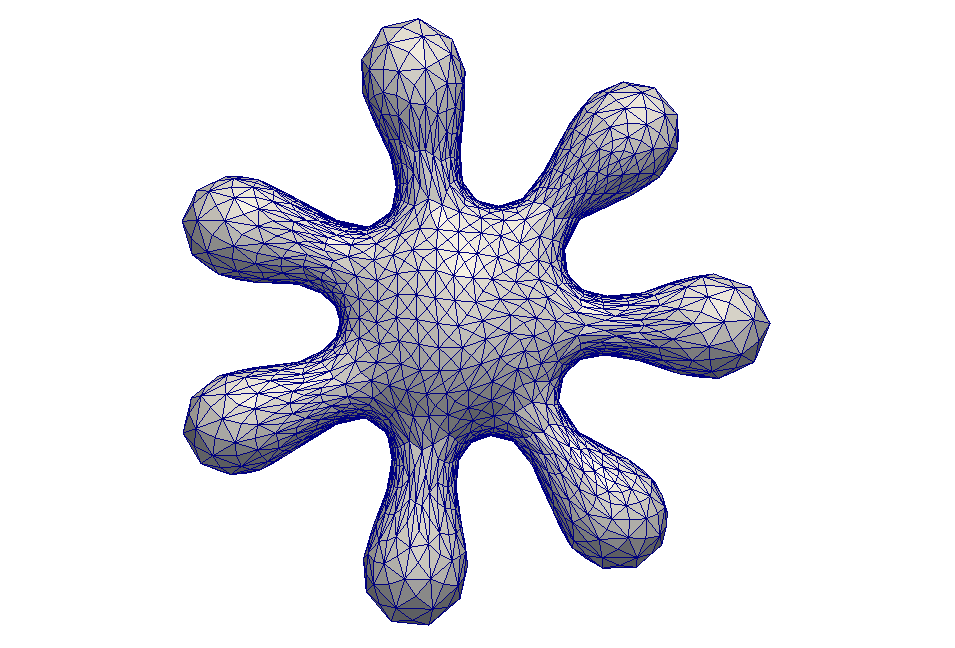}
\caption{(Color online)
Flow for a seven-arm figure with $\mathcal{V}_r = 0.38$. Here $\beta^* = 0.05$ and $M_0^* = 180$. The triangulations $\Gamma^h$ at times $t=0$ and $5$.
}
\label{fig:7arms}
\end{figure}

\begin{figure*}
\center
\includegraphics[angle=-0,width=\localwidth]{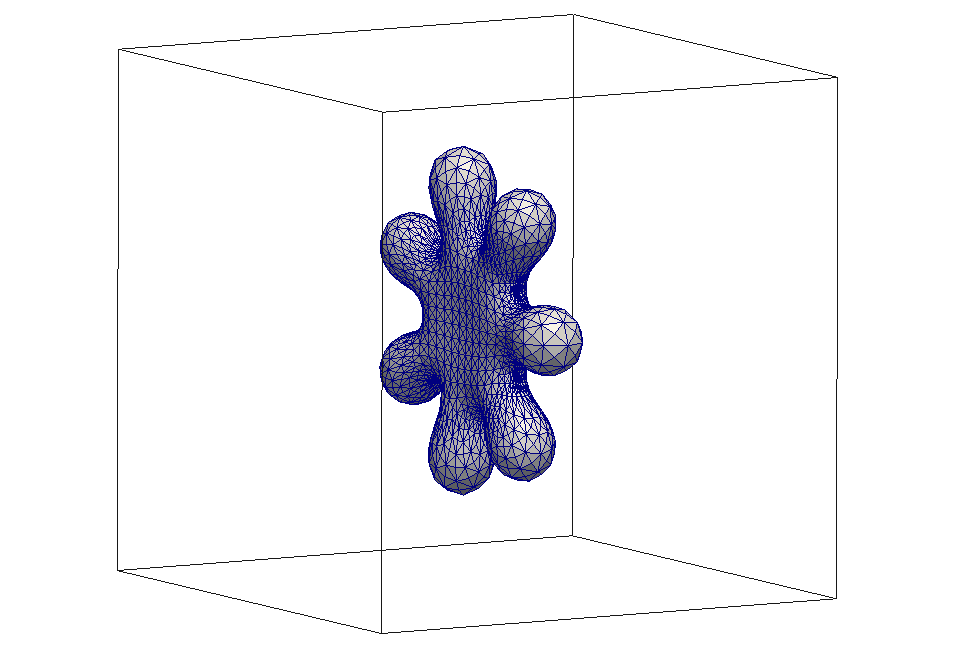} 
\includegraphics[angle=-0,width=\localwidth]{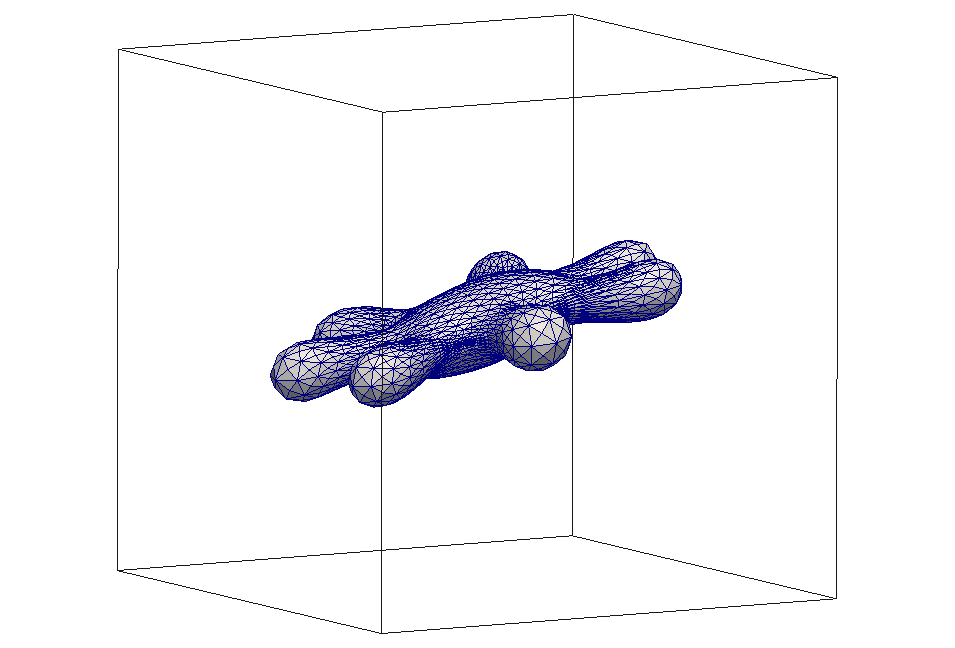} 
\includegraphics[angle=-0,width=\localwidth]{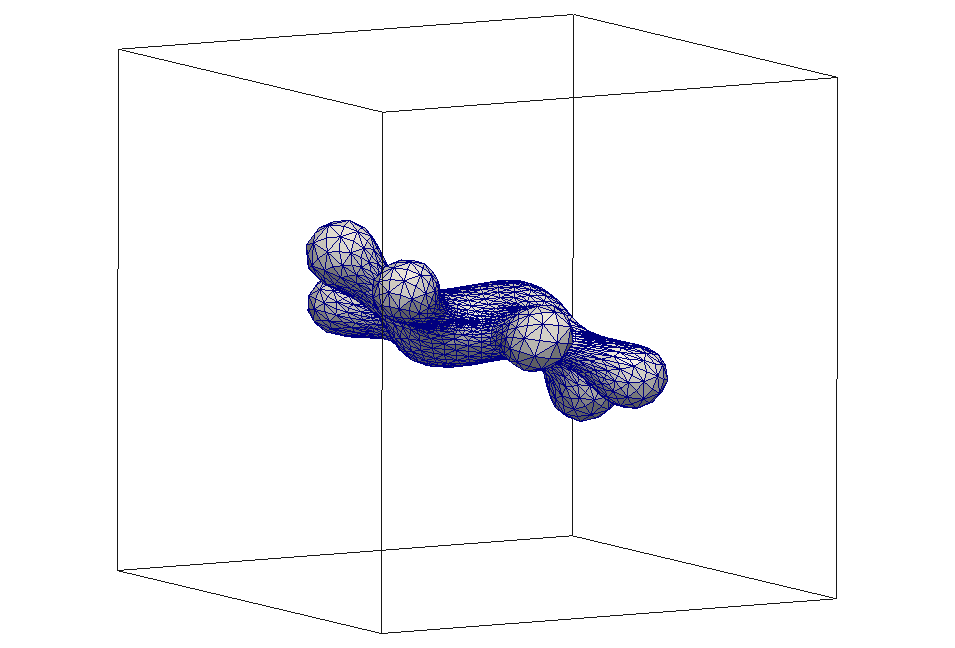} 
\includegraphics[angle=-0,width=\localwidth]{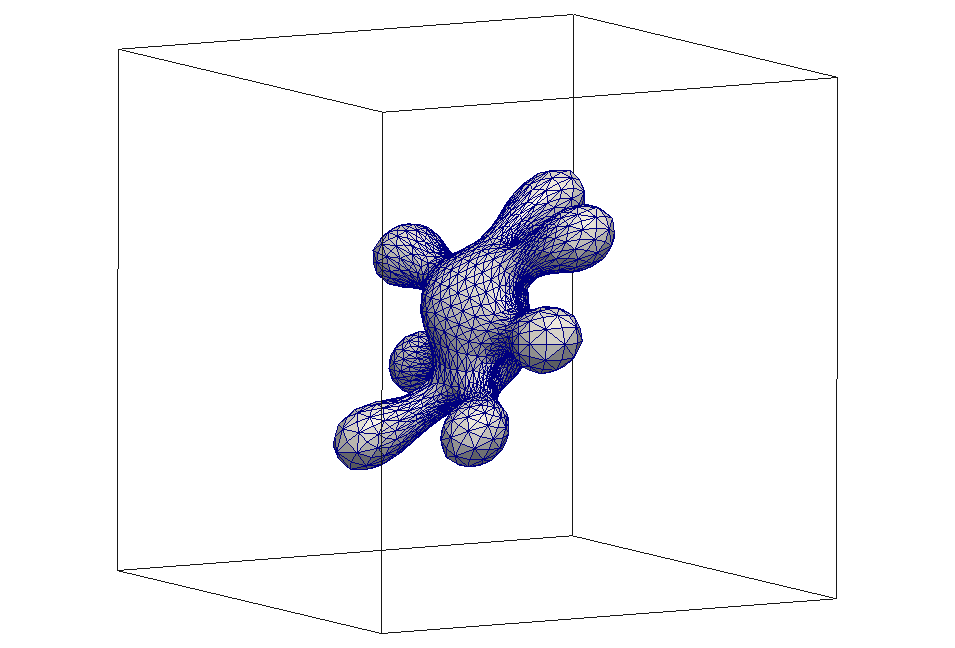} 
\\
\includegraphics[angle=-0,width=\localwidth]{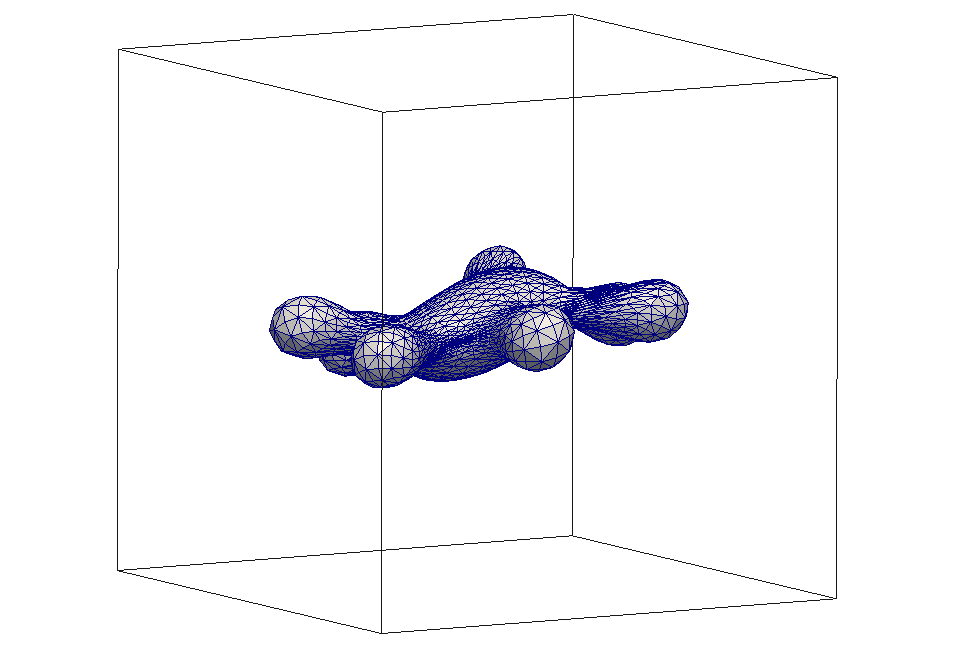} 
\includegraphics[angle=-0,width=\localwidth]{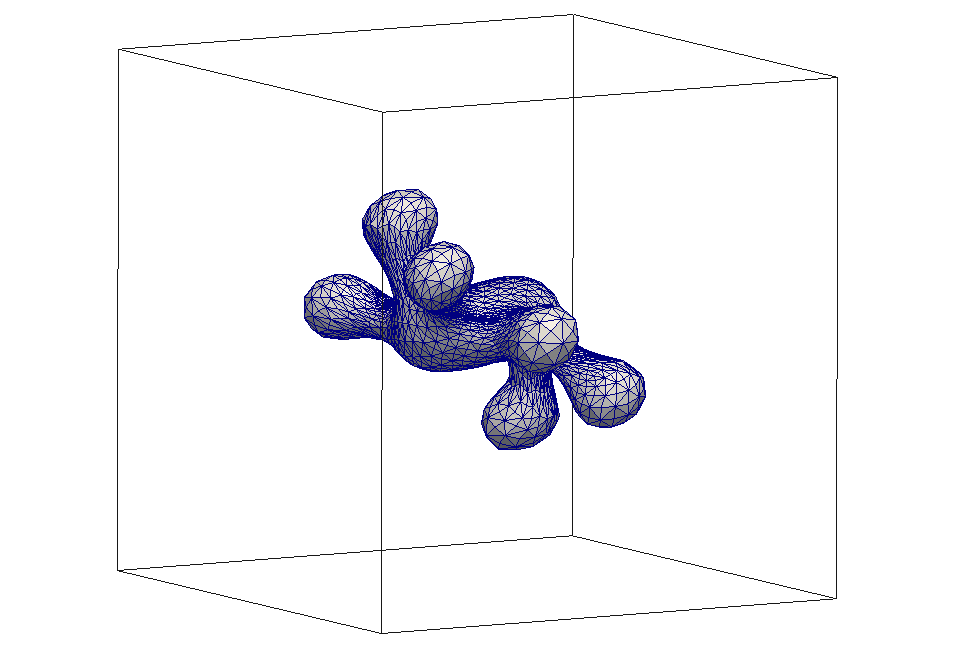} 
\includegraphics[angle=-0,width=\localwidth]{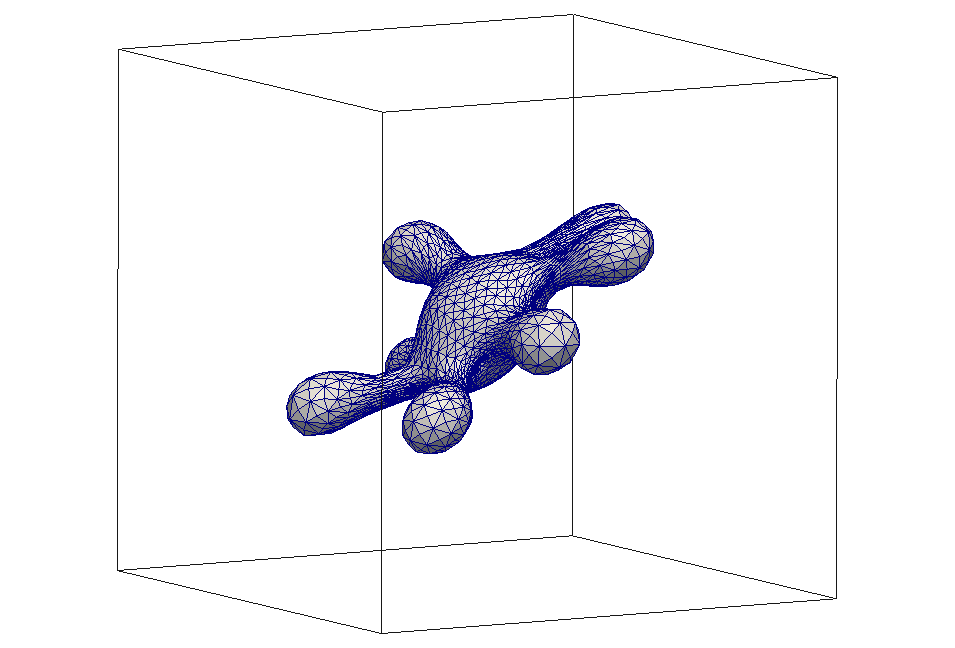} 
\includegraphics[angle=-0,width=\localwidth]{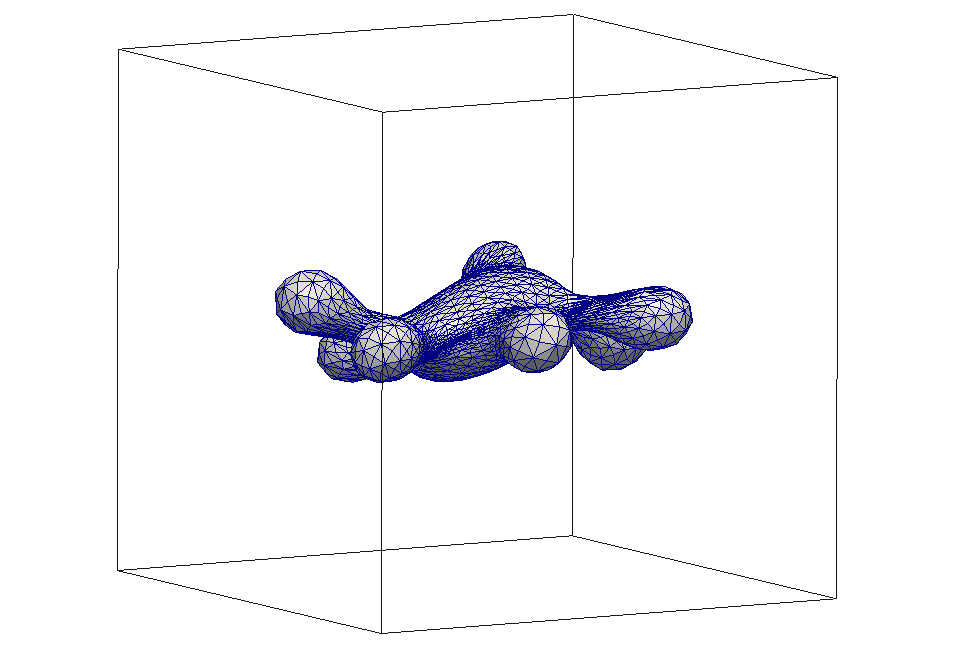} 
\caption{(Color online)
Shear flow for a budding shape with $\Lambda= \mu_\Gamma^* = 1$. Here
$\beta^* = 0.05$ and $M_0^* = 180$.
The plots show the interface $\Gamma^h$ within $\overline\Omega$
at times $t=0,\ 2.5,\ 5,\ 7.5,\ 10,\ 12.5,\ 15,\ 17.5$.
}
\label{fig:7armshear1}
\end{figure*}

\begin{figure*}
\center
\includegraphics[angle=-0,width=\localwidth]{figures/7armsADE_r0_m1_m1_t0} 
\includegraphics[angle=-0,width=\localwidth]{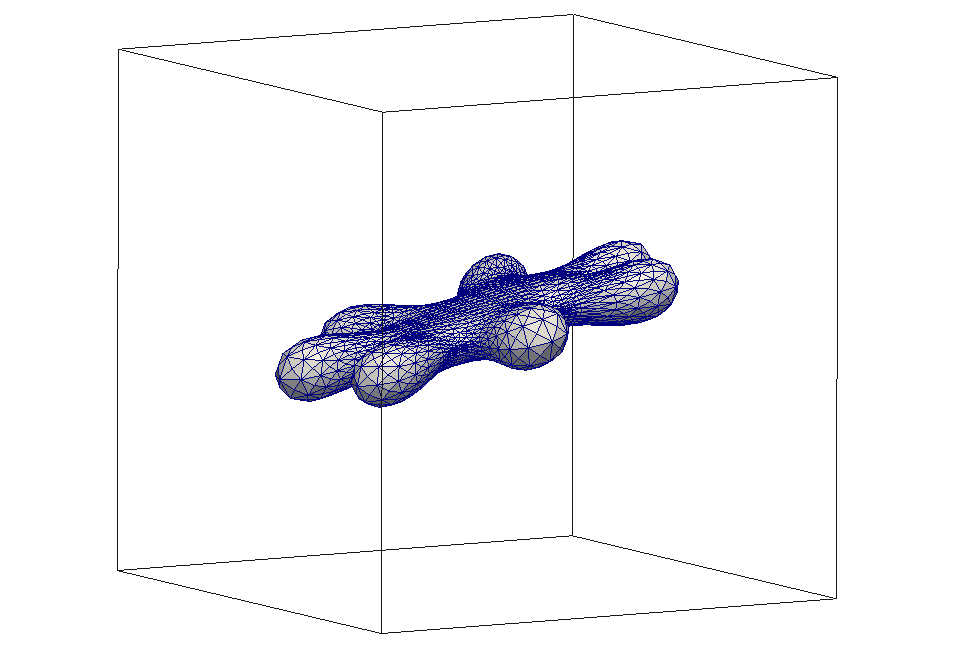} 
\includegraphics[angle=-0,width=\localwidth]{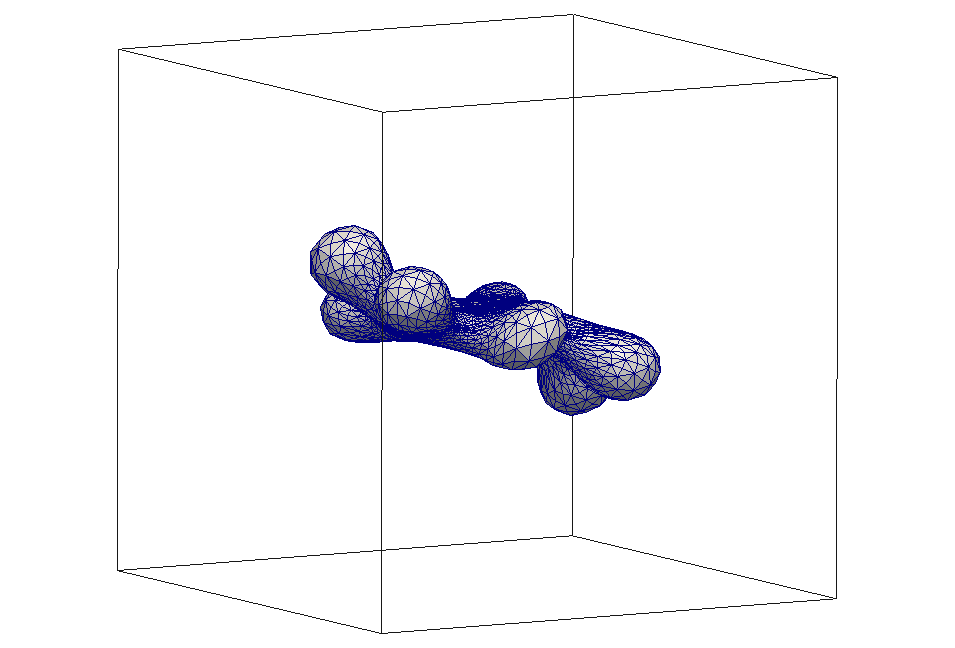} 
\includegraphics[angle=-0,width=\localwidth]{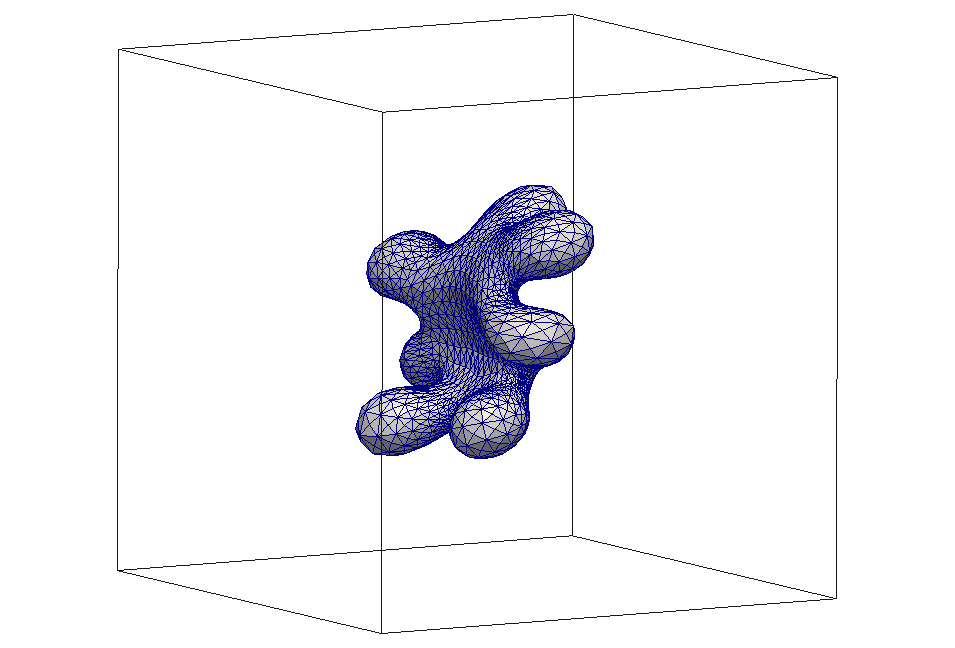} 
\\
\includegraphics[angle=-0,width=\localwidth]{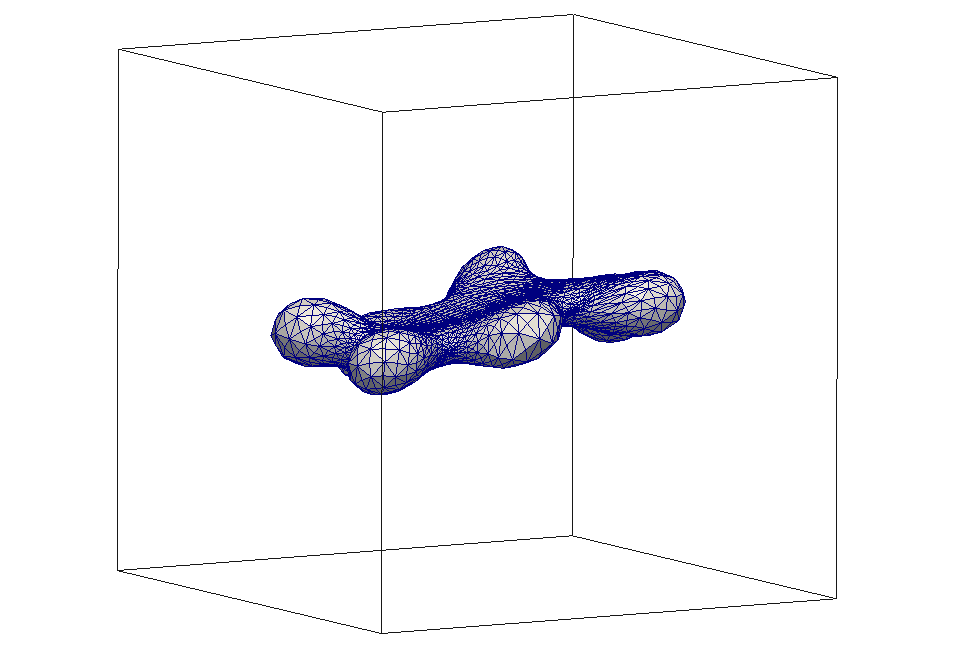} 
\includegraphics[angle=-0,width=\localwidth]{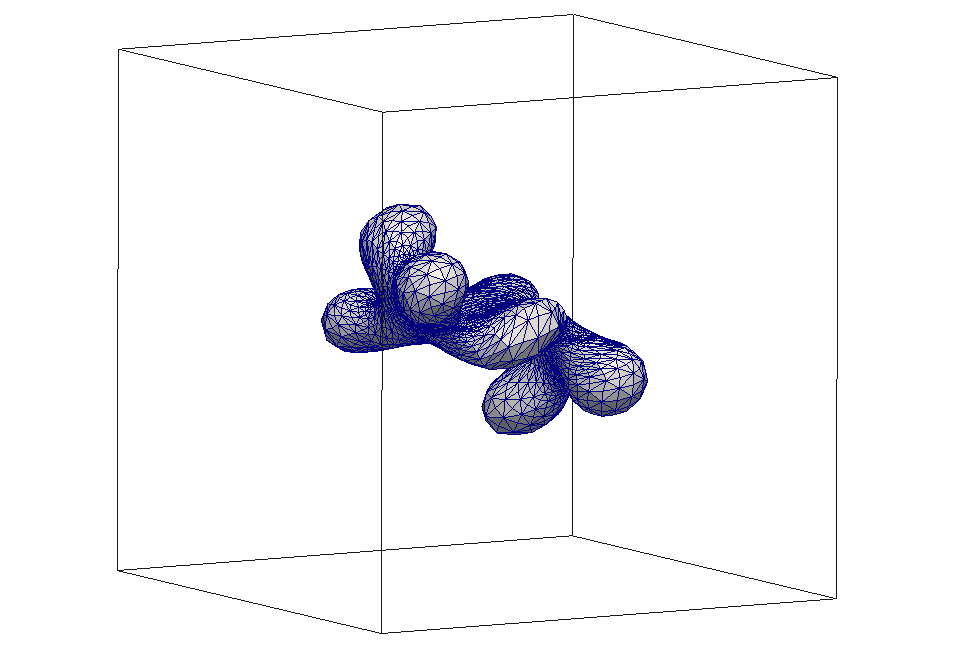} 
\includegraphics[angle=-0,width=\localwidth]{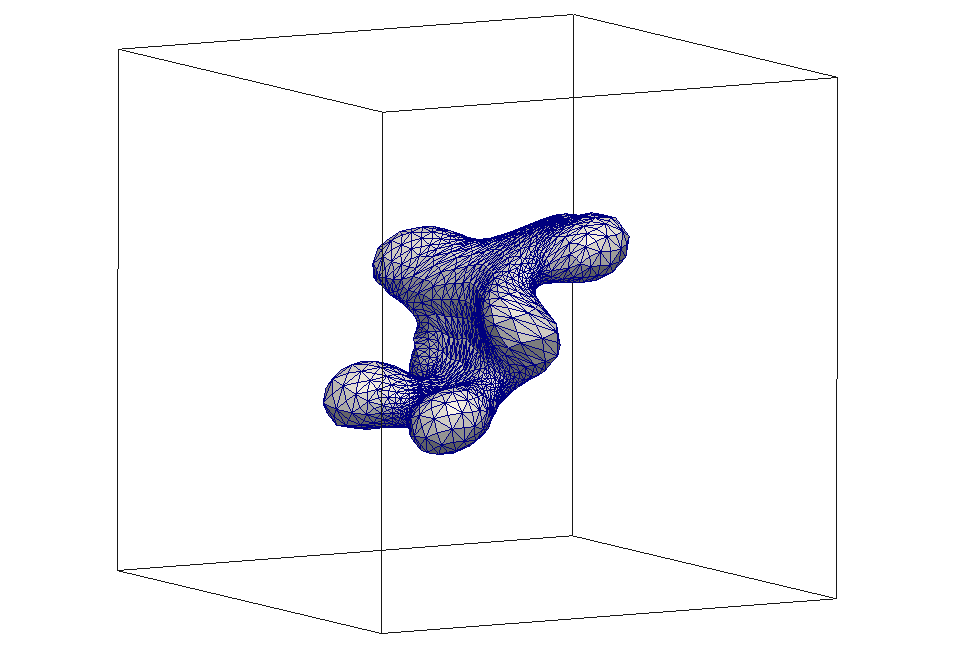} 
\includegraphics[angle=-0,width=\localwidth]{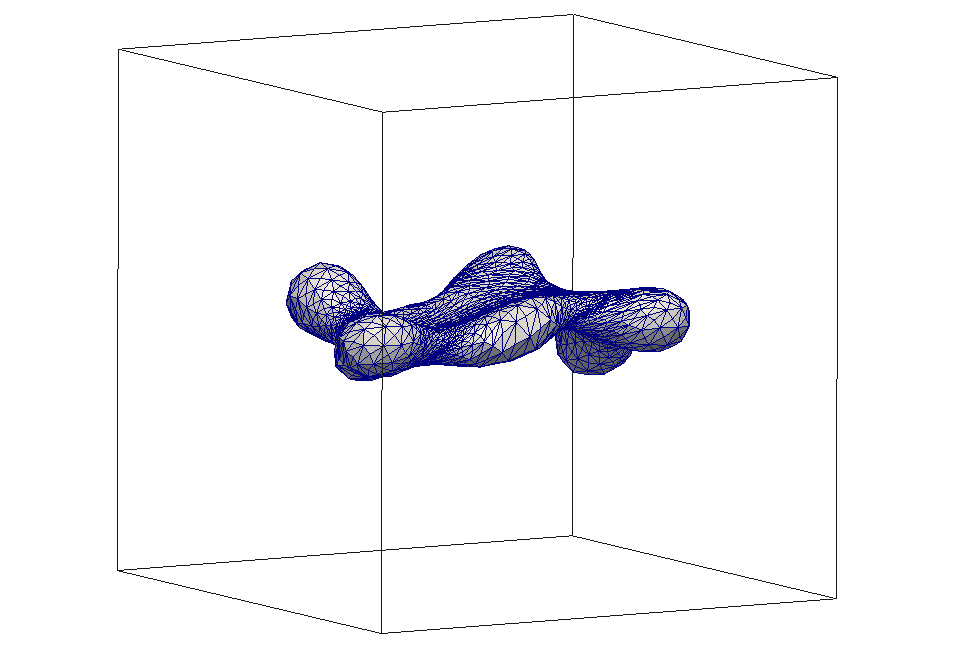} 
\caption{(Color online)
Same as Figure~\ref{fig:7armshear1} with $\beta^* = 0$.
}
\label{fig:7armshear1_noade}
\end{figure*}

We consider simulations for a scaled version of the final shape from 
\citet[Fig.\ 23]{willmore} with reduced volume $\mathcal{V}_r = 0.38$
and $\mathcal{A}(0) = 10.54$, so that $S = 0.92$, 
inside the domain $\overline\Omega = [-2,2]^3$.
We set $\Lambda = \mu_\Gamma^*=\alpha^*=1$.
In order to maintain the seven-arm shape during the evolution we set 
$\beta^* = 0.05$ and $M_0^* = 180$.
The first experiment is for no-slip boundary conditions on $\partial\Omega$ and
shows that the seven arms grow slightly, see Figure~\ref{fig:7arms}.
If we use the shear flow boundary conditions, 
on the other hand, we observe the behaviour in Figure~\ref{fig:7armshear1},
where we have changed the value of $\alpha^*$ to $0.05$. 
The vesicle can be seen tumbling, with a tumbling period of about $7$, with 
the seven arms remaining intact throughout.
Repeating the same experiment with $\beta^*=0$ yields the simulation in
Figure~\ref{fig:7armshear1_noade}. Not surprisingly, some of the arms of the
vesicle are disappearing.
We also tried to investigate whether the arms enhance or inhibit the tumbling
behaviour of the vesicle. To this end, we repeated the simulation in 
Figure~\ref{fig:7armshear1} for an ellipsoidal vesicle with the same reduced
volume and surface area. This vesicle also exhibited TU with a tumbling period
of about $7$, so there was no significant change to the behaviour in
Figure~\ref{fig:7armshear1}.

\section{Conclusions}
We have introduced a parametric finite element method for the evolution of 
bilayer membranes by coupling a general curvature elasticity model for the 
membrane to (Navier--)Stokes systems in the two bulk phases and to a surface 
(Navier--)Stokes system. The model is based on work by \citet{ArroyoS09},
which we generalized such that area difference elasticity effects (ADE) 
are taken into account. 
Our main purpose was to study the influence of the area difference
elasticity and of the spontaneous curvature on the evolution of the 
membrane. 
In contrast to most other works, we discretized the full
bulk \mbox{(Navier--)}Stokes systems coupled to the surface 
(Navier--)Stokes system 
and for the first time coupled this to a bending energy involving 
ADE and spontaneous curvature. 

The numerical simulations led to the following findings.

\begin{itemize}
\item The proposed numerical method conserves the volume enclosed by the 
membrane and the surface area of the membrane to a high precision.
\item The transition from a tank treading (TT) motion to a transition motion 
(TR) and to a tumbling (TU) motion depended strongly on
the surface viscosity. We observed that the surface viscosity alone with no 
viscosity contrast between inner and outer fluid can lead to
a transition from tank treading to a TR-motion and to
tumbling. Similar observations have been reported by
\cite{NoguchiG05a} using a particle-based method. 
\item The surface viscosity at which a transition between the different 
motions TT, TR and TU occur, strongly depends on the spontaneous
curvature and on the initial alignment of the vesicle. 
In particular, we observed that for negative spontaneous
curvature and an initial biconcave vesicle aligned such that the shortest axis
is in the shear flow direction all transitions 
occurred for larger values of the 
surface viscosity. For this alignment, and for positive spontaneous curvature, 
we observed that tumbling 
occurred already for much smaller values of the surface viscosity. 
The reverse was true for an alternative alignment.
Here we
recall that our sign convention for curvature means that spheres have negative
mean curvature.
\item In some cases, shear flow can lead to drastic shape changes, 
in particular for the ADE-model. For example, we observed the transition of a 
budded pear-like shape to a cup-like stomatocyte shape in shear flow if an 
ADE-model was used for the curvature elasticity.
\item The ADE-model can also lead to starfish-type 
shapes with several arms, see e.g.\ \cite{WintzDS96, Seifert97}.
In computations for a seven-arm starfish for a model involving an 
ADE 
type energy, we observed that in shear flow the overall
structure seems to be quite robust. In particular, the seven arms deformed 
but remained present even in a tumbling motion. However, arms tend to 
disappear if the area difference elasticity term is neglected. 
\end{itemize}
Thus we have shown that the proposed numerical method is a robust tool to 
simulate bilayer membranes for quite general models which in particular take 
the full hydrodynamics and a curvature model involving area difference 
elasticity and spontaneous curvature into account.  

\bigskip
\noindent 
{\bf Acknowledgement.} 
The authors gratefully acknowledge the support of the Deutsche
Forschungsgemeinschaft via the SPP 1506 entitled ``Transport processes at
fluidic interfaces'' and of the Regensburger Universit\"atsstiftung
Hans Vielberth.


\def\soft#1{\leavevmode\setbox0=\hbox{h}\dimen7=\ht0\advance \dimen7
  by-1ex\relax\if t#1\relax\rlap{\raise.6\dimen7
  \hbox{\kern.3ex\char'47}}#1\relax\else\if T#1\relax
  \rlap{\raise.5\dimen7\hbox{\kern1.3ex\char'47}}#1\relax \else\if
  d#1\relax\rlap{\raise.5\dimen7\hbox{\kern.9ex \char'47}}#1\relax\else\if
  D#1\relax\rlap{\raise.5\dimen7 \hbox{\kern1.4ex\char'47}}#1\relax\else\if
  l#1\relax \rlap{\raise.5\dimen7\hbox{\kern.4ex\char'47}}#1\relax \else\if
  L#1\relax\rlap{\raise.5\dimen7\hbox{\kern.7ex
  \char'47}}#1\relax\else\message{accent \string\soft \space #1 not
  defined!}#1\relax\fi\fi\fi\fi\fi\fi}
\end{document}